\documentclass[conference]{IEEEtran}
\IEEEoverridecommandlockouts

\usepackage{cite}
\usepackage{amsmath,amssymb,amsfonts}
\usepackage{graphicx}
\usepackage{textcomp}
\usepackage{xcolor}
\usepackage[most]{tcolorbox}
\usepackage{comment}
\usepackage{makecell} 
\usepackage{pifont}
\usepackage{amsfonts}
\usepackage{amsmath}

\usepackage{array}
\usepackage{balance}
\usepackage{booktabs}
\usepackage{colortbl}
\usepackage{comment}
\usepackage{csvsimple}
\usepackage{epsfig}
\usepackage{float}
\usepackage{flushend}
\usepackage{framed}
\usepackage{glossaries}
\usepackage{graphicx}
\usepackage{hhline}
\usepackage{hyperref}
\usepackage{latexsym}
\usepackage{listings}
\usepackage{makecell}
\usepackage{marvosym}
\usepackage{mathrsfs}
\usepackage{multirow}
\usepackage{paralist}
\usepackage{pgfplots}
\usepackage{pifont}
\usepackage{soul}
\usepackage{subfigure}
\usepackage{textcomp}
\usepackage{threeparttable}
\usepackage{tikz}
\usepackage{tikz-cd}
\usepackage{verbatim}
\usepackage{wrapfig}
\usepackage{xcolor}
\usepackage{xparse}
\usepackage{xspace}
\usepackage[color,matrix,arrow,all]{xy}
\usepackage{tabularx}
\usepackage{enumitem}
\usepackage{longtable}
\usepackage{hhline} 

\usepackage[algo2e]{algorithm2e} 

\usepackage[export]{adjustbox}

\definecolor{green}{RGB}{0,128,0}

\definecolor{yellow}{RGB}{255,200,18}

\newcommand{\revision}[1]{\textcolor{black}{{#1}}}

\newcommand{\bi}{\begin{itemize}}
\newcommand{\ei}{\end{itemize}}

\newcommand{\be}{\begin{enumerate}}
\newcommand{\ee}{\end{enumerate}}
\newcommand{\beqn}{\begin{eqnarray*}}
\newcommand{\eeqn}{\end{eqnarray*}}

\newcommand{\stitle}[1]{\vspace*{1.2pt}\noindent{\bf #1}}

\newcommand{\eg}{{e.g.,}\xspace}

\NewDocumentCommand{\mourad}{ mO{} }{\textcolor{blue}{\textsuperscript{\textit{Mourad}}\textsf{\textbf{\small[#1]}}}}
\NewDocumentCommand{\hani}{ mO{} }{\textcolor{blue}{\textsuperscript{\textit{Hani}}\textsf{\textbf{\small[#1]}}}}
\NewDocumentCommand{\mohamed}{ mO{} }{\textcolor{orange}{\textsuperscript{\textit{Mohamed}}\textsf{\textbf{\small[#1]}}}}
\NewDocumentCommand{\shahmeer}{ mO{} }{\textcolor{blue}{\textsuperscript{\textit{Shahmeer}}\textsf{\textbf{\small[#1]}}}}
\NewDocumentCommand{\mike}{ mO{} }{\textcolor{blue}{\textsuperscript{\textit{Mike}}\textsf{\textbf{\small[#1]}}}}

\newcommand{\system}{\textsc{HCT-QA}\xspace}
\newcommand{\hct}{{HCT}\xspace}
\newcommand{\hcts}{{HCTs}\xspace}
\newcommand{\nltosql}{{NL-to-SQL}\xspace}

\makeglossaries

\newglossaryentry{refNamePlural}
{
    name=\textit{HCTs},
    description={}
}

\newglossaryentry{refNameSingleton}
{
    name=\textit{HCT},
    description={}
}
\begin{document}

\def\BibTeX{{\rm B\kern-.05em{\sc i\kern-.025em b}\kern-.08em
    T\kern-.1667em\lower.7ex\hbox{E}\kern-.125emX}}

\title{[Experiment, Analysis, and Benchmark] \system: A Benchmark for Question Answering on Human-Centric Tables}

\author{
\IEEEauthorblockN{Mohammad S. Ahmad}
\IEEEauthorblockA{\textit{QCRI, HBKU} \\
Doha, Qatar \\
mohammadshahmeerah@hbku.edu.qa}
\and
\IEEEauthorblockN{Zan Naeem\textsuperscript{*}}
\IEEEauthorblockA{\textit{Meta} \\
Palo Alto, USA \\
znaeem@meta.com}
\and
\IEEEauthorblockN{Michael Aupetit}
\IEEEauthorblockA{\textit{QCRI, HBKU} \\
Doha, Qatar \\
maupetit@hbku.edu.qa}
\and
\IEEEauthorblockN{Ahmed Elmagarmid}
\IEEEauthorblockA{\textit{QCRI, HBKU} \\
Doha, Qatar \\
aelmagarmid@hbku.edu.qa}
\and
\IEEEauthorblockN{Mohamed Eltabakh}
\IEEEauthorblockA{\textit{QCRI, HBKU} \\
Doha, Qatar \\
meltabakh@hbku.edu.qa}
\and
\IEEEauthorblockN{Xiaosong Ma\textsuperscript{*}}\thanks{\textsuperscript{*} Work done while at QCRI.}
\IEEEauthorblockA{\textit{MBZUAI} \\
Abu Dhabi, UAE \\
Xiaosong.Ma@mbzuai.ac.ae}
\and
\IEEEauthorblockN{Mourad Ouzzani}
\IEEEauthorblockA{\textit{QCRI, HBKU} \\
Doha, Qatar \\
mouzzani@hbku.edu.qa}
\and
\IEEEauthorblockN{Chaoyi Ruan\textsuperscript{*}}
\IEEEauthorblockA{\textit{NUS} \\
Singapore \\
ruancy@comp.nus.edu.sg}
\and
\IEEEauthorblockN{Hani Al-Sayeh}
\IEEEauthorblockA{\textit{QCRI, HBKU} \\
Doha, Qatar \\
halsayeh@hbku.edu.qa}
}

\maketitle

\newcommand\msa[1]{\textcolor{magenta}{#1}}
\newcommand\mac[1]{\textcolor{orange}{MA: #1}} 
\newcommand\ma[1]{\textcolor{orange}{#1}} 
\newcommand\ttt[1]{\texttt{#1}}



\begin{abstract}
Tabular data embedded in PDF files, web pages, and other types of documents is prevalent in various domains.
These tables, which we call human-centric tables (\textit{HCTs} for short), are dense in information but often exhibit complex structural and semantic 
layouts. 
To query these \hcts, 
some existing solutions focus on transforming them 
into relational formats. However, they fail to handle the diverse and complex layouts of \hcts, making them not amenable to easy querying with SQL-based approaches. Another emerging option is to use Large Language Models (LLMs) and Vision Language Models (VLMs). 
However, there is a lack of standard evaluation benchmarks to measure and compare the performance of models to query \hcts using natural language.
To address this gap, 
we propose the Human-Centric Tables Question-Answering extensive benchmark (\textit{HCT-QA}) consisting of thousands of \hcts with several thousands of natural language questions with their respective answers. 
More specifically, HCT-QA 
includes 1,880 real-world \hcts with 9,835 QA pairs in addition to 4,679 synthetic \hcts with 67.7K QA pairs.
Also, we show through 
extensive experiments the performance 
of 25 and 9 different LLMS and VLMs, respectively, in an answering \system's questions. 
In addition, 
we show how finetuning an LLM on \system improves F1 scores by up to 25\ percentage points 
compared to the off-the-shelf model.   
Compared to existing benchmarks, \system stands out for its 
broad complexity and diversity of covered \hcts and generated questions, 
its comprehensive metadata enabling deeper insight and analysis, and its novel synthetic data and QA generator.
%
\end{abstract}

\begin{IEEEkeywords}
Human-Centric Tables, LLMs, Data Querying
\end{IEEEkeywords}

\section{Introduction}
\label{section:introduction}

Answering natural language (NL) questions  
over structured and semi-structured data is a fundamental problem 
\cite{wang2020natural,song2025fevisqa,acharya2022question,peng2024live}. The task of \textbf{Table QA} focuses on retrieving or computing answers to NL questions grounded in tabular data~\cite{tapas-paper, yu-etal-2018-spider, valuenet-icde-paper, garnl2sql-paper, ratsql-paper, catsql-paper}. 
Early studies 
used primarily relational tables 
from 
Wikipedia and 
widely used databases~\cite{yu-etal-2018-spider, birdbench-paper, wikitablequestions_2017}. 
Over time, research has evolved in two 
directions: 
(1)~Direct  
question answering using neural models~\cite{tapas-paper, tapex-paper, omnitab-paper}, and 
(2) 
translating NL questions into SQL queries (\nltosql)~\cite{NLtoSQLsurvey2024, valuenet-icde-paper, garnl2sql-paper, ratsql-paper, catsql-paper}. 
Despite increasing attention to this area, 
existing efforts remain concentrated on relational or relational-like table structures.

\begin{figure*}[t]
 \centering
    \includegraphics[width=0.82\textwidth]{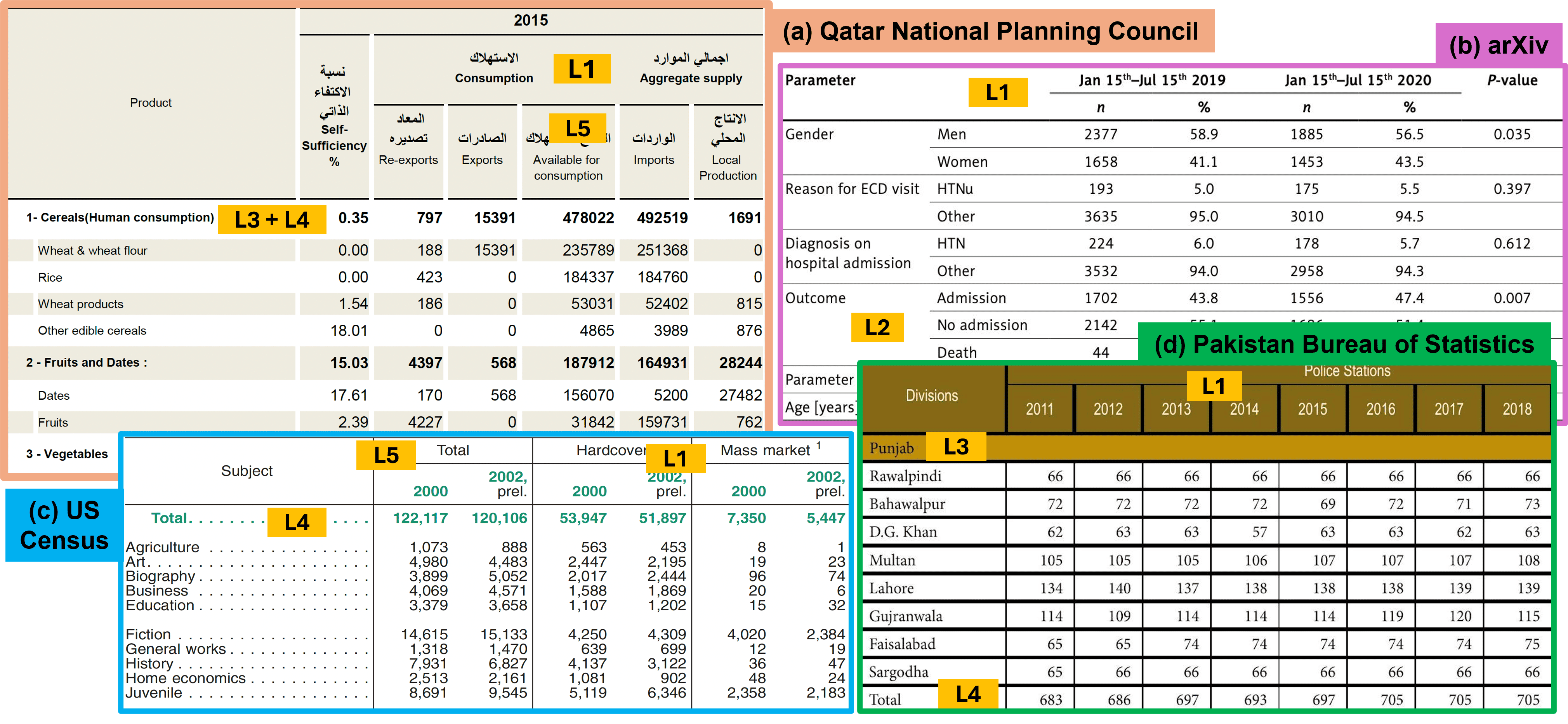}
    \vspace{-7pt}
      \caption{Examples of Human-Centric Tables (\hcts) from different real-world data sources exhibiting various levels of structural and semantic complexity including column nesting (L1), row nesting (L2), row group labels (L3), row aggregation (L4), and column aggregation (L5)}
      \vspace{-11pt}
     \label{fig:HCT-examples}
 \end{figure*}

Tabular data embedded 
in PDFs files, 
web pages, and other 
types of documents 
is 
widespread across many domains.
Figure~\ref{fig:HCT-examples} illustrates 
examples of such data obtained from the Qatar National Planning Council~\cite{qnpc}, scientific 
publication archives~\cite{arxiv}, the US Census~\cite{uscensus}, and the Pakistan Bureau of Statistics~\cite{pakcensus}. 
These tables employ intricate structural and visual layouts to 
convey large volumes of 
information 
in a concise representation to support 
strategic decision-making. 

Unlike traditional relational tables, which have a simple layout, document-embedded tables are primarily designed for human readability
rather than automated processing. We refer to this form of tabular data 
as \textbf{Human-Centric Tables (\hcts)}. \hcts represent a distinct data modality, characterized by intricate layouts, rich semantic content, and significant practical value. 
However, systematically processing \hcts for tasks such as information extraction, interpretation, 
and question answering remains a major challenge.

With the rapid advances in multimodal document understanding~\cite{shen2024ambiguous, lin2025querying, lu2012dataset, li2023toward}, 
the scope of Table QA is progressively extending beyond 
relational 
tables~\cite{catsql-paper, garnl2sql-paper, mitsopoulou2025analysis}. 
This evolution warrants the need for not only systems capable of robust 
question answering over \hcts, but also 
methodologies and benchmarks that enable rigorous evaluation and identification of existing limitations of these systems while answering questions on \hcts. 
Although a few benchmarks have been recently introduced, 
they suffer from key shortcomings—most notably, the absence of 
metadata-driven analyses (i.e., analyses incorporating \hcts and question characteristics),
and limited scalability 
due to the reliance on extensive 
manual effort. 

\textbf{Properties and Challenges of \hcts.}
\hcts 
exhibit highly diverse structural features 
designed for human interpretation. 
Figure~\ref{fig:HCT-examples} showcases common 
structural properties 
observed in \hcts 
including 
\textit{column nesting} (L1), \textit{row nesting} (L2), and \textit{row group labels} (L3), as well as \textit{embedded aggregations} that may appear \textit{row-wise} (L4) or \textit{column-wise} (L5).
Visual cues-such as variations in text or row color (e.g., L3 in Table (d) and L4 in Table (c)), bold or emphasized text (e.g., L4 in Table (a) and Table (c)), and the positioning of elements within cells (e.g., L2 in Table (b)) convey additional semantic information. 

These structural and visual complexities make direct question answering with neural QA models 
challenging. Also, as 
\hcts are not inherently relational, SQL queries cannot be executed on them directly.
Existing efforts, such as AutoTables~\cite{AutoTables_57}, have attempted to transform HCT-like tables into relational formats. However, such approaches are limited to 
relatively simple layouts, often struggle on 
complex real-world \hcts. 
Consequently, the pipeline that first converts \hcts to relational tables and then applies NL to SQL transformations 
remains fragile and error-prone 
\cite{kim2020natural, catsql-paper, garnl2sql-paper, mitsopoulou2025analysis}. 

\textbf{New Opportunities with LLMs and VLMs.}
Recent advances in Large Language Models (LLMs) and Vision Language Models (VLMs)  pave the way for a new  {\em LLM-oriented} approach for querying \hcts. 
This approach can 
(1)~efficiently handle \hcts alongside other human-centric data formats (e.g., charts 
and images), 
(2)~avoid the complex and error-prone NL-to-SQL pipeline, and 
(3)~directly 
understand NL questions.
However, a systematic evaluation to identify the strengths, weaknesses, and potential improvements of LLM-oriented approaches remains largely an uncharted area. 
To address this gap, we propose \system, a novel benchmark designed to assess LLMs and VLMs as the engines for 
querying \hcts.
\system 
can be also used to evaluated other types of data processing engines designed to answer NL questions on \hcts.

\begin{table*}[t]
\centering
\caption{Comparison of major Table QA benchmarks with \system.}
\vspace{-7pt}
\label{tab:benchmark_comparison}
\resizebox{\textwidth}{!}{%
\begin{tabular}{lcccccccccc}
\toprule
\textbf{Benchmark Name} & 
\textbf{Data Type} &  
\makecell{\textbf{Num HCT}\\\textbf{Tables}} & 
\textbf{Table Modalities} &  
\textbf{Num Sources} & 
\textbf{Num QA Pairs} & 
\makecell{\textbf{Table}\\\textbf{Metadata}} &
\makecell{\textbf{QA}\\\textbf{Metadata}} &
\makecell{\textbf{Avg. Questions}\\\textbf{Per Table}} &  
\makecell{\textbf{Question Syntactic}\\\textbf{Similarity}} \\ 
\midrule

HiTab \revision{\cite{cheng-etal-2022-hitab}} & Tables & 3,597 & JSON & 3 & 10,672 & \ding{55} & \ding{55} & 3 & 0.45 \\

MultiHiertt \revision{\cite{zhao-etal-2022-multihiertt}} & Tables + Text & 9,842 & HTML & 2 & 10,440 & \ding{55} & \ding{55} & N/A & 0.23 \\

TableVQA-Bench \revision{\cite{tablevqa_2024}} & Tables & 205 & HTML, Image & 1 & 250 & \ding{55} & \ding{55} & 1.2 & 0.186 \\

TAT-QA \revision{\cite{tatqa2021}} & Tables + Text & 2,757 & 2D List & 1 & 12,650 & \ding{55} & \ding{55} & N/A & 0.2879 \\[3pt]

\textbf{HCT-QA (Ours)} & \textbf{Tables} & \textbf{6,559} & \makecell{\textbf{HTML, CSV,}\\\textbf{Image, MD}} & \textbf{4 + Synthetic} & \textbf{77,582} & \checkmark & \checkmark & \textbf{11} & \textbf{0.325} \\

\bottomrule
\end{tabular}
}
\vspace{-11pt}
\end{table*}

{\textbf{Challenges and Contributions.}}
Devising the proposed benchmark involves several challenges:
(1)~Collecting and generating \hcts with an extensive diversity of 
complex layouts (Figure~\ref{fig:HCT-examples}). 
(2)~Constructing an extensive variety of NL questions ranging from simple selection to complex aggregation and ranking questions.
(3)~Obtaining 
accurate ground truth answers, 
and developing meaningful evaluation metrics to compare them to LLMs' answers. 
%
(4)~Exploring the impact of LLM-specific strategies such as finetuning 
on answering questions on HCTs. 
%
%
The key contributions of \system are:
\begin{itemize}
\item A large set of \hcts and question-answer pairs (QA) consisting of 1,880 \hcts from four real-world data sources spanning diverse domains. \hcts are provided in various formats including images, CSV,  HTML, and Markdown. We also provide 9,835 QA pairs crafted by experts and verified by annotators (Section~\ref{section:tableProperties}).

\item Extensive manually-obtained metadata of the \hcts and QA pairs to capture the most common properties 
of \hcts and questions. This metadata allows for a deeper understanding of LLM/VLM 
performance and a better identification of performance gaps 
(Section~\ref{section:queryProperties}).

\item A novel synthetic \hct generator to complement the real-world datasets. We used the generator to produce 4,679 \hcts and 67,747 template-based QA pairs across seven semantic domains. The generator offers flexible configuration  making it a valuable and practical resource for the research community  (Section~\ref{section:synthetic}).

\item An extensive evaluation of $25$ LLMs and $9$ VLMs of different sizes ranging from $3B$ to $100B+$ parameters. 
Our evaluation reveals that large closed-weight models perform the best overall, the promise of processing \hct images directly using VLMs, and the effectiveness of finetuning some models using \system (Section \ref{section:evaluation2}). 

\end{itemize}


Our benchmark is publicly available on HuggingFace~\cite{HCT_QA_hf}. All experimentation code and the synthetic HCT generator are publicly available on our GitHub repository~\cite{hctqa_repo}.

\section{Related work}
\label{section:relatework}

Several research and commercial efforts addressed  
document understanding and information extraction from complex documents 
\cite{shen2024ambiguous, lin2025querying, li2023toward}. 
However, none of these fully address the core question {\em What are the opportunities and challenges of using LLM-based approaches to query \hcts?}

\textbf{Table extraction techniques}
in several commercial tools~\cite{aspose,azure_document_intelligence,innodata,docugami} and recent research~\cite{burdick2020table, tableformer_2022, robustTableDetection_Ma_2022, omniParser_2024,  smolDocling_2025, Khan2021ICDAR_13, Oro2008ICDAR_14, 
Qiao2021ICDAR_22} focus on extracting tables similar to \hcts from PDFs into machine-readable formats such as CSV 
and Markdown. However, they do not provide end-users any systematic support for querying, analyzing, or discovering the extracted tables. 
Also, our analysis shows that VLMs can 
bypass this extraction step by operating directly on the image version of \hcts.

\textbf{Table QA approaches} 
focus on relational tables as they 
use \nltosql to answer the questions~\cite{Weir2020ACM_25, Basik2018ACM_26, Wang2019_27, Brunner2021ValueNetAN_28, sen2020VLDB_29,liu2024surveynl2sqllargelanguage}. However, SQL queries cannot be executed on tables with complex layouts like \hcts. 
As mentioned above, AutoTables~\cite{AutoTables_57} fail to achieve high accuracy on complex real-world \hcts.
Other approaches 
tried to perform repairs and transformations on non-relational tables but still do not present a means to convert a \hct into a relational table~\cite{10.1431751.3231766, 10.147407790.3407831}. 
Such a multi-step, database-oriented pipeline-spanning \hcts extraction, schema discovery, and \nltosql transformation introduces several error-prone phases and could be an overkill for practical applications. Other non-\nltosql related approaches such as direct inference using deep learning models have again been restricted to 
handling relational 
tables~\cite{zhang-etal-2025-tablellm, tapas-paper, tapex-paper, chemmengath2021topictransferabletablequestion}.

\textbf{Table QA benchmarks} mostly focus on flat Wikipedia tables which, when compared to \hcts, are significantly simpler~\cite{tablebench_2025, wikitablequestions_2017, fetaqa_2021, tabfact_2020, ottqa_2021, hybridqa_2021, zhang2024reactable, zhu2024autotqa}. 
FinQA ~\cite{finqa_2022} consist of tables from the finance domain 
but specifically exclude tables with complex nested structures. 
There are also several benchmarks that cover only one source or domain 
and have a very small number of QA pairs and tables such as 
AIT-QA\cite{aitqa_2021} with only 515 QA pairs on 116 tables and TableVQA-Bench~\cite{tablevqa_2024} with only 250 QA pairs on 205 \hcts.
Some benchmarks focus on multi-modal QA where questions are on text, tables, images or other modalities~\cite{spiqa_neurips2024, tatdqa2022, zhao-etal-2022-multihiertt, scitat_2024, urban2023caesura, MMLongbenchdoc_neurips2024}. 
Compared to these benchmarks, \system focuses on complex-layout \hcts 
from 
diverse real-world sources, provides tens of thousands 
of QA pairs on thousands of \hcts 
without requiring metadata, 
and provides a synthetic \hct and QA pair generator for scalability. 

In Table \ref{tab:benchmark_comparison}, we present the key distinguishing dimensions of the related benchmarks.
It can be observed that each of them 
suffers from 
key limitations like 
a limited number of \hcts or QA pairs, a small number of questions per table, etc. 
Also, as the table shows, 
the most distinctive advantage of \system 
lies in its comprehensive table- and QA-level metadata, which enables fine-grained and metadata-driven analysis.  

In addition, to assess the diversity of questions, we introduced two statistics: the \textit{average number of questions} per table and the \textit{question syntactic similarity}.
The syntactic similarity is computed at the table level (for tables with more than one question) by calculating the pairwise word-level Jaccard similarity between all question pairs, followed by averaging the results.
As shown, \system exhibits a higher average number of questions per table and greater lexical diversity. 

Zooming-in on HiTab~\cite{cheng-etal-2022-hitab} (the closest to \system), 
we see that it includes \hcts from three sources and provides only 10,672 QA pairs.
In contrast, \system encompasses nearly twice as many \hcts drawn from a broader range of sources and offers over seven times more QA pairs.
Also, \system includes the visual representations (images) of the \hcts, thereby enabling evaluation 
of VLMs. 
Moreover, a closer examination of samples from existing benchmarks revealed several data quality issues. 
For instance, after manually inspecting 206 randomly selected QA samples from the HiTab benchmark, we found that 22\% of the ground-truth answers were incorrect, 8\% of the questions were irrelevant to their associated tables, 37\% contained minor grammatical errors, and 12\% included typographical mistakes.
In developing \system, 
we placed a strong emphasis on ensuring data quality through a comprehensive and rigorous verification process.

\section{Real-world datasets}
\label{section:real_data}
\begin{figure}[t]
    \centering
    \includegraphics[width=.95\linewidth]{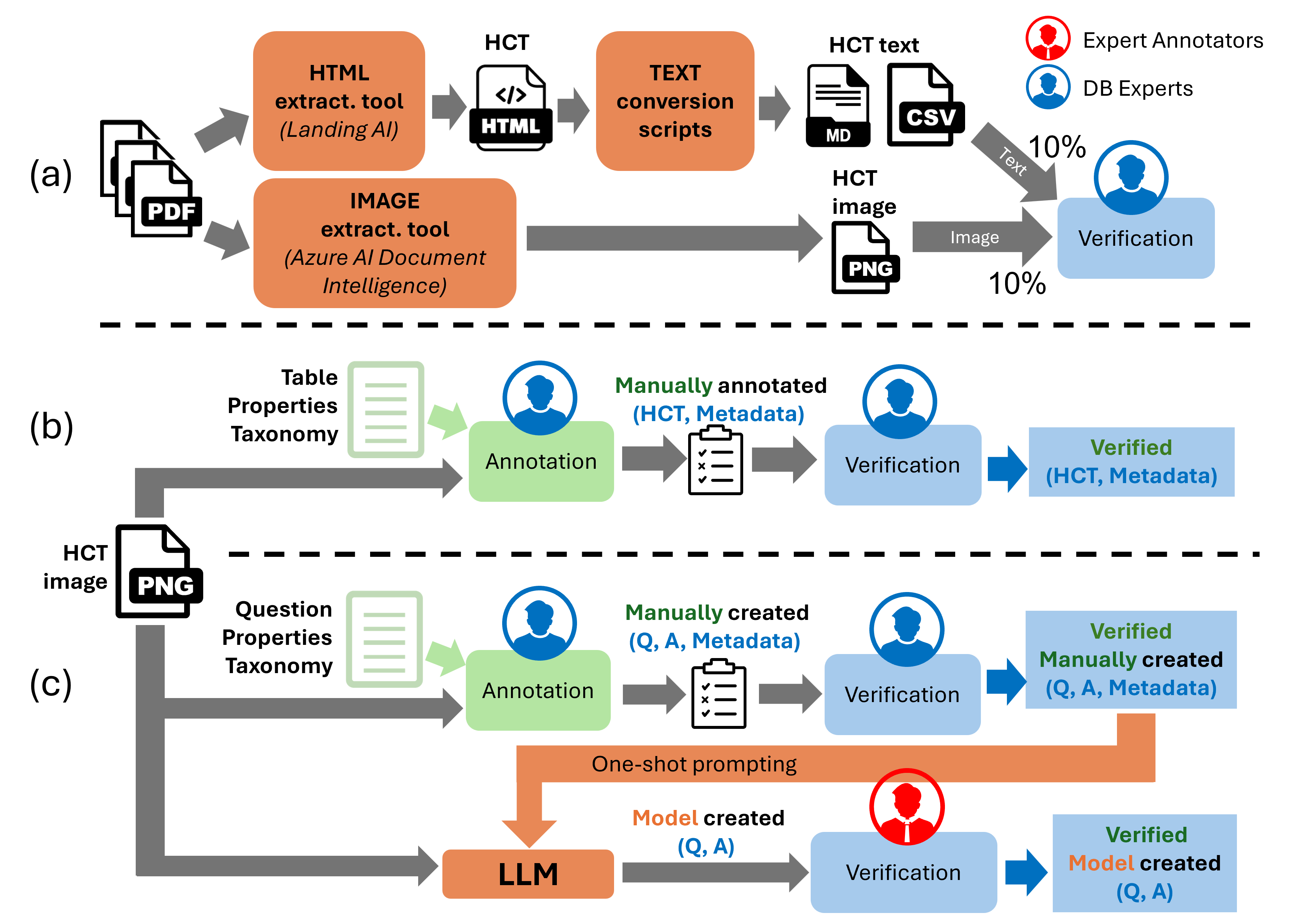}
    \vspace{-5pt}
    \caption{Preparing the real-world dataset. (a) Extracting \hcts. 
    (b) Annotating \hcts with table properties. 
    (c) Generating QA pairs 
    with question properties. 
    }
    \label{fig:preparing_the_real-world_dataset}
\vspace{-10pt}
\end{figure}

In this section, we present the real-world dataset. 
Each of its element 
consists of a quadruplet in the form of
$(T_{HCT},Q_{NL},A, M)$,  
where $T_{HCT}$ is an \hct object, $Q_{NL}$ is a NL  question,  $A$ is the ground truth answer, and $M$ is rich metadata about $T_{HCT}$ and $Q_{NL}$. 
In the following, 
we first present the details of the \hcts 
collected from thousands of PDFs across diverse domains (Section~\ref{section:tableProperties}), 
and then present the details of the QA collection process (Section~\ref{section:queryProperties}). 
Figure \ref{fig:preparing_the_real-world_dataset} illustrates the 
dataset preparation process.

\begin{table}[ht]
\centering
\renewcommand{\arraystretch}{1.2}
\caption{Definitions of \hct (table) properties.}
\vspace{-7pt}
\label{tab:hct_property_definitions}
\large  
\resizebox{1.0\linewidth}{!}{
\begin{tabular}{p{0.18\textwidth}p{0.82\textwidth}}

\hline
\rowcolor{green!30} \textbf{Property Name} & \textbf{Definition} \\
\hline

\hline
\rowcolor{green!20} Column Nesting & Headers/column names are spread across more than one level (row). \\
\hline
\, Balanced * & The depth of each branch within the nesting is the same throughout the table. \\
\hline
\,  Unbalanced & The depth of each branch within the nesting is not the same. \\
\hline
\,  Symmetric* & Child column names repeat uniformly across table columns where nesting is present. 
\\
\hline
\,  Asymmetric* & Child 
column names are different across the table in columns where nesting is present. \\
\hline
\rowcolor{green!20} Row Nesting & A cell spans multiple rows but is displayed as one merged cell. \\
\hline
\,  Balanced* & The depth of each row branch is the same. \\
\hline
\,  Unbalanced & The depth of each row branch is not the same. \\
\hline
\,  Symmetric* & Child cell values repeat across the table uniformly in rows where nesting is present. 
\\
\hline
\,  Asymmetric* & Child 
cell values are different across the table in rows where nesting is present. \\
\hline
\rowcolor{green!20} Col. Aggregation & Columns that aggregate all or a subset of other columns. \\
\hline
\,  Global* & A column that aggregates all other numerical columns in the table (excluding local-group aggregation columns). \\
\hline
\,  Local* 
& A column that aggregates a subset of numerical columns in the table. \\
\hline
\,  Explicit* & An aggregation column where its name clearly states it's an aggregation (e.g., total). 
\\
\hline
\,  Implicit* & An aggregation column where its name is ambiguous about whether it's an aggregation. \\
\hline
\rowcolor{green!20} Row Aggregation & Rows that aggregate all or a subset of other rows. \\
\hline
\,  Global* & A row aggregating all other numerical rows in the table (excluding local-group aggregation rows). \\
\hline
\,  Local* & A row that aggregates a subset of numerical rows in the table. \\
\hline
\,  Explicit* & A row where some textual cell value clearly states it's an aggregation (e.g., total) 
\\
\hline
\,  Implicit* & A row where all textual cell values are ambiguous about whether it's an aggregation. \\
\hline
\rowcolor{green!20} Others &\\
\hline
\,  Row Group Label & A row that acts as a heading/title for either all or a subset of rows below or above it. \\
\hline
\,  Split Header Cell & One cell that is split diagonally (visually) and contains two values. \\
\hline
\,  Stand. Rel. Table* & Table displays none of the other HCT properties (nesting, aggregations, labels). \\
\hline
\end{tabular}
}
\vspace{-5pt}
\end{table}


\subsection{Real-world HCTs} 
\label{section:tableProperties}

\textbf{Data Sources.}
We collected thousands of PDFs from diverse sources, including  
    {Qatar National Planning Council (QNPC)}~
    \cite{qnpc}
    {Research Papers} from \textit{ArXiv.org}, \textit{biorxiv.org}, and \textit{medarch.org}, 
    {US Census} covering PDF reports from the US Census Bureau~\cite{uscensus} 
    and {Pakistan Census} from the official website of the Pakistan government~\cite{pakcensus}. 
    These sources cover 
    various domains including  education, land usage, machine learning, data management, finance, astrophysics, materials science, cryptography, astrophysics, medicine, and security.   
    In total, we extracted \textbf{1,880} HCTs. 

\begin{figure*}[!t]
\vspace{-5pt}
\begin{center}
\resizebox{1\linewidth}{!}{
\begin{tabular}{cc}
\includegraphics[width=0.45\linewidth]
{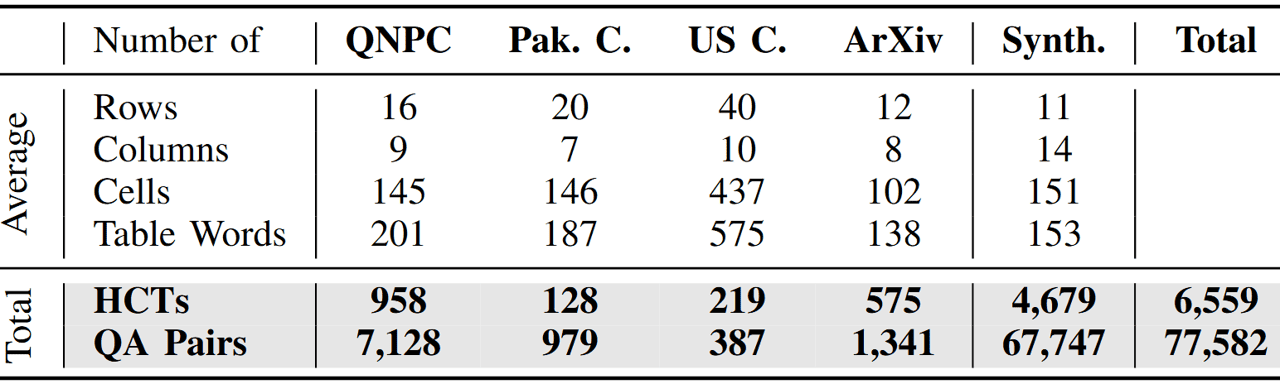}&\includegraphics[width=0.4\linewidth, height=100pt]
{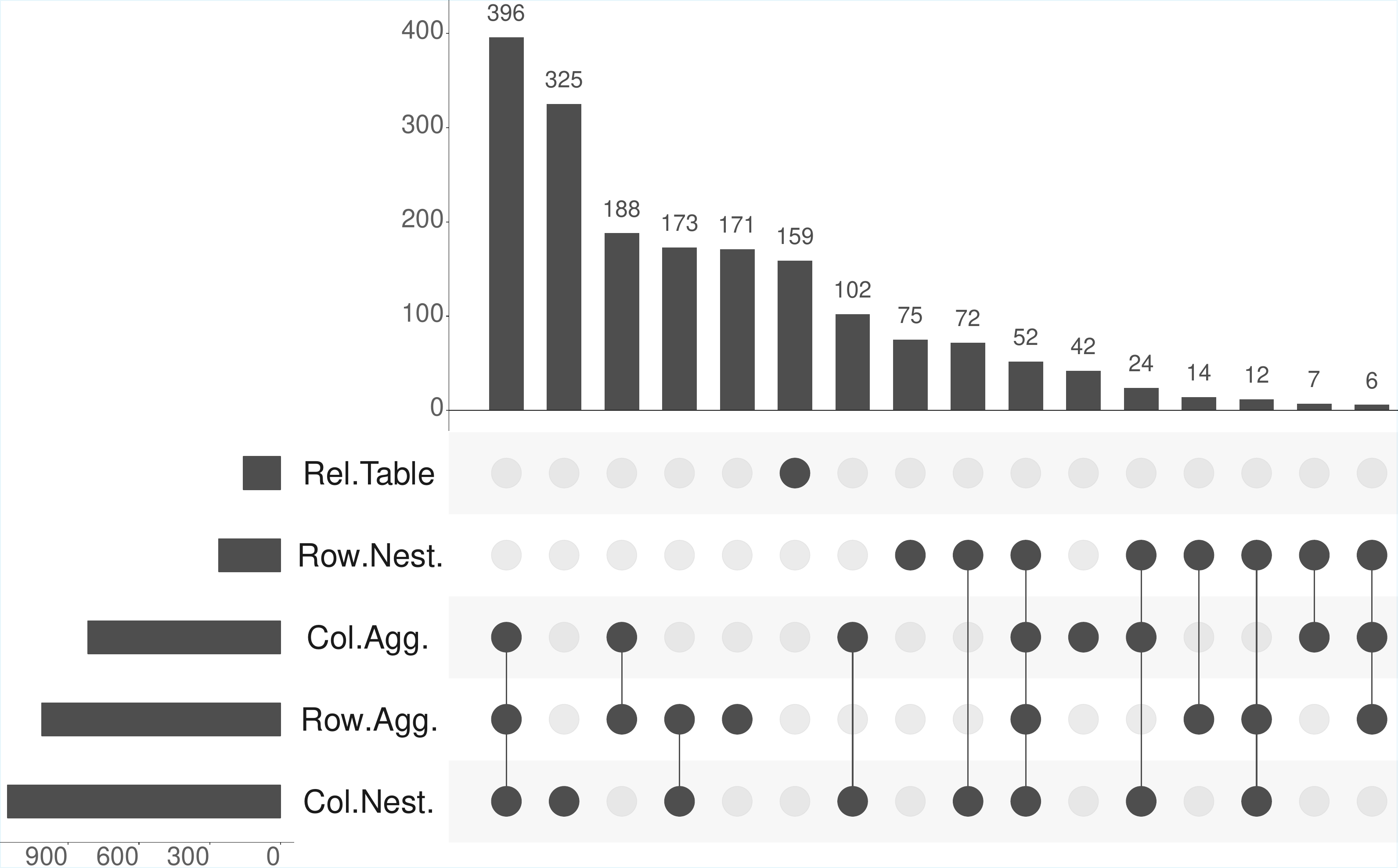}\\
{\footnotesize (a) Statistics of \hcts and QA pairs} & {\footnotesize (b) Main properties of real-world \hcts} \\
\includegraphics[width=0.4\linewidth, height=100pt]
{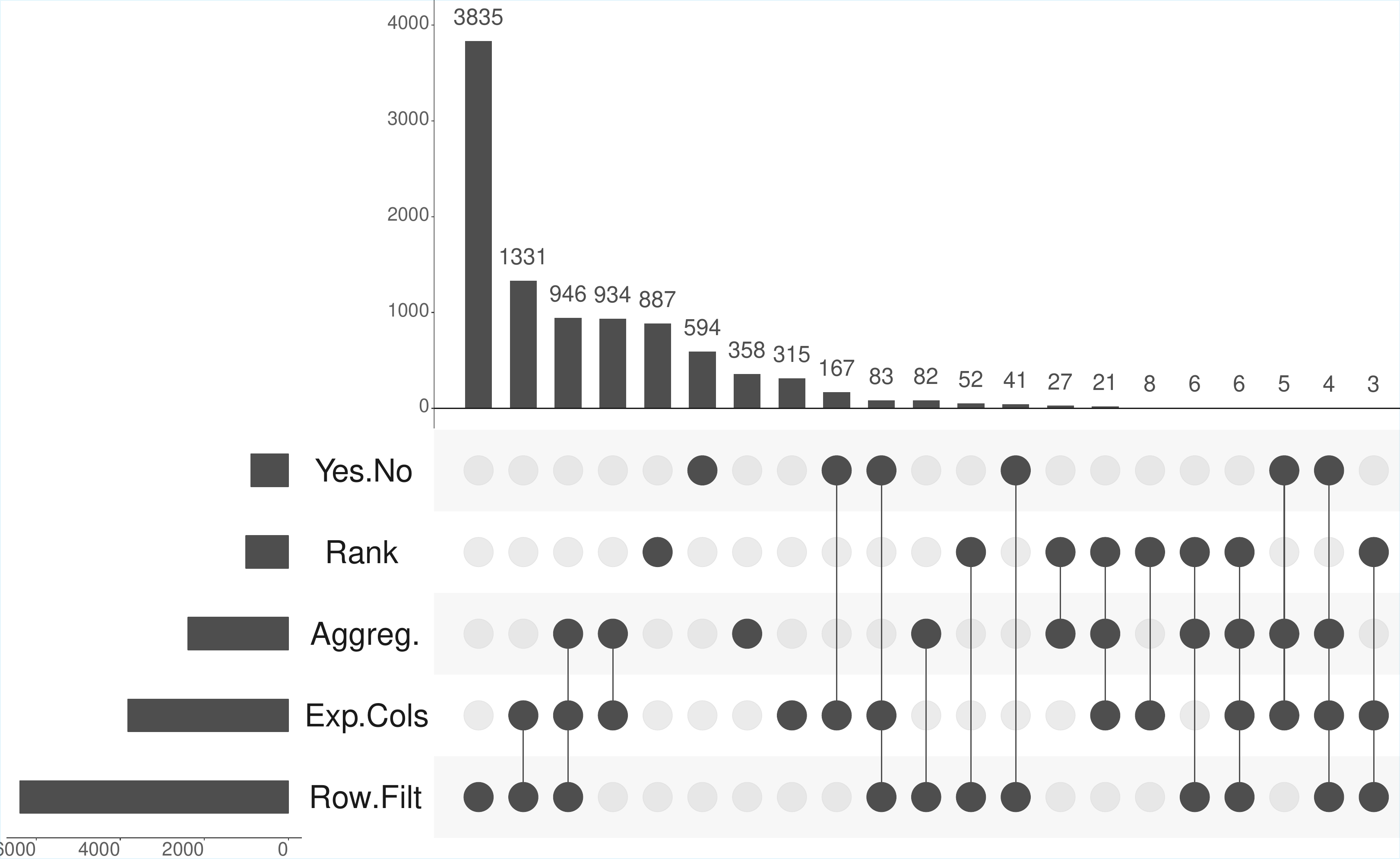}
&\includegraphics[width=0.4\linewidth, height=100pt]{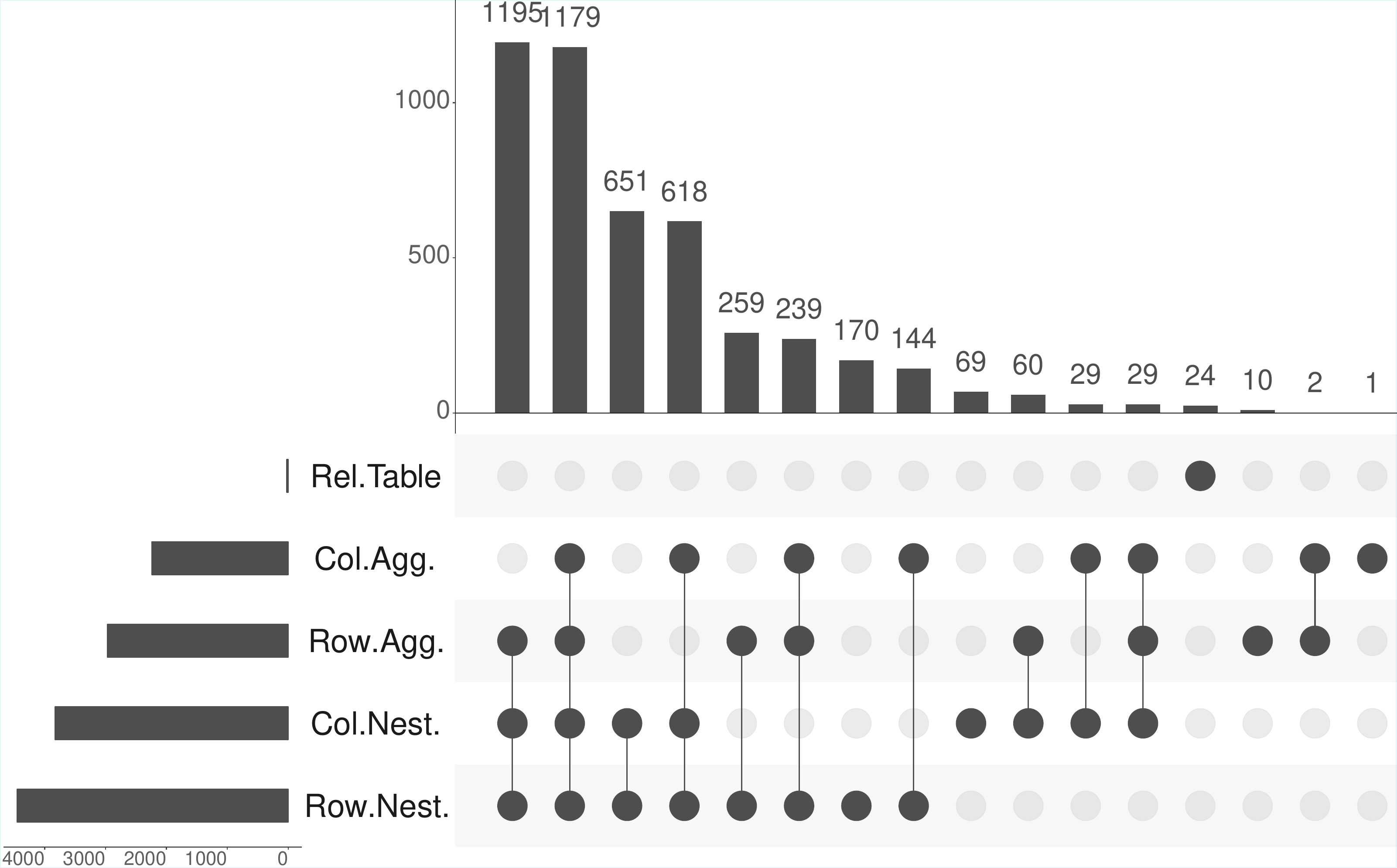}\\
{\footnotesize (c) Main properties of real-world QA pairs} & {\footnotesize (d) Main properties of synthetic \hcts} \\
\end{tabular}
}
\vspace{-10pt}
\caption{Main properties of the \hcts and QA pairs in \system.
}
\label{fig:upset_table_props}
\end{center}
\vspace{-22pt}
\end{figure*}

\textbf{Extraction and Transformation.}
We employed the Landing AI tool~\cite{landingai} to generate an HTML representation of the \hcts, 
from which we further derived the corresponding CSV and Markdown formats.
As part of the extraction pipeline, we conducted manual verification on over 10\% of 
the extracted \hcts 
to ensure the tool's ability to capture the structural complexities 
such as column nesting, row nesting, row groupings, and embedded aggregations. The verified sample demonstrated 100\% accuracy.
Also, 
we utilized Azure AI Document Intelligence~\cite{azure_document_intelligence} to extract \hct 
images.
A similar 10\% manual verification confirmed the absence of errors.
Figure~\ref{fig:upset_table_props}(a) presents statistics on each source. 

\textbf{Characterization of \hct Properties.}
Across the broad range of \hcts we investigated, we observed that certain structural 
layouts are highly prevalent such as column nesting, row nesting, and embedded aggregations whereas others, 
like nested tables, occur less frequently and are considered outliers.
In this study, we focus 
on 
common structural patterns  and exclude \hcts exhibiting rare or irregular layouts. 
The dominant layout patterns (referred to as \textit{properties}) are summarized in Table~\ref{tab:hct_property_definitions}.
The top-level properties are highlighted in green, 
while additional sub-properties are 
specified for cases where finer distinctions are expected to influence the LLM/VLM’s comprehension 
of 
\hcts, and hence impact the QA accuracy.
For instance, if an \hct exhibits column nesting (L1 in Figure~\ref{fig:HCT-examples}), we further specify whether the nesting is \textit{balanced}—meaning all child nodes occur at the same hierarchical depth—or \textit{unbalanced}, where nesting levels vary. 
Referring to Figure~\ref{fig:HCT-examples}, Tables~(b), (c), and (d) demonstrate balanced column nesting, whereas Table~(a) illustrates an unbalanced column nesting.
We also record whether the nesting is \textit{symmetric}—where column names and layouts are consistent across parent nodes—or \textit{asymmetric}, where they differ.
A similar level of detail is captured for the row nesting property (L2 in Figure~\ref{fig:HCT-examples}).
Embedded aggregation columns and rows also represent a common structural characteristic across  \hcts. 
For these cases, we track whether the aggregation is \textit{global} (spanning all columns or rows) or \textit{local} (restricted to a subset), and whether it is \textit{explicit}, e.g., “Total,” “Sum,” and “Avg.”, or \textit{implicit}, e.g., “Fruits,” denoting a collection of fruit categories. 

Manually creating 
fine-grained metadata for each \hct object represents a distinctive feature \system, enabling more in-depth analyses of LLM and VLM performance. 
Figure~\ref{fig:upset_table_props}(a) presents aggregated statistics illustrating the distribution of tables across various property combinations.
As shown, the combination of \{\textit{column nesting}, \textit{row aggregation}, \textit{column aggregation}\} is the most frequent, appearing in 396 \hcts and 
\textit{column nesting} is the most prevalent structural property overall, occurring in more than 900 \hcts.
To support the systematic annotation 
of this metadata, we developed a custom annotation interface \cite{hctqa_repo} that allows human experts to visually inspect each \hct and select the corresponding layout properties.
Each annotation is subsequently reviewed and verified by a second annotator to ensure accuracy and consistency. 

\subsection{Questions and Ground-Truth Answers Generation} 
\label{section:queryProperties}

A question over an \hct object refers to a NL question 
aimed at retrieving specific information—either directly (e.g., selecting particular value(s)) 
or through computation (e.g., expressions or aggregations).
Our focus is on single-\hct questions.
For the real-world dataset, we adopt two 
approaches for question generation: 
\revision{(1)~Manually-generated questions -- these are produced by the authors of this paper who are database experts with several years of combined experience in academia and industry, and 
(2)~model-generated questions -- these are produced by an LLM. 
}
All QA pairs, regardless of their source, are subsequently subject to independent human verification to 
maintain high dataset quality.

\begin{figure}[h]
 \centering
 \vspace{-2pt}
    \includegraphics[width=0.44\textwidth]{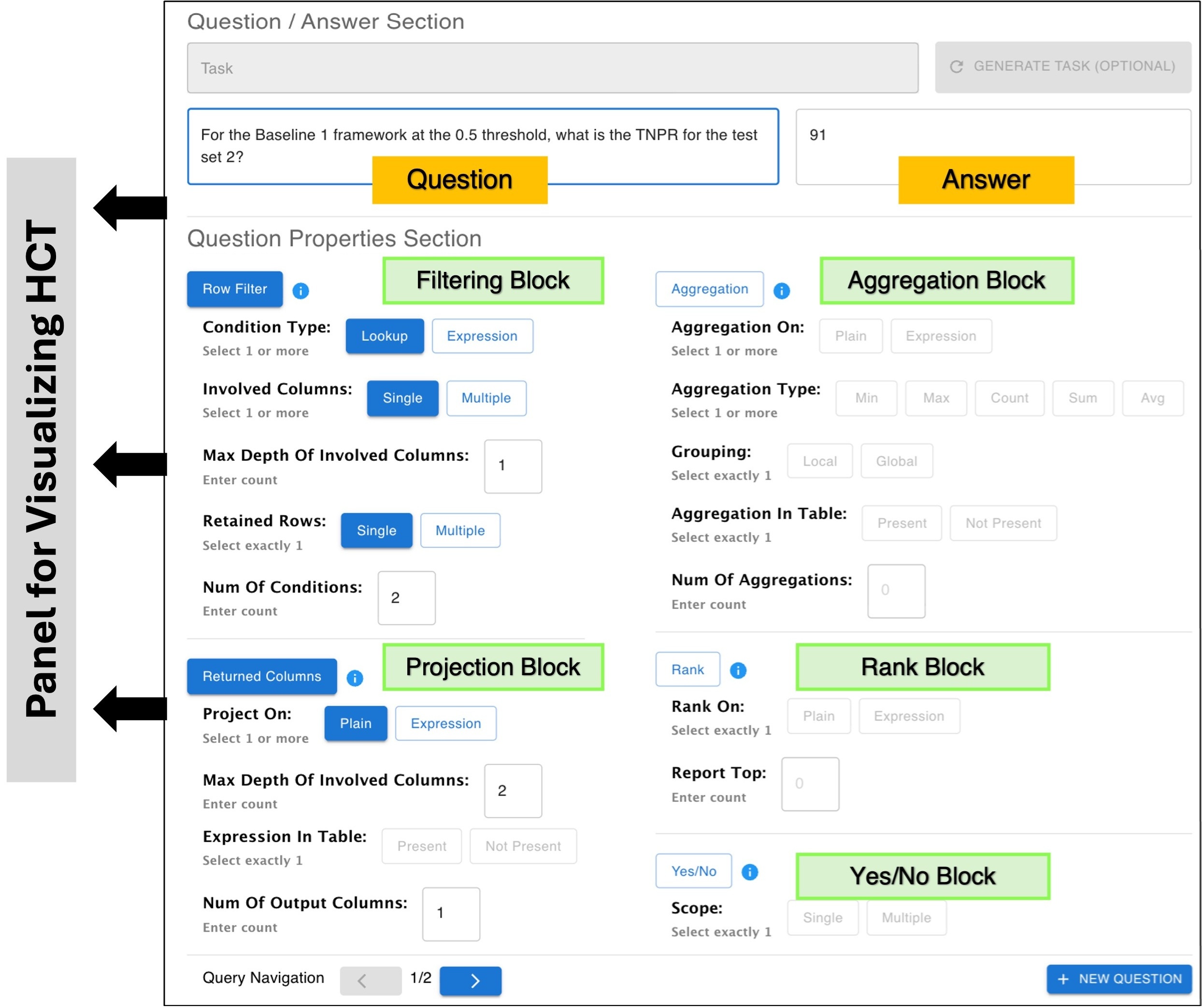}
      \vspace{-8pt}
      \caption{Question interface for the properties of  manually-generated questions}
     \vspace{-4pt}
     \label{fig:HCT-query-interface}
 \end{figure}

\textbf{Characterization of Question Properties.}
We designed a set of properties, inspired by common SQL operations, to systematically characterize each question. These properties capture both the type and complexity of operations—such as filtering, aggregation, rank, and projection—required to derive the answer. 
For each operation, additional sub-properties are maintained to capture more details on the operation's complexity. 
For example, the question properties include the number of filtering conditions, 
whether the projected columns include expressions or not, 
the number and type of aggregators (e.g., min, max, and avg.) as well as if they are local (a specific subset of values in the \hct) or global (values across the \hct), and the ranking on plain column or expression. 
These properties provide valuable insights into the types of questions that models tend to find most challenging. 
To facilitate capturing these properties, we designed a custom interface as depicted 
in Figure~\ref{fig:HCT-query-interface}. These properties are only provided for the manually-generated questions. 

\textbf{Manually-Generated QA Pairs.}
\revision{To generate a question and its answer, we use the annotation interface 
shown in Figure~\ref{fig:HCT-query-interface}.} 
This interface enables the user to visualize the table image, 
formulate a question with its associated answer, and annotate the relevant question properties.
Also, the interface presents real-time statistics on the types of questions 
generated, encouraging the user to create a balanced distribution of questions 
involving diverse operators, expressions, and aggregations.
Each question undergoes verification by two independent 
\revision{persons}
to ensure quality and consistency. We guarantee that every \hct includes 
at least one manually-generated question. 
In total, we produced \textbf{3,248} such questions with their properties
based on real-world \hcts, with each question requiring approximately \textbf{1.5} to \textbf{3} minutes to complete. 

\textbf{Model-Generated QA Pairs.}
To expand our question set beyond, 
we use OpenAI’s GPT~\cite{openai2023gpt4-turbo} to 
automatically generate additional questions. The model is provided 
with the HCT image, a sample manually-generated QA pair from the previous step
(using one-shot prompting), and detailed instructions specifying the required constraints, such as the target operations and the desired level of question complexity.
The model then produces a set of eight QA pairs per \hct, 
enabling efficient large-scale question generation. 
Using this approach, we generated \textbf{6,587} additional QA pairs 
on real-world \hcts in approximately \textbf{30} minutes.

To ensure the accuracy and clarity of model-generated questions, we established a rigorous human verification and correction pipeline. Each model-generated QA pair is reviewed by a team of annotators who (i)~assess the quality of the question as \textit{good}, \textit{ambiguous} 
or \textit{nonsensical} and (ii)~correct the corresponding answer if it is incorrect or poorly formatted. 
Questions that are rated as nonsensical are excluded from the dataset, while those marked as ambiguous are re-written. 
\revision{Our annotation team comprised of 14 undergrad students majoring in Computer Science who had experience with databases. Annotators were given a training session by the authors on the task description and did supervised mock exercises of the task.}
Each QA 
pair is evaluated by two annotators \revision{independently}. In cases of disagreement, a third annotator 
makes the final decision.
To monitor the consistency and attentiveness of annotators, we embedded a set of pre-validated, manually-generated questions within their evaluation batches. Annotators who misclassified more than one of these verified questions (e.g., labeling them as ambiguous or nonsensical or incorrectly flagging their answers) were excluded from the annotation pool. \revision{Only a single annotator failed our hidden test questions and was replaced.}

Our model-based pipeline for QA generation and verification was highly effective and scalable. Overall, only 4\% of model-generated questions were marked as nonsensical, 24\% as ambiguous, and 72\% as good. 
Moreover, the model's answers required minor corrections for nearly 50\% of the questions. 
Combining the manually-generated and model-generated questions, we ended up with a total of \textbf{9,835} QA pairs over the  \textbf{1,880} \hcts. 
In Figure~\ref{fig:upset_table_props}(a), we present the distribution of the questions across the different datasets. Moreover, 
in Figure~\ref{fig:upset_table_props}(b), we present aggregated distribution of the question properties over all questions. 
For example, roughly \textbf{2,000} questions involve aggregations, while \textbf{1,000} questions involve ranking, and there are \textbf{594} yes/no questions. 
\revision{Overall, Figure~\ref{fig:upset_table_props}(b) shows a good distribution of NL questions on real-world \hcts ranging from simple yes/no questions to complex combination of aggregation, ranking, column expression, etc.}

\revision{\textbf{Manual Effort.} Referring to Figure \ref{fig:preparing_the_real-world_dataset}, to prepare 1,880 real-world \hcts, a combined 150 manual work hours (MWH) were spent as follows: 
15 MWH for \hcts filtering to meet certain standards (e.g., visual quality and no out-of-scope properties), 
15 MWH for sample verification (10\% from text and image modalities), 
60 MWH for annotating \hct properties, and 80 MWH for 2-way 
verification of the annotated \hcts and their properties. 
To prepare 9,835 QA pairs, 1020 MWH were spent as follows: 
150 MWH for manually creating and annotating QA pairs and their properties,
270 MWH for 2-way verification of the manually created and annotated QA pairs, and 
600 MWH to verify model generated QA pairs. 
The creation and verification of the manual QA pairs was done by the authors, whereas the model generated QA pair verification was done by a team of 14 annotators. Combined, this entire effort took around 1,170 MWH.}

\section{HCT-QA Synthetic  Generator}
\label{section:synthetic}



\begin{figure}[t]
    \centering
    \begin{tabular}{c}
        \includegraphics[width=0.94\linewidth]{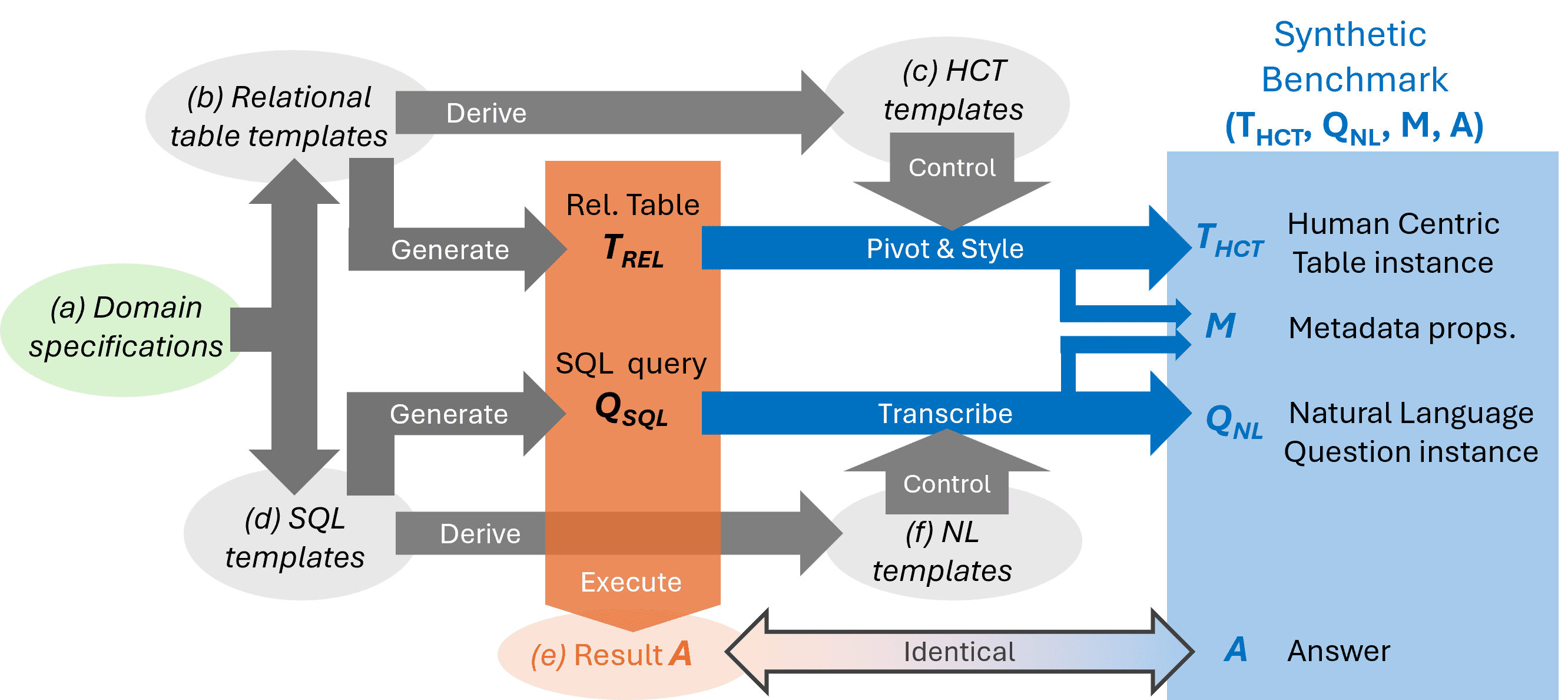}
    \end{tabular}
    \vspace{-7pt}
    \caption{HCT-QA Synthetic Data Generator workflow
  }
  \vspace{-13pt}
    \label{fig:synthetic_data_generator_overview}
\end{figure}

To  
enable large-scale generation of \hcts with their corresponding QA pairs, 
we built a synthetic generator comprising two components: 
a \textbf{\hct Data Generator} (Section~\ref{section:synthetic-HCT}) and 
a \textbf{QA Generator} (Section~\ref{section:synthetic-query}). 
This generator serves as a valuable resource for the research community, facilitating the exploration of 
research challenges 
on \hcts.
The generator is highly configurable, allowing it to capture nearly all common \hct and question properties defined earlier. 
\revision{It can also be configured to create QA pairs with multiple operators, which would be difficult to create  manually at scale.} 

\revision {The main goal of the HCT-QA Synthetic Generator is to reduce the time and labor cost required to create \hcts and QA pairs on them. The entire process of manually extracting real-world \hcts, creating manually generated or human-verified model-generated QA pairs, and labeling the \hct and question properties takes around 4 minutes per QA pair. 
In contrast, creating the domain vocabulary for an entire family of synthetic \hcts, specifying the related table and question templates takes about 1 hour in total with the  generation of a synthetic \hct instance and a related QA pair from that family taking only less than a second. 
Thus, the synthetic generator enables scaling up the generation process to several thousands \hcts and QA pairs in a few hours, while still controlling their diversity and complexity through the template design.} 
\revision{As we will show throughout this section, the HCT-QA Synthetic Generator has the following four properties:}
\begin{enumerate}
    \item \revision{Scalability: It can be easily
    extended to cover 
    new domains and generate thousands of HCTs and QA pairs with minimal efforts.}
    \item \revision{Guaranteed semantic correctness: Based on a template-based approach, semantic fidelity can be ensured across the constructed HCTs and their generated NL questions.} 
    \item \revision{Broad coverage of question complexity: 
    NL questions are generated using various SQL templates spanning different levels of sophistication and operator complexities.}
    \item \revision{Disambiguation: The generated NL questions have a clear interpretation in terms of what they are referring to in the table regardless of their complexity.}
    
\end{enumerate}

The synthetic \hct and QA generation process (Figure~\ref{fig:synthetic_data_generator_overview}) begins with domain specifications (a) to create custom \textit{table templates} 
(b) and \textit{question templates} 
(d) 
from which, inspired by SQL, 
sample instances of relational tables $T_{REL}$ and SQL queries $Q_{SQL}$ are generated, respectively.
Executing $Q_{SQL}$ on $T_{REL}$ gives ground-truth results $A$ (e). 
We derive \hct templates (c) from relational table templates to control how to pivot and style the same $T_{REL}$ instance into different variants of \hcts instances $T_{HCT}$. 
We derive NL templates (f) from SQL templates that shape the transcription of $Q_{SQL}$ queries into NL  questions $Q_{NL}$. 
Results $A$ (e) to $(T_{REL}, Q_{SQL})$ instances are used as answers to $(T_{HCT}, Q_{NL})$ instances as they are identical by design. \revision{Metadata $M$ characterizing \hcts and question properties (Section~\ref{section:real_data}) are computed too, forming the $(T_{HCT},Q_{NL},A,M)$ benchmark quadruplets.}


\begin{figure}[t]
    \centering
    \setlength{\tabcolsep}{1pt}
    \vspace{-4pt}
    \begin{tabular}{c}    
        \includegraphics[width=.95\linewidth]{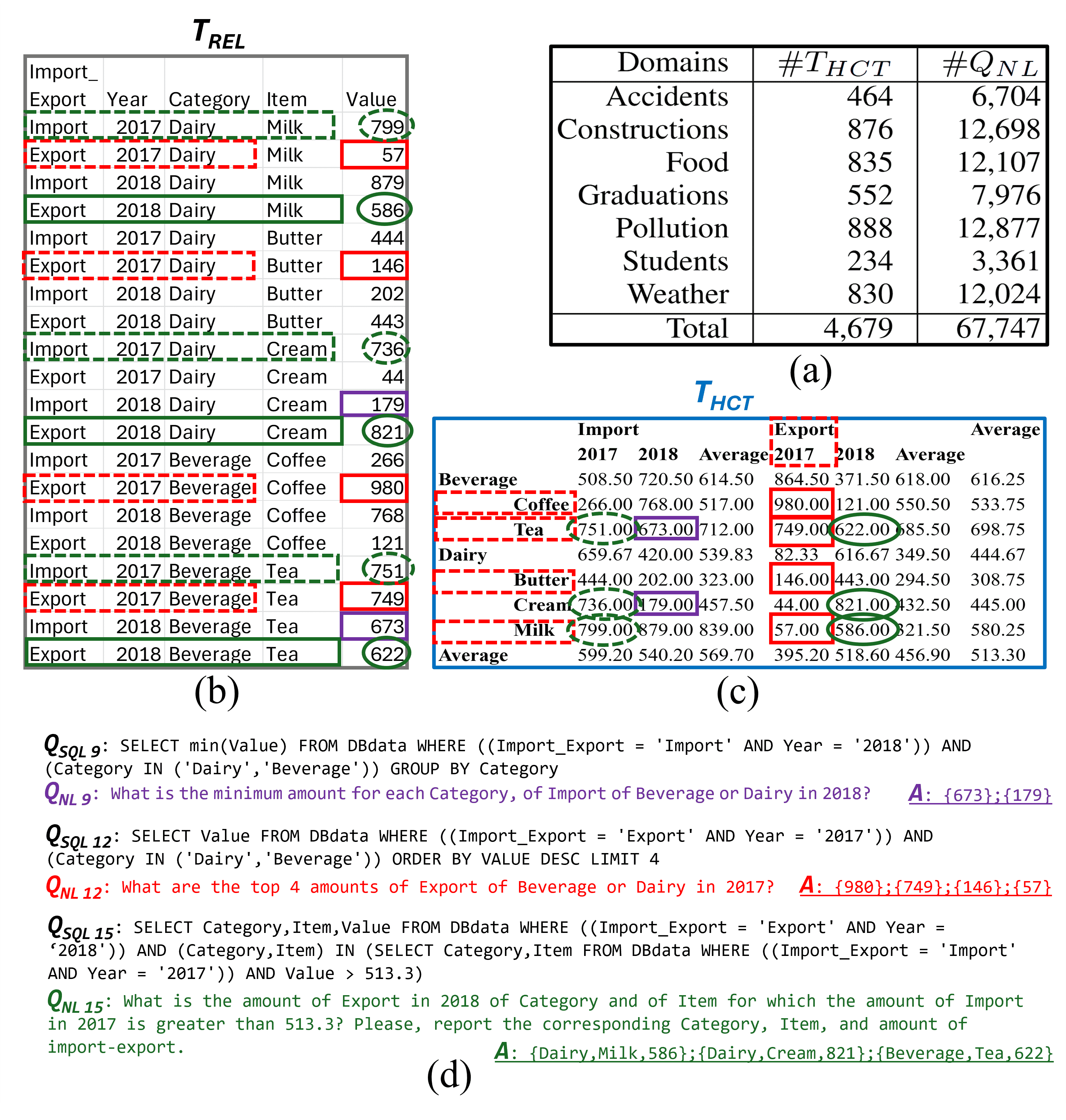}
    \end{tabular}
    \vspace{-10pt}
    \caption{(a) Statistics of the synthetic \hcts and QAs for $7$ domains. An example of a relational table $T_{REL}$ (b) and a generated 
    $T_{HCT}$ from it (c). (d) Examples of synthetic QAs, 
    where $Q_{SQL}$ is transformed into $Q_{NL}$ using custom templates, and executed on $T_{REL}$ to get the ground-truth answer $A$. }
    \vspace{-10pt}
    \label{fig:synthetic_data_generator}
\end{figure}

\subsection{Synthetic \hct Data Generator}
\label{section:synthetic-HCT}







The \textit{HCT Data Generator} generates \hcts in multiple formats such as CSV, HTML, PDF, and PNG, 
and supports various stylistic features such as row header indentation, various background colors and fonts, and line separations. Through a combination of configuration files and templates, the generator can reproduce 
$15$  (marked with ``*" in Table~\ref{tab:hct_property_definitions}) of the $19$ \hct 
properties. 
The remaining 4 properties 
are not supported in this version due to their complex irregularities. 

\textbf{Domain Specifications (Figure ~\ref{fig:synthetic_data_generator_overview}(a))}. 
We begin by defining 
a set of realistic 
application domains, 
such as \textit{import–export 
statistics across countries and months over several
years}. 
Each domain is associated with 
\revision{a \textit{domain vocabulary} consisting of 
(1)~a JSON configuration file listing attribute names and their set of possible values}
and \revision{(2)}~catalog tables
listing a comprehensive set of attributes \revision{from the vocabulary}, their data types, and possible value ranges.
Various \hcts are then created by 
different combinations of attributes and value ranges.
For example, in a ``Food import–export” domain (Figure~\ref{fig:synthetic_data_generator}(b)), user-defined attributes such as {\footnotesize{
\texttt{Year: \{2015,...,2023\}}, \texttt{Import$–$Export Op.: \{Import, Export\}}, \texttt{Food Category: \{Meat, Dairy\}},}}  and 
{\footnotesize{
\texttt{Food Item: \{Chicken, Beef, Cream, Milk, Yogurt\}}
}}
are 
used to generate \hcts describing food import-export activities over 
years.

\textbf{Types of Attributes (Figure ~\ref{fig:synthetic_data_generator_overview}(a)).} 
We distinguish between two types of attributes, 
\textit{independent} 
attributes such as 
{\small{\ttt{Year}}} and 
{\small{\ttt{Import-Export}}} 
for which all pairs of values are possible,  e.g., 
{\small{(\ttt{2022}, \ttt{Import})}}, and 
\textit{hierarchical} attributes such as 
{\small{\ttt{(Country, State, City)}}} and 
{\small{\ttt{(Food category, Food item)}}}, 
where attribute values follow an inherent order and only certain combinations are meaningful.
For example, combinations such as 
{\small{(\ttt{California}, \ttt{Boston})}} 
are invalid. 
\textit{Independent} attributes can appear as either row or column headers at any nesting level. 
In contrast, \textit{hierarchical} attributes must respect 
logical order — parent attributes, such as \ttt{Country}, must precede child ones, such as {\small{\ttt{City}}}, from left to right or top to bottom within headers.
Moreover, all attributes belonging to the same hierarchy must be placed together, either entirely in the row headers or entirely in the column headers, e.g., an \hct cannot have {\small{\ttt{Country}}} 
as a row header and {\small{\ttt{City}}} as a column header.
More details 
are provided in~\cite{hctqa_repo}.

\textbf{Relational Table Templates (Figure~\ref{fig:synthetic_data_generator_overview}(b)).} 
We manually define \textit{Table templates} in JSON format that give details about which attributes and values to pick for a given scenario to form a family of relational tables.  
For example, for the \textit{import-export} scenario, the data generator uses the table template to 
create the relational table ($T_{REL}$) depicted in Figure~\ref{fig:synthetic_data_generator}(b). 
The template defines a target field, e.g., the \ttt{Value} numerical attribute in Figure~\ref{fig:synthetic_data_generator}(b), 
to contain the fine-grained value (automatically generated) for each combination over the rest of the attributes. 
The generator 
enforces uniqueness constraints on the target field to prevent ambiguity when evaluating model-predicted answers against ground-truth values.
Relational tables serve two essential purposes:
(1)~They enable systematic generation of \hcts through pivoting operations~\cite{pivotTable2005,pivottabler}; and
(2)~they are used to construct the ground truth answers by executing SQL queries on them without 
human intervention.

\textbf{\hct Templates (Figure~\ref{fig:synthetic_data_generator_overview}(c)).} 
These templates define the final construction rules for producing 
\hcts instances from the domain specifications and structural combinations set in the relational table template from which they are automatically generated. They are highly configurable,  for example, \revision{one} can specify
the depth of row and column headers,
the assignment of attributes to each header, 
and the inclusion and type of aggregation functions
and whether the aggregation occurs at a global or local level.
They also capture layout specifications such as nesting balance and symmetry, 
as well as stylistic elements, e.g., 
line separations, and indentation. 

As shown 
in Figure~\ref{fig:synthetic_data_generator}(c), an example 
\hct instance ($T_{HCT}$) is generated from its corresponding relational table ($T_{REL}$).
In this case, attributes \ttt{(Import\_Export, Year)} form nested column headers, while \ttt{(Category, Items)} define nested row headers.
The column nesting is balanced and symmetric, whereas the row nesting is balanced but asymmetric.
This instance includes both local and global aggregation rows and columns (average). 
Food \ttt{Category} row headers (\ttt{Beverage} and \ttt{Dairy}) are indented and their row contains implicit local average  aggregation of 
food \ttt{Items} (\textit{e.g.}, \ttt{Coffee} and \ttt{Tea} for \ttt{Beverage}), while global aggregation of food \ttt{Category} and column aggregations are all explicit ("Average" appears in the header). 
These properties are 
automatically recorded in the 
metadata $M$.
\revision{It is worth mentioning that with the relational table and the HCT templates, the Synthetic HCT Data Generator only requires the domain vocabulary to generate HCTs for 
new domains.}

\textbf{Statistics on Synthetic \hcts.} 
In our benchmark and as summarized in Figure~\ref{fig:synthetic_data_generator}(a), 
we covered seven distinct domains and generated hundreds of \hcts per domain with a total of \textbf{4,679} synthetic \hcts. 
These \hcts exhibit different complexities and layouts as presented in Figure~\ref{fig:upset_table_props}(c).

\subsection{Synthetic Question \& Answer Generator}
\label{section:synthetic-query}


Our synthetic QA Generator coupled with the \hct Generator, allows us to  scale \system 
to produce tens of thousands of QA on thousands of synthetic \hcts with $100\%$ accuracy.




\begin{figure}
\vspace{-5pt}
 \centering
    \includegraphics[width=0.85\linewidth]{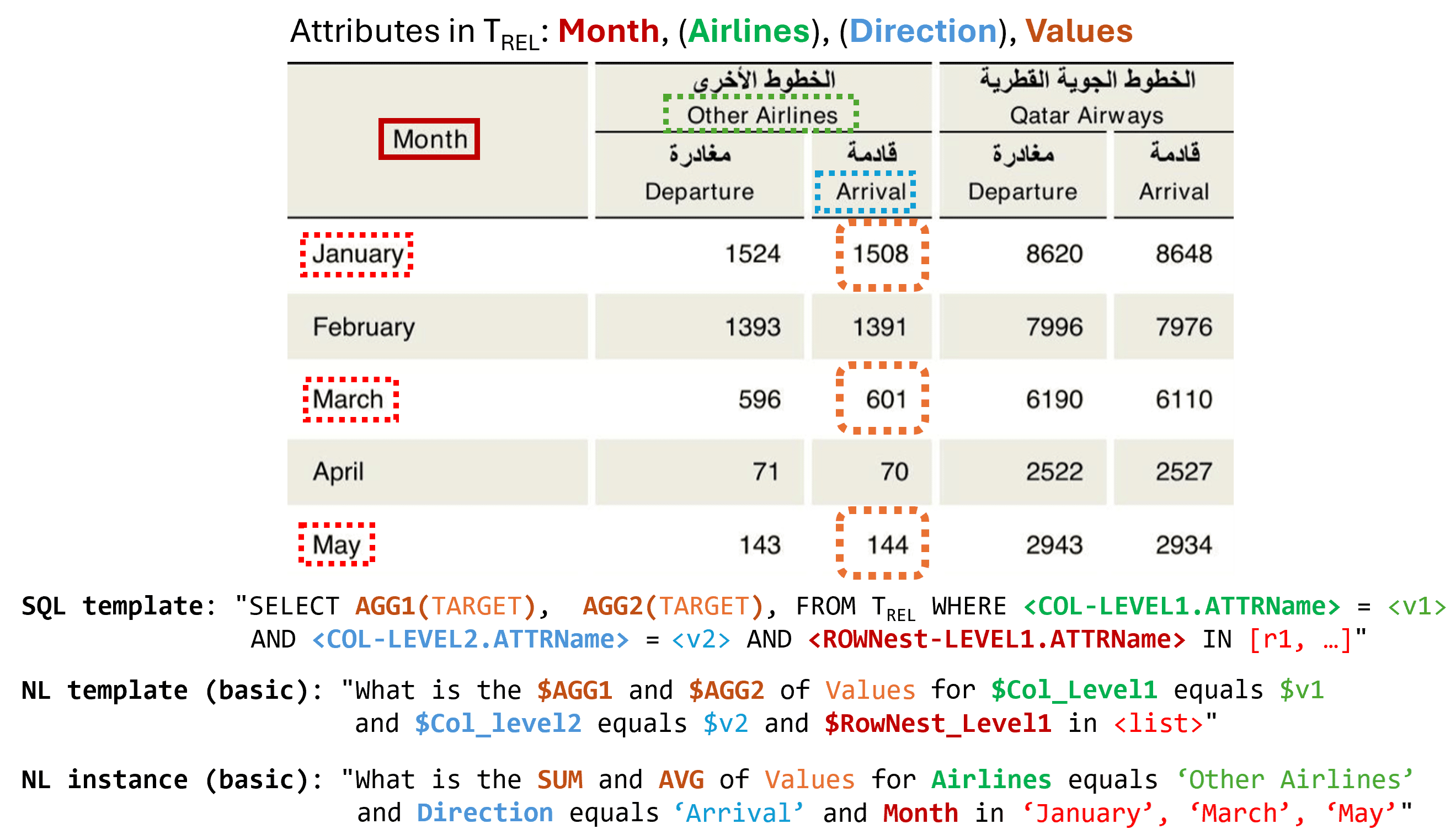}
      \vspace{-9pt}
      \caption{Example of a SQL query template, a basic NL template, and its instantiation on the \hct (top) for synthetic query generation. 
      }
      \vspace{-14pt}
     \label{fig:SQL-template}
 \end{figure}
 
\textbf{SQL Query Templates (Figure \ref{fig:synthetic_data_generator_overview}(d).} 
The QA 
Generator builds upon the same domain specifications and table templates described in Section~\ref{section:synthetic-HCT}.
\revision{Table \ref{sql_templates}} lists 15 SQL query templates, each representing a distinct level of complexity. These templates are used to instantiate SQL queries with varying structures and 
operations \revision{from simple or multiple rows and columns selection (SQL$_{1,2,3,6}$), to multiple aggregations over one row or one column (SQL$_{4,7}$), when the same aggregation is already displayed in the \hct (SQL$_5$), or grouped by any attribute levels or only by the top level one, and with or without reporting the attributes (SQL$_{8,9,10,11}$). We also generate sorting and top-k queries (SQL$_{12,13}$) as well as selections with operations (SQL$_{14}$) and a selection conditioned on another selection (SQL$_{15}$).} 




As an example, Figure~\ref{fig:SQL-template} presents a SQL query template \revision{(SQL$_7$ in Table \ref{sql_templates})} 
to generate global aggregation queries (i.e., without sub-groups) on the relational representation $T_{REL}$ (not shown) of an \hct. 
The template applies one equality condition to each attribute involved in column nesting, 
allowing it to naturally extend to multi-level nesting scenarios. 
When applied to the presented \hct, these conditions trace through the structure, one level at a time, to reach a specific column (green \& blue colors of the WHERE clauses correspond to the same color boxes in the \hct column headers). 
Additionally, the template introduces IN conditions on attributes associated with row nesting. 
The \hct 
in Figure~\ref{fig:SQL-template} contains no row nesting, and hence a maximum of one condition is possible (i.e., on the \texttt{Month} column). 
In contrast, the \hct in Figure~\ref{fig:synthetic_data_generator}(c) contains row nesting, and the template allows putting conditions on the corresponding columns (i.e., on the \texttt{Category} and \texttt{Item} columns in $T_{REL}$).   
These 
conditions 
contribute to increasing the query’s 
complexity.
Finally, the template applies one or more aggregation functions
over the selected output values.


\textbf{Ground Truth Generation (Figure ~\ref{fig:synthetic_data_generator_overview}(e).} 
The relational table $T_{REL}$ and the generated SQL queries are imported 
into a relational DBMS for execution and the generation of the ground-truth 
answer $A$, thus forming the triplet ($T_{REL},Q_{SQL},A$).  
In Figure~\ref{fig:synthetic_data_generator}(d), we illustrate three SQL query instances generated from different SQL query templates \revision{(SQL$_9$, SQL$_{12}$, and SQL$_{15}$ in Table \ref{sql_templates})}. 






\textbf{Natural Language Templates (Figures~\ref{fig:synthetic_data_generator_overview}(d-f))}. 
For each SQL query template, we manually define a NL 
template to transcribe $Q_{SQL}$ into $Q_{NL}$ specific to the \hct domain.
To illustrate the need for these custom NL templates, let's consider the SQL query template in Figure~\ref{fig:SQL-template} that would be transcribed into a (basic) NL template as: 
\begin{quote}
\footnotesize\ttfamily
"What is the \$AGG1 and \$AGG2 of Values for \$Col\_Level1 equals \$v1 and \$Col\_level2 equals \$v2 and \$RowNest\_Level1 IN <list>"
\end{quote}
\noindent Then, a (basic) NL question instance, given some values from the \hct in Figure~\ref{fig:SQL-template}, would be:
\begin{quote}
\footnotesize\ttfamily
"What is the SUM and AVG of Values for Airlines equals `Other Airlines' and Direction equals `Arrival' and Month IN [`January', `March', `May']"
\end{quote}

Obviously, this basic direct transcription does not form a seamless NL question. 
Hence, we customize the NL templates for each of the seven domains covered in our generator 
(Figure~\ref{fig:synthetic_data_generator}(a)), generating more natural questions. We illustrate a few examples in Figure~\ref{fig:synthetic_data_generator}(d). 
\revision{Similar to the construction of synthetic HCTs, the Synthetic QA generator is domain-agnostic. It relies solely on specifying a domain by its vocabulary describing attributes (e.g. \ttt{Month}) and values (\ttt{May}, \ttt{June}...) in a JSON format, and the manual design of $15$ $Q_{NL}$ templates based on this vocabulary and simple rules to ensure English syntax. Each $Q_{NL}$ template transcribes in natural language one of the $15$ $Q_{SQL}$ templates shown in Table \ref{sql_templates}, ensuring non-repetitive queries. The full list of $15$ $Q_{NL}$ templates for each of the $7$ domains and the rules to build them, is available in Appendix D in the extended report~\cite{hctqa_repo}.} 
As each $Q_{NL}$ template corresponds to a $Q_{SQL}$ template with specific query properties, we automatically record them in the metadata $M$.

\begin{table*}[ht]
\vspace{-5pt}
\caption{
%
\revision{SQL templates of Synthetic HCT-QA. 
$r|c$: depth of row$|$column (R$|$C) headers of the \hct. 
$Attr_C^*$: all attributes of C. 
$Attr_R^1$: left-most attribute of R. 
$Num$: avg. of selected column.
$Dir\stackrel{\text{rand}}{\sim}  \{$ASC, DESC$\}$, $k\stackrel{\text{rand}}{\sim} \{2,3,4,5\}$, and $Op\stackrel{\text{rand}}{\sim}\{<,>\}$.}} 

\label{sql_templates}
\centering
\vspace{-8pt}

\scriptsize 
\begin{tabular}{|l|l|}
 \hline
 & \revision{SQL templates}\\   
 \hline
\revision{SQL$_1$} & \revision{{\bf SELECT} Value {\bf FROM} Data {\bf WHERE} (Attr$_R$ = Val$_R$)$^r$ {\bf AND} (Attr$_C$ = Val$_C$)$^c$}\\ \hline
\revision{SQL$_2$} & \revision{{\bf SELECT} Value {\bf FROM} Data {\bf WHERE} (Attr$_R$ {\bf IN} ListVal$_R$)$^r$ {\bf AND} (Attr$_C$ = Val$_C$)$^c$}\\ \hline
\revision{SQL$_3$} & \revision{{\bf SELECT} Value {\bf FROM} Data {\bf WHERE} (Attr$_R$ = Val$_R$)$^r$ {\bf AND} (Attr$_C$ {\bf IN} ListVal$_C$)$^c$} \\ \hline
\revision{SQL$_4$} & \revision{{\bf SELECT} AGG$_1$(Value), AGG$_2$(Value) {\bf FROM} Data {\bf WHERE} (Attr$_R$ = Val$_R$) {\bf AND} (Attr$_C$ {\bf IN} ListVal$_C$)}\\ \hline
\revision{SQL$_5$} & \revision{{\bf SELECT} AGG$_*$(Value) {\bf FROM} Data {\bf WHERE}  (Attr$_R$ = Val$_R$)$^r$ {\bf AND} (Attr$_C$ {\bf IN} ListVal$_C$)$^c$}\\ \hline
\revision{SQL$_6$} & \revision{{\bf SELECT} Value {\bf FROM} Data {\bf WHERE} (Attr$_R$ {\bf IN} ListVal$_R$)$^r$ {\bf AND} (Attr$_C$ {\bf IN} ListVal$_C$)$^c$}\\ \hline
\revision{SQL$_7$} & \revision{{\bf SELECT} AGG$_1$(Value), AGG$_2$(Value) {\bf FROM} Data {\bf WHERE} (Attr$_R$ {\bf IN} ListVal$_R$)$^r$ {\bf AND} (Attr$_C$ = Val$_C$)$^c$} \\ \hline
\revision{SQL$_8$} & \revision{{\bf SELECT} Attr$_C^*$, AGG$_1$(Value), AGG$_2$(Value) {\bf FROM} Data {\bf WHERE} (Attr$_R$ {\bf IN} ListVal$_R$)$^r$ {\bf AND} (Attr$_C$ {\bf IN} ListVal$_C$)$^c$ {\bf GROUP BY} Attr$_C^*$}\\ \hline
\revision{SQL$_9$} & \revision{{\bf SELECT} MIN(Value) {\bf FROM} Data {\bf WHERE} (Attr$_R$ {\bf IN} ListVal$_R$)$^r$ {\bf AND} (Attr$_C$ = Val$_C$)$^c$ {\bf GROUP BY} Attr$_R^1$} \\ \hline
\revision{SQL$_{10}$} & \revision{{\bf SELECT} Attr$_R^1$, AGG$_1$(Value) {\bf FROM} Data {\bf WHERE} (Attr$_R$ {\bf IN} ListVal$_R$)$^r$ {\bf AND} (Attr$_C$ = Val$_C$)$^c$ {\bf GROUP BY} Attr$_R^1$} \\ \hline
\revision{SQL$_{11}$} & \revision{{\bf SELECT} Attr$_R^1$, Attr$_C^*$, AGG$_1$(Value) {\bf FROM} Data {\bf WHERE} (Attr$_R$ {\bf IN} ListVal$_R$)$^r$ {\bf AND} (Attr$_C$ {\bf IN} ListVal$_C$)$^c$ {\bf GROUP BY} Attr$_R^1$, Attr$_C^*$} \\ \hline
\revision{SQL$_{12}$} & \revision{{\bf SELECT} Value {\bf FROM} Data {\bf WHERE} (Attr$_R$ {\bf IN} ListVal$_R$)$^1$ {\bf AND} (Attr$_C$ = Val$_C$)$^c$ {\bf ORDER BY VALUE} Dir {\bf LIMIT} k} \\ \hline
\revision{SQL$_{13}$} & \revision{{\bf SELECT} Value {\bf FROM} Data {\bf WHERE} (Attr$_R$ {\bf IN} ListVal$_R$)$^1$ {\bf AND} (Attr$_C$ = Val$_C$)$^c$ {\bf ORDER BY VALUE} Dir} \\ \hline
\revision{SQL$_{14}$} & \revision{{\bf SELECT} Attr$_R^*$ {\bf FROM} Data {\bf WHERE} (Attr$_{C}$ = Val$_{C}$)$^c$ {\bf AND} Value Op Num} \\ \hline
\revision{SQL$_{15}$} & \revision{{\bf SELECT} Attr$_{R}^*$, Value {\bf FROM} Data {\bf WHERE} (Attr$_{C\neq C_{14}}$ = Val$_{C\neq C_{14}}$)$^c$ {\bf AND} 
(Attr$_R^*$ {\bf IN} Result(SQL$_{14}))$} \\ \hline


\end{tabular}
\vspace{-10pt}
\end{table*}

\textbf{Statistics on Synthetic Questions}. 
We generated \textbf{67,747} QA 
pairs on the $4,679$  synthetic \hcts across seven domains. 

\textbf{Scalability}.
Thanks to manually defined JSON templates, our  Synthetic Data Generator can be 
augmented with minimal effort. The domain specifications can be augmented at will, adding attribute names and their list of values, to generate \hcts in any domain of interest. Also, the relational table templates allow controlling the diversity and complexity of the generated \hcts.
SQL query templates can be 
easily configurable due to their computational nature. The corresponding NL templates require minimum 
effort to adapt to a new domain, using a small set of syntax rules.
This template-based automation allows unlimited \hct-QA 
accurate data generation in minimal time 
without the need for costly verification.

\revision{\textbf{Manual Effort.} A key advantage of the synthetic data generator is that, once the tool is 
implemented—a process that required several hundred hours of development—adding new SQL 
or natural language templates for a specific domain typically takes no more than two to three hours.}

\section{Experimental evaluation}
\label{section:evaluation2}

While \system is designed to evaluate data processing engines during answering NL questions on \hcts, 
our evaluation focus on 
%
%
LLMs and VLMs. 
%
To this end, we explain our experimental setup in Section \ref{evaluation-experimental-setup} and then answer seven research questions (RQs) experimentally in Section \ref{section:analysis-and-findings}.

\subsection{Experimental setup}
\label{evaluation-experimental-setup}

\textbf{Models.} We tested a large variety of open- and closed-weight LLMs and VLMs (Table~\ref{tab:model_performance}). All open-weight models were downloaded from HuggingFace.

\textbf{Datasets.} 
There are 1,880 real-world \hcts from four data sources along with 9,835 QA pairs 
and 4,679 synthetic \hcts with 67,747 QA pairs (total of \textbf{6,559} \hcts with \textbf{77,582} QA pairs). 
For the results in Table~\ref{tab:model_performance}, the text modality \hcts were represented in HTML format and the vision-text modality \hcts were images encoded in base64.


\textbf{Prompting Approach and Hyperparameters.}
We experimented with \mbox{zero-shot} and \mbox{one-shot} prompting with no clear indication that the latter improved model performance. Thus,  all subsequent experiments  use \mbox{zero-shot} prompting. 
We ask the model to "only answer the question using information from the \hct" and allow it to refuse to answer the question if it deems it unanswerable.
For inference, the temperature was set to 0, 
and the maximum output tokens to 256. For finetuning, the learning rate was 5.0e-07 and the number of epochs was 2. The finetuning config file can be found in~\cite{hctqa_repo}. We post-process the model responses using a rule-based approach to extract the actual answer if one was provided. All inference and finetuning experiments were run on 8 NVIDIA A100 GPUs. \revision{Inference  which required around 77k prompts took a total of 129 hours, averaging  1 hour, 1.5 hours, and 22 hours per small, medium, and cloud model, respectively. Additionally, our finetuning experiments took 30 hours.} 

\textbf{Evaluation Metrics.} 
We use two  metrics for 
model accuracy: (i)~\textit{F1 Score}--commonly adopted in several table QA benchmarks, e.g.,  ~\cite{yu-etal-2018-spider, hybridqa_2021, tatqa2021, zhao-etal-2022-multihiertt, cheng-etal-2022-hitab}, and (ii)~\textit{Complete Containment (CC Score)}--a binary score that evaluates to 1 iff the recall is 100\%, otherwise  it is 0. 
\revision{The two metrics are complementary: 
F1 rewards partial correctness of the answer while CC rewards an answer only if it contains the ground truth in its entirety. 
For example, assume the ground truth is \textit{[``Apple", ``300"]} and three distinct models output:  
\textit{M1 [``Apple", ``300"]}, \textit{M2 [``Apple"]}, and \textit{M3 [``Apple", ``300", ``Orange'', ``10'']}, then their scores will be:
M1 (F1= 1.0, CC= 1), M2 (F1= 0.66, CC= 0), and M3 (F1= 0.66, CC= 1).}

\begin{table*}[h]
\centering
\vspace{-5pt}
\caption{Overall performance of models on \system}
\vspace{-8pt}
\label{tab:model_performance}
\resizebox{\textwidth}{!}{%
\begin{tabular}{l c cc|cc|cc|cc|cc|cc}
\toprule
\multirow{2}{*}{\textbf{Model Name}} &
\multirow{2}{*}{\textbf{Model Param Size}} &
\multicolumn{2}{c}{\textbf{QNPC}} &
\multicolumn{2}{c}{\textbf{PAK}} &
\multicolumn{2}{c}{\textbf{US CENSUS}} &
\multicolumn{2}{c}{\textbf{arXiv}} &
\multicolumn{2}{c}{\textbf{Synthetic Tables}} &
\multicolumn{2}{c}{\textbf{Average}} \\
& & \textbf{F1} & \textbf{CC Score} & \textbf{F1} & \textbf{CC Score} & \textbf{F1} & \textbf{CC Score} & \textbf{F1} & \textbf{CC Score} & \textbf{F1} & \textbf{CC Score} & \textbf{F1} & \textbf{CC Score} \\
\midrule

        \rowcolor{gray!20} \multicolumn{14}{c}{\textbf{TEXT-ONLY LLMs}} \\
        \rowcolor{blue!20} \multicolumn{14}{c}{\textbf{Large Models -- 100B+ (Closed Weight)}} \\
        ChatGPT 3.5 & 175B & 0.413 & 0.315 & 0.445 & 0.356 & 0.385 & 0.333 & 0.459 & 0.386 & 0.345 & 0.279 & \textbf{0.409} & \textbf{0.334} \\
ChatGPT 4o & Unknown & 0.663 & 0.571 & 0.720 & 0.639 & 0.650 & 0.583 & 0.719 & 0.654 & 0.563 & 0.365 & \textbf{0.663} & \textbf{0.562} \\
ChatGPT 4.1 & Unknown & 0.621 & 0.545 & 0.640 & 0.576 & 0.650 & 0.600 & 0.667 & 0.605 & 0.543 & 0.415 & \textbf{0.624} & \textbf{0.548} \\
        \textit{Average} & - & \textit{0.566} & \textit{0.477} & \textit{0.602} & \textit{0.524} & \textit{0.562} & \textit{0.505} & \textit{0.615} & \textit{0.548} & \textit{0.484} & \textit{0.353} & \textit{0.566} & \textit{0.481} \\
        \rowcolor{blue!20} \multicolumn{14}{c}{\textbf{Medium Models -- 27B–72B (Open Weight)}} \\
Llama-3.1-70B-Instruct & 70B & 0.586 & 0.484 & 0.639 & 0.551 & 0.585 & 0.519 & 0.632 & 0.555 & 0.512 & 0.323 & \textbf{0.591} & \textbf{0.486} \\
Llama-3.3-70B-Instruct & 70B & 0.573 & 0.497 & 0.636 & 0.571 & 0.548 & 0.530 & 0.645 & 0.595 & 0.499 & 0.361 & \textbf{0.580} & \textbf{0.511} \\
Qwen2.5-72B-Instruct & 72B & 0.621 & 0.532 & 0.673 & 0.593 & 0.589 & 0.540 & 0.678 & 0.612 & 0.582 & 0.339 & \textbf{0.629} & \textbf{0.523} \\
Qwen3-32B & 32B & 0.523 & 0.435 & 0.578 & 0.491 & 0.538 & 0.479 & 0.576 & 0.503 & 0.458 & 0.227 & \textbf{0.535} & \textbf{0.427} \\
gemma-2-27b-it & 27B & 0.549 & 0.444 & 0.617 & 0.514 & 0.509 & 0.466 & 0.631 & 0.552 & 0.433 & 0.207 & \textbf{0.548} & \textbf{0.437} \\
gemma-3-27b-it & 27B & 0.535 & 0.445 & 0.598 & 0.514 & 0.527 & 0.469 & 0.622 & 0.560 & 0.476 & 0.275 & \textbf{0.552} & \textbf{0.453} \\
\textit{Average (Medium)} & - & \textit{0.565} & \textit{0.473} & \textit{0.624} & \textit{0.539} & \textit{0.550} & \textit{0.500} & \textit{0.631} & \textit{0.563} & \textit{0.493} & \textit{0.289} & \textit{0.572} & \textit{0.473} \\

        \rowcolor{blue!20} \multicolumn{14}{c}{\textbf{Small Models -- 4B–20B (Open Weight)}} \\
        
gemma-2-9b-it & 9B & 0.475 & 0.360 & 0.537 & 0.413 & 0.409 & 0.357 & 0.497 & 0.422 & 0.383 & 0.171 & \textbf{0.460} & \textbf{0.345} \\
gemma-3-12b-it & 12B & 0.498 & 0.371 & 0.545 & 0.412 & 0.465 & 0.401 & 0.543 & 0.443 & 0.449 & 0.195 & \textbf{0.500} & \textbf{0.364} \\
Qwen1.5-7B-Chat & 7B & 0.220 & 0.173 & 0.268 & 0.220 & 0.194 & 0.182 & 0.314 & 0.268 & 0.124 & 0.059 & \textbf{0.224} & \textbf{0.180} \\
Qwen2-7B-Instruct & 7B & 0.333 & 0.273 & 0.356 & 0.305 & 0.283 & 0.289 & 0.375 & 0.341 & 0.295 & 0.107 & \textbf{0.328} & \textbf{0.263} \\
Qwen2.5-7B-Instruct & 7B & 0.417 & 0.343 & 0.454 & 0.384 & 0.361 & 0.383 & 0.483 & 0.445 & 0.334 & 0.124 & \textbf{0.410} & \textbf{0.336} \\
Qwen3-14B & 14B & 0.490 & 0.430 & 0.535 & 0.482 & 0.458 & 0.454 & 0.577 & 0.547 & 0.394 & 0.156 & \textbf{0.491} & \textbf{0.414} \\
Mistral-7B-Instruct-v0.1 & 7B & 0.071 & 0.062 & 0.077 & 0.065 & 0.032 & 0.027 & 0.143 & 0.129 & 0.062 & 0.040 & \textbf{0.077} & \textbf{0.065} \\
Mistral-7B-Instruct-v0.2 & 7B & 0.119 & 0.108 & 0.120 & 0.123 & 0.137 & 0.124 & 0.169 & 0.173 & 0.059 & 0.035 & \textbf{0.121} & \textbf{0.112} \\
Mistral-7B-Instruct-v0.3 & 7B & 0.295 & 0.219 & 0.321 & 0.255 & 0.253 & 0.222 & 0.342 & 0.298 & 0.151 & 0.068 & \textbf{0.273} & \textbf{0.213} \\
Mathstral-7B-v0.1 & 7B & 0.326 & 0.266 & 0.387 & 0.320 & 0.184 & 0.169 & 0.362 & 0.329 & 0.215 & 0.097 & \textbf{0.295} & \textbf{0.236} \\
Meta-Llama-3-8B-Instruct & 8B & 0.344 & 0.241 & 0.374 & 0.268 & 0.334 & 0.302 & 0.376 & 0.301 & 0.262 & 0.100 & \textbf{0.338} & \textbf{0.242} \\
Llama-3.1-8B-Instruct & 8B & 0.363 & 0.257 & 0.393 & 0.278 & 0.327 & 0.290 & 0.416 & 0.342 & 0.310 & 0.119 & \textbf{0.362} & \textbf{0.257} \\
Phi-3-mini-4k-instruct & 4B & 0.233 & 0.167 & 0.265 & 0.185 & 0.223 & 0.198 & 0.339 & 0.267 & 0.211 & 0.107 & \textbf{0.254} & \textbf{0.185} \\
Phi-3.5-vision-instruct & 4B & 0.274 & 0.193 & 0.235 & 0.170 & 0.190 & 0.143 & 0.245 & 0.189 & 0.108 & 0.059 & \textbf{0.211} & \textbf{0.151} \\
Phi-4-mini-instruct & 4B & 0.225 & 0.156 & 0.131 & 0.086 & 0.044 & 0.037 & 0.154 & 0.116 & 0.062 & 0.021 & \textbf{0.123} & \textbf{0.083} \\
Fanar-C-9B-Instruct & 9B & 0.431 & 0.322 & 0.463 & 0.351 & 0.473 & 0.416 & 0.522 & 0.451 & 0.437 & 0.217 & \textbf{0.465} & \textbf{0.351} \\
\textit{Average (Small)} & - & \textit{0.320} & \textit{0.246} & \textit{0.341} & \textit{0.270} & \textit{0.273} & \textit{0.249} & \textit{0.366} & \textit{0.316} & \textit{0.241} & \textit{0.105} & \textit{0.308} & \textit{0.237} \\

        \midrule

        \rowcolor{gray!20} \multicolumn{14}{c}{\textbf{VISION-TEXT VLMs}} \\
        \rowcolor{blue!20} \multicolumn{14}{c}{\textbf{Large Models -- 100B+ (Closed Weight)}} \\
ChatGPT 4o Vision & Unknown & 0.543 & 0.513 & 0.638 & 0.613 & 0.348 & 0.328 & 0.618 & 0.597 & - & - & \textbf{0.537} & \textbf{0.513} \\

        \rowcolor{blue!20} \multicolumn{14}{c}{\textbf{Small Models -- 3B–12B (Open Weight)}} \\
Pixtral-12B-2409 & 12B & 0.378 & 0.289 & 0.471 & 0.372 & 0.300 & 0.273 & 0.457 & 0.411 & 0.364 & 0.161 & \textbf{0.394} & \textbf{0.301} \\
InternVL2-4B & 4B & 0.236 & 0.181 & 0.281 & 0.199 & 0.148 & 0.162 & 0.305 & 0.297 & 0.265 & 0.102 & \textbf{0.247} & \textbf{0.188} \\
InternVL2\_5-8B-MPO & 8B & 0.193 & 0.129 & 0.270 & 0.196 & 0.177 & 0.170 & 0.349 & 0.288 & 0.208 & 0.076 & \textbf{0.240} & \textbf{0.172} \\
Phi-3-vision-128k-instruct & 4B & 0.214 & 0.187 & 0.280 & 0.252 & 0.210 & 0.199 & 0.339 & 0.317 & 0.241 & 0.135 & \textbf{0.257} & \textbf{0.218} \\
Phi-3.5-vision-instruct & 4B & 0.208 & 0.138 & 0.314 & 0.227 & 0.148 & 0.120 & 0.332 & 0.269 & 0.265 & 0.133 & \textbf{0.253} & \textbf{0.177} \\
llava-1.5-7b-hf & 7B & 0.067 & 0.061 & 0.096 & 0.092 & 0.067 & 0.057 & 0.042 & 0.039 & 0.030 & 0.006 & \textbf{0.061} & \textbf{0.051} \\
llava-v1.6-vicuna-7b-hf & 7B & 0.123 & 0.091 & 0.157 & 0.118 & 0.095 & 0.077 & 0.133 & 0.107 & 0.115 & 0.032 & \textbf{0.125} & \textbf{0.085} \\
Molmo-7B-D-0924 & 7B & 0.130 & 0.067 & 0.164 & 0.100 & 0.092 & 0.075 & 0.147 & 0.103 & 0.156 & 0.027 & \textbf{0.138} & \textbf{0.074} \\

\textit{Average (Small)} & - & \textit{0.194} & \textit{0.143} & \textit{0.254} & \textit{0.194} & \textit{0.155} & \textit{0.142} & \textit{0.263} & \textit{0.229} & \textit{0.205} & \textit{0.084} & \textit{0.214} & \textit{0.158} \\

        \bottomrule
        \hline
    \end{tabular}
    }
\\[-2pt]
\vspace{2mm} 
{\footnotesize\textit{Note: ChatGPT-4o Vision was not run on synthetic \hct images due to API constraints}}
\vspace{-16pt}
\end{table*}

\subsection{Analysis and findings}
\label{section:analysis-and-findings}



\textbf{(RQ1) How do language models perform overall on \system?} 
As shown in Table~\ref{tab:model_performance}, the relatively recent versions of large closed-weight LLMs like ChatGPT 4o and ChatGPT 4.1 on the text modality perform reasonably well, achieving F1 scores of 66\% and 62\% respectively. This shows that even for state-of-the-art LLMs, there is ample room for improvement on this task. Despite ChatGPT 4o being the best performing model overall, we see that its predecessor ChatGPT 3.5 scores an average F1 of 40\% which is less \revision{than} all of the medium sized open-weight LLMs. This highlights the progress more recent LLMs have made in their reasoning abilities despite their much smaller size. Furthermore, Qwen2.5-72B-Instruct achieves an average F1 of 62.9\%, which is only 3.4 percentage points less than ChatGPT 4o's score, despite its size of 72B parameters being considerably smaller than the OpenAI model. \revision{Similarly, Qwen2.5-72B-Instruct's CC Score of 52\% is only 4 percentage points less that ChatGPT 4o CC Score of 56\%.} On average, large LLMs achieve an F1 of 56\% compared to 57\% for medium-sized LLMs. Even, when excluding the lagging ChatGPT 3.5, large LLMs have an average F1 score of 64\% which is only 7 percentage points higher than the medium-sized LLMs' average. This showcases the potential for medium-sized LLMs on this task. 

With an average F1 of 30\%, small LLMs lag behind their larger competitors.
Even among small LLMs, gemma-3-12b-it's average F1 score of 50\% outperforms ChatGPT 3.5 by nearly 10 percentage points. \revision{However, we see a considerable drop in gemma-3-12b-it's CC Score; only 36.4\% which is far below the average for medium and large models. This indicates that it does get a lot of the questions partially correct but struggles to provide the fully correct answer.} Moreover, small VLMs with an average F1 of 21.4\% are far worse off than small LLMs. But, 
among the VLMs, Pixtral-12B outperforms many LLMs such as Llama-3.1-8B-Instruct, Mistral-7B-Instruct-v0.3, and Qwen2-7B-Instruct. This indicates that VLMs do have the potential to compete with similarly sized LLMs on \system, possibly by utilizing the visual features of \hcts that may be lost in a textual representation. Overall, there remains clear room for improvement for both LLMs and VLMs.
\revision{Furthermore, the F1 scores of models on synthetic \hcts is, on average, slightly less than on real-world \hcts. This shows that our synthetic generator produces \hcts and QA pairs that are challenging for all models.
However, there is a considerable drop in the CC Scores on synthetic \hcts. This could be due to many of the synthetic QA pairs requiring answers with many more values than the QA pairs on real-world \hcts--as the number of values in the answer increases, the models' ability to generate a completely correct answer significantly degrades.
For example, the \textit{Qwen2.5-72B-Instruct} model under the synthetic HCT tables  has the  highest F1 score while its CC score is relatively low.}

\begin{figure}[t]
    \centering
    \includegraphics[width=.9\linewidth]{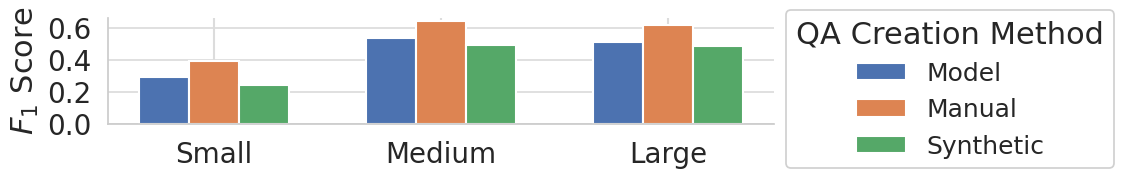}
    \vspace{-12pt}
    \caption{\revision{Model performance by size across different QA creation types}}
    \vspace{-6pt}
    \label{fig:question_creation_type_scores_barplot} 
\end{figure}

\revision{In Figure~\ref{fig:question_creation_type_scores_barplot} we also compare how models perform on QA pairs created using our three methods.
Synthetic questions appear to be the most challenging which is expected as the majority require \textit{Row Filtering}, \textit{Expression on Columns}, and \textit{Aggregation} together. On average, models of all sizes perform best on the manually-generated QA pairs. This indicates that our model-generated and synthetic questions provide a slightly harder challenge for 
the evaluated models.} 

\textbf{(RQ2) Are models improving over time?} 
There appears to be no consistent trend with regards to newer generations of models outperforming their predecessors. The Qwen family of small LLMs shows continuous improvement as Qwen1.5-7B, Qwen2-7B, Qwen2.5-7B, and Qwen3-14B all outperform their predecessor with average F1 scores of 22.4\%, 32.8\%, 41\%, and 49.1\% respectively. This shows that the Qwen family is improving the performance of their small LLMs by an average of 9\% per generation. \revision{The same can be seen with their CC Scores as Qwen3-14B is 23 percentage points better than Qwen1.5-7B-Chat. This shows that the models are getting much better at 
giving the complete correct answer.} We see similar trends with the Mistral 7B LLMs which also improve by nearly 9\% per generation and Gemma LLMs that improve by 4\%. 
Conversely,  the Phi family of small LLMs shows consistent deterioration in each successive generation with Phi-3-mini-4k-instruct achieving an average F1 score of 25.4\% which is nearly double that of Phi-4-mini-instruct, despite the latter being released 8 months later. The ChatGPT family shows a mixed trend with a significant improvement of 26 percentage points between ChatGPT 3.5 and ChatGPT 4.0, but then ChatGPT 4.1 under-performs its predecessor by 4 percentage points. There are also cases like gemma-2-27b-it and gemma-3-27b-it which shows the newer generation improving by a mere 0.4 percentage points in terms of average F1. We also see a similar trend with small VLMs as Phi-3-vision-128k-instruct outperforms its successor, Phi-3.5-vision-instruct, by 0.4 percentage points. Thus, we cannot say that newer versions of models 
outperform their older versions across the board. 

\begin{table}[t]
    \centering
    \scriptsize
    \caption{Model F1 scores across different \hcts input formats}
    \vspace{-12ptc}
    \label{tab:model_performance_format_variation_f1}
    \resizebox{\linewidth}{!}{%
    \begin{tabular}{lcccc}
        \toprule
        \textbf{Model} & \textbf{CSV} & \textbf{Markdown} & \textbf{HTML} & \textbf{Text Obfuscated HTML} \\
        \midrule
ChatGPT 4.1 & 0.484 & 0.529 & 0.543 & 0.499 \\
ChatGPT 4o & 0.541 & 0.567 & 0.563 & 0.531 \\
Qwen2.5-72B-Instruct & 0.422 & 0.453 & 0.582 & 0.508 \\
Llama-3.3-70B-Instruct & 0.431 & 0.443 & 0.499 & 0.398 \\
Qwen2.5-7B-Instruct & 0.301 & 0.311 & 0.334 & 0.358 \\
Llama-3.1-8B-Instruct & 0.191 & 0.262 & 0.311 & 0.277 \\
\textit{Average} & \textit{0.395} & \textit{0.426} & \textit{0.472} & \textit{0.42} \\

        \bottomrule
    \end{tabular}%
    }
    \vspace{-10pt}
\end{table}

\begin{table}[t]
    \centering
    \caption{Finetuned model performances across different test sets.}
    \vspace{-6pt}
    \label{tab:finetune_model_performance_f1}
    \resizebox{\linewidth}{!}{
    \begin{tabular}{l c ccc}
        \toprule
        \multirow{2}{*}{\textbf{Model Name}} & 
        \multirow{2}{*}{\textbf{Finetuning Set}} & 
        \multicolumn{3}{c}{\textbf{Test Sets (F1 Score)}} \\
        \cmidrule(lr){3-5}
        & & \textbf{Real HCTs} & \textbf{Synthetic HCTs} & \textbf{HiTab~\cite{cheng-etal-2022-hitab}} \\
        \midrule
Llama-3.1-8B-Instruct & None & 0.351 & 0.333 & 0.538 \\
Llama-3.1-8B-Instruct & Real HCTs & 0.593 & 0.534 & 0.606 \\
Llama-3.1-8B-Instruct & Synth. HCTs & 0.471 & 0.879 & 0.561 \\
Llama-3.1-8B-Instruct & Real \& Synth. HCTs & 0.595 & 0.886 & 0.616 \\
        \bottomrule
    \end{tabular}
    }
    \vspace{-12pt}
\end{table}


\textbf{(RQ3) Does the input format of the \hct matter in text-modality?} 
For the text modality LLMs, we tested the effect of representing the \hct in CSV and Markdown formats instead of HTML. The results are shown in Table~\ref{tab:model_performance_format_variation_f1}.
The HTML format outperforms CSV and Markdown formats by an average F1 score difference of 7.7 and 4.6 percentage points, respectively. This result is intuitive as HTML is a more expressive format that can better retain the complex structure of the \hcts. Some models are more sensitive than others to the table format. Qwen2.5-72B-Instruct performs 13 percentage points better with the HTML format than with Markdown and shows a similar difference between the HTML and CSV formats as well. Whereas, ChatGPT 4o shows a relative indifference to the table format as there is only a 0.2 percentage point difference in F1 scores between the Markdown and HTML formats, with the former performing minutely better. Our results show that while large LLMs like ChatGPT can perform well with both Markdown and HTML, medium and small sized LLMs perform much better with an HTML representation of the \hct.

We also explore the effect of semantically coherent and meaningful text within the \hcts. 
We created \textit{Text Obfuscated-HTML} versions of the \hct HTMLs where we (i) replace all column names with ambiguous terms 
(ii) replace all text cell values in the \hcts with similar ambiguous terms, and 
(iii) replace the corresponding QA pairs with the respective ambiguous terms from the obfuscated table. 
Although the complexity of the questions and table structure remains unchanged, we see that the text obfuscation negatively impacts model performance. LLMs on average, experience a decrease in F1 score of 5 percentage points on Text Obfuscated-HTML format when compared to the original HTMLs. Even a large model like ChatGPT 4.1 cannot compensate for the lack of semantics as its F1 score drops by 4.4 percentage points. 

\textbf{(RQ4) Does finetuning models improve performance?}
We finetuned the Llama-3.1-8B-Instruct model on three FT sets: 
(i)~\textit{Real HCTs} which has 7,867 (80\% of total) \hct-QA triplets on real world \hcts only, (ii)~\textit{Synth. HCTs} which has 54,186 (80\% of total) \hct-QA triplets on synthetic \hcts only, 
(iii)~\textit{Real + Synth. HCTs} which combines both the ~\textit{Real HCTs} and ~\textit{Synth. HCTs} sets (62,053 in total). 
We then evaluate the models on the respective \textit{Real HCTs} and \textit{Synth. HCTs} test sets. We ensure that no \hct in the test set is seen during finetuning. 
Furthermore, to test the generalization of finetuning in \system we test on the HiTab dataset~\cite{cheng-etal-2022-hitab}. 
The results are shown in Table~\ref{tab:finetune_model_performance_f1}. 
Finetuning on the \textit{Real + Synth. HCTs} set drastically increases the performance of the model across the board. The F1 score improvement, compared to the non-finetuned model is 24.1 percentage points on real \hcts from \system, 55 percentage points on synthetic \hcts from \system, and nearly 8 percentage points on the out of domain HiTab dataset. Furthermore, finetuning on the \textit{Synth. HCTs} set yields an F1 score improvement of 12 percentage points on the real \hcts test set. This reinforces the usefulness of our synthetic data generator since  training only on synthetic \hcts generalizes well to real-world \hcts.

\iftrue 
\begin{table}[t]
\setlength{\tabcolsep}{3pt} 
\scriptsize 
    \centering
    \vspace{2pt}
    \caption{Effect of table properties on F1. Only properties with significant effect ($p<0.05$) are shown}
    \vspace{-8pt}
    \label{tab:regression_results}
    \begin{tabular}{lclc}
        \hline
        Table Properties & Coef. (p-value) & Table Properties & Coef. (p-value) \\
        \hline
        Symmetric Row Nesting & -0.103 (0.016) &  Global Column Aggreg. & -0.035 (0.028)\\
        Asymmetric Row Nesting & -0.142 (0.001) & Implicit Col. Aggreg. & 0.081 (0.001) \\
        Balanced Col. Nesting & -0.270 (0.022) & Implicit Row Aggreg. & -0.065 (0.007)\\
        Unbalanced Col. Nesting & -0.315 (0.009) & Row Group Label & -0.069 (0.000)\\
        \hline
    \end{tabular}
    \vspace{-12pt}
\end{table}
\fi

\begin{figure}[t]
    \centering
    \begin{tabular}{m{0.1cm}m{8.2cm}}
    (a)&\includegraphics[width=\linewidth]{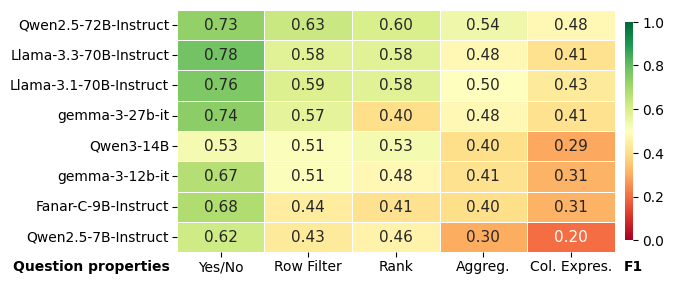}\\
    (b)&\includegraphics[width=\linewidth]{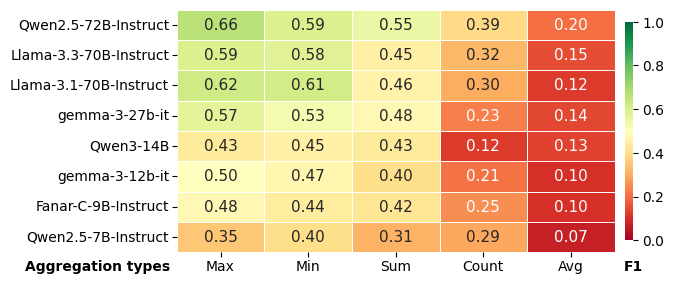}\\
    \end{tabular}
    \vspace{-10pt}
    \caption{\revision{F1 scores for models across a subset of 
    question properties~(a) and aggregation types~(b). Models are ordered by decreasing size from top to bottom, and by decreasing average scores from left to right.}}
    \vspace{-10pt}
    \label{fig:heamtap_per_question_prop_scores}
\end{figure}

\textbf{(RQ5) How do \hct properties affect model performance?} 
Table~\ref{tab:regression_results} shows the regression coefficient of various \hct properties against F1 scores. 
We can see that the presence of most of these properties, except \textit{Implicit Column Aggregation Terms}, decreases the F1 score on average. We also see a clear trend in models being able to better understand symmetrical and balanced nesting patterns compared to asymmetrical and unbalanced ones: \textit{Symmetric Row Nesting} decreases F1 by~10 percentage points whereas \textit{Asymmetric Row Nesting} decreases it by~14 percentage points. The same trend is seen when comparing \textit{Balanced Column Nesting} with \textit{Unbalanced Column Nesting} with the former having a decrease of 27 percentage points on F1 against 31.5 percentage points for the latter. Column and row nesting increases the \hct difficulty more than any type of aggregate rows or columns, having the largest negative impact on F1 scores. 


\textbf{(RQ6) How do question properties affect model performance?} 
Our question metadata allows for fine-grained analysis to understand what types of questions different models excel at and struggle with. Figure~\ref{fig:heamtap_per_question_prop_scores} shows 
such an analysis on a subset of models and metadata. Heatmap (A) shows that all models struggle with questions requiring aggregations or expressions on columns. 
We can also see that Qwen3-14B, despite being better than models like Fanar-C-9B and gemma-3-27b at questions involving ranking and row filtering, pales in comparison to both on yes/no questions. Furthermore, Qwen2.5-72B shows good performance on aggregation questions compared to 
other models.
\revision{In general, models perform better with yes/no and row filter questions as this type of question is answered directly from the data cells, similarly to scan, filter and project stateless operators in databases (i.e., a single row/cell is enough to answer the question), while ranking is more challenging for the models because, similar to the \texttt{ORDER BY} stateful operator in databases, it requires a full view of the data. Aggregation is even more challenging because it operates on multiple cells, which can be in different locations of the \hct. 
In Figure~\ref{fig:heamtap_per_question_prop_scores}), we observe that, for example, end users would be better off using gemma-3-27b than the larger Llama-3.3-70B and Llama-3.1-70B models for questions requiring aggregation, row filtering, and yes/no answers, as their performance is nearly the same, while gemma-3-27b is much more efficient due to its smaller size.} 

\begin{figure}[t]
    \centering
    \includegraphics[width=\linewidth]{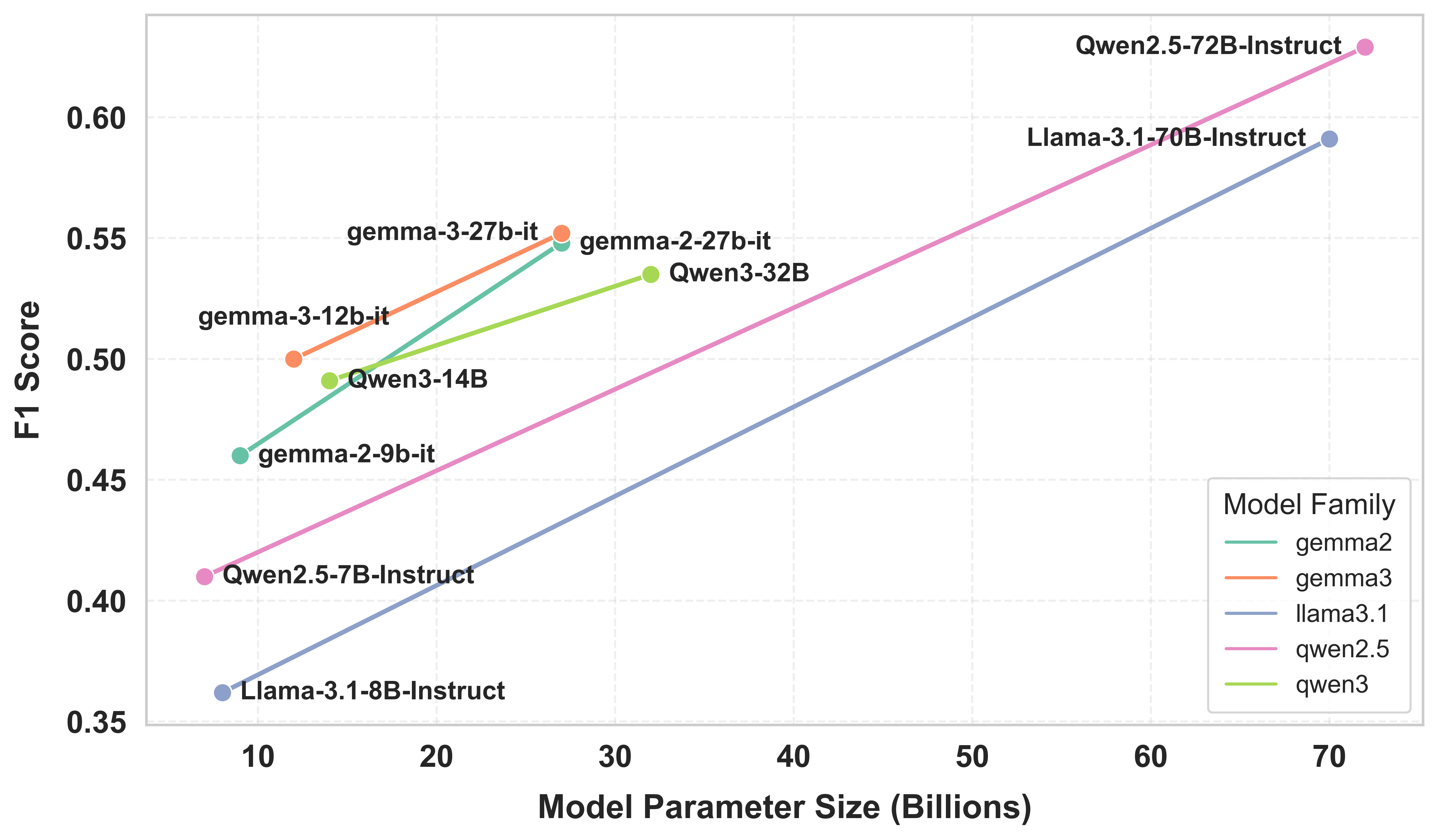}
    \vspace{-10pt}
    \caption{F1 score across model size (parameters) within model variants}
    \vspace{-10pt}
    \label{fig:size_instruct_progression_effects}
\end{figure}

We go one step deeper and analyze what types of aggregations these models struggle with in particular. Heatmap (B) \revision{in Figure~\ref{fig:heamtap_per_question_prop_scores}} shows the F1 scores of the models on the five different types of aggregation. We can clearly see that all models struggle significantly with averaging, presumably because this involves a multi-step calculation. We can also see that both Llama-3.1-70B and Llama-3.3-70B are particularly good at min and max questions but struggle more with sum questions. Whereas, Qwen2.5-72B is the only model that performs reasonably well on sum questions which would explain how it outperforms the other models on aggregation questions in general. As expected, all models struggle with questions involving aggregation and particularly those that require averaging values. The metadata provided in \system allows for such granular analysis to be done for each individual model so that specific areas for improvement can be identified. 
\textbf{(RQ7) How are model size and performance correlated?} 
Figure~\ref{fig:size_instruct_progression_effects} shows how F1 score changes between different size variants of the same model. As expected, larger models consistently outperform their smaller counterparts. The gemma-2 models have the sharpest increase with $0.40\%$ 
F1 per 1B parameters (F1$/$B). Even when scaling to larger sizes like with Qwen2.5 and Llama-3.1 we see an increase of $0.34\%$ F1$/$B 
and $0.37\%$ F1$/$B 
, respectively. 
Hence, it is worth exploring fine-tuning smaller models for this task rather than simply relying on larger, resource-hungry off-the-shelf models.



\section{Conclusion and future work}
\label{section-conclusion}
In this paper, we presented \system, a comprehensive benchmark designed to evaluate language models on question answering over human-centric tables (\hcts), typically embedded within documents and web pages.

Our benchmark reveals several key insights.
First, both LLMs and VLMs exhibit strong potential for question answering on \hcts. 
Second, finetuning LLMs substantially enhances performance, 
even when finetuned on only synthetic data, models demonstrated strong generalization to real-world \hcts.
Finally, even the best-performing models (e.g., ChatGPT 4o) continue to face challenges with complex queries (e.g., involving aggregations) on complex \hct structures. 
\revision{
Futhermore, our benchmark can be beneficial to other researchers in at least two ways:
(i)~Using the synthetic data generator to create training and test \hcts and QA pairs for new thematic domains in a cheap and quick manner. This is especially useful for training, finetuning, and evaluating new models.
(ii)~Adding  question templates to the generator to cover more types of questions.}

\stitle{Future Work.} \revision{Our benchmark can be extended along several complementary directions.}
\revision{First, we plan to broaden the scope of supported queries by enabling cross-\hct question answering, e.g.,  queries involving joins or unions across multiple \hcts. We also aim to incorporate queries that require more advanced OLAP operations, including \texttt{CUBE} and \texttt{PIVOT}. Initial experiments with two medium-sized LLMs indicate that these models are currently unable to handle such operators, highlighting substantial room for improvement.}

\revision{Second, we intend to conduct deeper studies on fine-tuning strategies for both LLMs and VLMs to improve their structural understanding of \hcts. In addition, we will evaluate emerging systems that go beyond current LLM/VLM architectures, as these models may offer improved reasoning capabilities for complex tabular and visual data.}

\revision{Third, we plan to enhance the linguistic diversity of the synthetic NL questions. One direction is use question paraphrasing capabilities using LLMs, which would increase lexical variation in the generated QA pairs. This extension will require robust validation strategies to ensure semantic equivalence between the original and paraphrased questions.
Furthermore, the generator can be enriched with additional NL templates that capture implicit comparative terms such as \textit{most}, \textit{least}, \textit{highest}, and \textit{lowest}, instead of relying solely on explicit SQL-style operators such as \textit{MIN} and \textit{MAX}.}

\clearpage
\newpage

\section{AI-Generated Content Acknowledgment}
\label{ai_generated}



Here are the different places where we used AI models in our work:
\begin{itemize}

\item LLMS were used to generate natural language questions and answers for the benchmark dataset as explained in Sections~\ref{section:tableProperties}. 

\item The LandingLens from LandingAI tool was used for automated extraction and conversion of tabular data from PDF documents into HTML format (Section~\ref{section:tableProperties}).

\item Azure AI Document Intelligence was used to extract table images from PDF documents 
(Section~\ref{section:tableProperties}).

\end{itemize}

{\small{
\bibliographystyle{IEEEtran}
\bibliography{refs}

@misc{hctqa_repo,
  title     = {{HCT-QA: A Benchmark for Question Answering
on Human-Centric Tables}},
  year      = {2025},
  howpublished = {\url{https://github.com/qcri/HCTQA-Benchmark}}
}

@misc{HCT_QA_hf,
  title = {{HCTQA benchmark data.}},
  howpublished = {\url{https://huggingface.co/datasets/qcri-ai/HCTQA}},
  note = {Accessed: 2025-05-15}
}

@misc{landingai,
  author    = {{Landing AI}},
  title     = {{LandingLens: AI-powered data annotation and model training platform}},
  year      = {2025},
  howpublished = {\url{https://landing.ai}},
  note      = {Accessed: 2025-10-18}
}

@inproceedings{zhao-etal-2022-multihiertt,
    title = "{MultiHiertt: Numerical Reasoning over Multi Hierarchical Tabular and Textual Data}",
    author = "Zhao, Yilun  and
      Li, Yunxiang  and
      Li, Chenying  and
      Zhang, Rui",
    editor = "Muresan, Smaranda  and
      Nakov, Preslav  and
      Villavicencio, Aline",
    booktitle = "Proceedings of the 60th Annual Meeting of the Association for Computational Linguistics (Volume 1: Long Papers)",
    month = may,
    year = "2022",
    address = "Dublin, Ireland",
    publisher = "Association for Computational Linguistics",
    url = "https://aclanthology.org/2022.acl-long.454/",
    doi = "10.18653/v1/2022.acl-long.454",
    pages = "6588--6600"
}

@inproceedings{zhang-etal-2025-tablellm,
    title = {{TableLLM: Enabling Tabular Data Manipulation by LLMs in Real Office Usage Scenarios}},
    author = "Zhang, Xiaokang  and
      Luo, Sijia  and
      Zhang, Bohan  and
      Ma, Zeyao  and
      Zhang, Jing  and
      Li, Yang  and
      Li, Guanlin  and
      Yao, Zijun  and
      Xu, Kangli  and
      Zhou, Jinchang  and
      Zhang-Li, Daniel  and
      Yu, Jifan  and
      Zhao, Shu  and
      Li, Juanzi  and
      Tang, Jie",
    editor = "Che, Wanxiang  and
      Nabende, Joyce  and
      Shutova, Ekaterina  and
      Pilehvar, Mohammad Taher",
    booktitle = "Findings of the Association for Computational Linguistics: ACL 2025",
    month = jul,
    year = "2025",
    address = "Vienna, Austria",
    publisher = "Association for Computational Linguistics",
    url = "https://aclanthology.org/2025.findings-acl.538/",
    doi = "10.18653/v1/2025.findings-acl.538",
    pages = "10315--10344",
    ISBN = "979-8-89176-256-5"
}

@inproceedings{ratsql-paper,
    title = "{RAT-SQL: Relation-Aware Schema Encoding and Linking for Text-to-SQL Parsers}",
    author = "Wang, Bailin  and
      Shin, Richard  and
      Liu, Xiaodong  and
      Polozov, Oleksandr  and
      Richardson, Matthew",
    editor = "Jurafsky, Dan  and
      Chai, Joyce  and
      Schluter, Natalie  and
      Tetreault, Joel",
    booktitle = "Proceedings of the 58th Annual Meeting of the Association for Computational Linguistics",
    month = jul,
    year = "2020",
    address = "Online",
    publisher = "Association for Computational Linguistics",
    url = "https://aclanthology.org/2020.acl-main.677/",
    doi = "10.18653/v1/2020.acl-main.677",
    pages = "7567--7578"
}

@article{catsql-paper,
author = {Fu, Han and Liu, Chang and Wu, Bin and Li, Feifei and Tan, Jian and Sun, Jianling},
title = {{CatSQL: Towards Real World Natural Language to SQL Applications}},
year = {2023},
issue_date = {February 2023},
publisher = {VLDB Endowment},
volume = {16},
number = {6},
issn = {2150-8097},
url = {https://doi.org/10.14778/3583140.3583165},
doi = {10.14778/3583140.3583165},
journal = {Proc. VLDB Endow.},
month = feb,
pages = {1534–1547},
numpages = {14}
}

@INPROCEEDINGS{garnl2sql-paper,
  author={Fan, Yuankai and He, Zhenying and Ren, Tonghui and Guo, Dianjun and Chen, Lin and Zhu, Ruisi and Chen, Guanduo and Jing, Yinan and Zhang, Kai and Wang, X.Sean},
  booktitle={2023 IEEE 39th International Conference on Data Engineering (ICDE)}, 
  title={Gar: A Generate-and-Rank Approach for Natural Language to {SQL} Translation}, 
  year={2023},
  volume={},
  number={},
  pages={110-122},
  doi={10.1109/ICDE55515.2023.00016}}

@misc{valuenet-icde-paper,
      title={ValueNet: A Natural {Language}-to-{SQL} System that Learns from Database Information}, 
      author={Ursin Brunner and Kurt Stockinger},
      year={2021},
      eprint={2006.00888},
      archivePrefix={arXiv},
      primaryClass={cs.DB},
      url={https://arxiv.org/abs/2006.00888}, 
}

@article{birdbench-paper,
  title={Can llm already serve as a database interface? a big bench for large-scale database grounded text-to-sqls},
  author={Li, Jinyang and Hui, Binyuan and Qu, Ge and Yang, Jiaxi and Li, Binhua and Li, Bowen and Wang, Bailin and Qin, Bowen and Geng, Ruiying and Huo, Nan and others},
  journal={Advances in Neural Information Processing Systems},
  volume={36},
  year={2024}
}

@misc{omnitab-paper,
      title={OmniTab: Pretraining with Natural and Synthetic Data for Few-shot Table-based Question Answering}, 
      author={Zhengbao Jiang and Yi Mao and Pengcheng He and Graham Neubig and Weizhu Chen},
      year={2022},
      eprint={2207.03637},
      archivePrefix={arXiv},
      primaryClass={cs.CL},
      url={https://arxiv.org/abs/2207.03637}, 
}

@inproceedings{Khan2021ICDAR_13,
author = {Khan, Umar and Zahid, Sohaib and Ali, Muhammad Asad and Ul-Hasan, Adnan and Shafait, Faisal},
title = {{TabAug: Data Driven Augmentation for Enhanced Table Structure Recognition}},
year = {2021},
isbn = {978-3-030-86330-2},
publisher = {Springer-Verlag},
address = {Berlin, Heidelberg},
url = {https://doi.org/10.1007/978-3-030-86331-9_38},
doi = {10.1007/978-3-030-86331-9_38},
booktitle = {Document Analysis and Recognition – ICDAR 2021: 16th International Conference, Lausanne, Switzerland, September 5–10, 2021, Proceedings, Part II},
pages = {585–601},
numpages = {17},
location = {Lausanne, Switzerland}
}

@inproceedings{Oro2008ICDAR_14,
  author={Oro, Ermelinda and Ruffolo, Massimo},
  booktitle={2009 10th International Conference on Document Analysis and Recognition}, 
  title={PDF-TREX: An Approach for Recognizing and Extracting Tables from PDF Documents}, 
  year={2009},
  volume={},
  number={},
  pages={906-910},
  doi={10.1109/ICDAR.2009.12}
}

@misc{aspose,
  title = {{Aspose}},
  howpublished  = {https://docs.aspose.com/pdf},
  note = {Accessed: 2025-05-15}
}

@misc{qnpc,
  title = {{Qatar National Planning Council}},
  howpublished  = {https://www.npc.qa/en/statistics},
  note = {Accessed: 2025-05-15}
}

@misc{uscensus,
  title = {{US Census Bureau}},
  howpublished  = {https://www.census.gov/},
  note = {Accessed: 2025-05-15}
}

@misc{pakcensus,
  title = {{Pakistan Bureau of Statistics}},
  howpublished  = {https://www.pbs.gov.pk/},
  note = {Accessed: 2025-05-15}
}

@misc{arxiv,
  title = {{arXiv}},
  howpublished  = {https://arxiv.org/},
  note = {Accessed: 2025-05-15}
}

@misc{smolDocling_2025,
      title={{SmolDocling}: An ultra-compact vision-language model for end-to-end multi-modal document conversion}, 
      author={Ahmed Nassar and Andres Marafioti and Matteo Omenetti and Maksym Lysak and Nikolaos Livathinos and Christoph Auer and Lucas Morin and Rafael Teixeira de Lima and Yusik Kim and A. Said Gurbuz and Michele Dolfi and Miquel Farré and Peter W. J. Staar},
      year={2025},
      eprint={2503.11576},
      archivePrefix={arXiv},
      primaryClass={cs.CV},
      url={https://arxiv.org/abs/2503.11576}, 
      note={\url{https://arxiv.org/abs/2503.11576}}, 
}

@inproceedings{yu-etal-2018-spider,
    title = "{S}pider: A Large-Scale Human-Labeled Dataset for Complex and Cross-Domain Semantic Parsing and {Text}-to-{SQL} Task",
    author = "Yu, Tao  and
      Zhang, Rui  and
      Yang, Kai  and
      Yasunaga, Michihiro  and
      Wang, Dongxu  and
      Li, Zifan  and
      Ma, James  and
      Li, Irene  and
      Yao, Qingning  and
      Roman, Shanelle  and
      Zhang, Zilin  and
      Radev, Dragomir",
    editor = "Riloff, Ellen  and
      Chiang, David  and
      Hockenmaier, Julia  and
      Tsujii, Jun{'}ichi",
    booktitle = "Proceedings of the 2018 Conference on Empirical Methods in Natural Language Processing",
    month = oct # "-" # nov,
    year = "2018",
    address = "Brussels, Belgium",
    publisher = "Association for Computational Linguistics",
    url = "https://aclanthology.org/D18-1425/",
    doi = "10.18653/v1/D18-1425",
    pages = "3911--3921"
}

@inproceedings{cheng-etal-2022-hitab,
    title = "{H}i{T}ab: A Hierarchical Table Dataset for Question Answering and Natural Language Generation",
    author = "Cheng, Zhoujun  and
      Dong, Haoyu  and
      Wang, Zhiruo  and
      Jia, Ran  and
      Guo, Jiaqi  and
      Gao, Yan  and
      Han, Shi  and
      Lou, Jian-Guang  and
      Zhang, Dongmei",
    editor = "Muresan, Smaranda  and
      Nakov, Preslav  and
      Villavicencio, Aline",
    booktitle = "Proceedings of the 60th Annual Meeting of the Association for Computational Linguistics (Volume 1: Long Papers)",
    month = may,
    year = "2022",
    address = "Dublin, Ireland",
    publisher = "Association for Computational Linguistics",
    url = "https://aclanthology.org/2022.acl-long.78/",
    doi = "10.18653/v1/2022.acl-long.78",
    pages = "1094--1110"
}

@INPROCEEDINGS{omniParser_2024,
  author={Wan, Jianqiang and Song, Sibo and Yu, Wenwen and Liu, Yuliang and Cheng, Wenqing and Huang, Fei and Bai, Xiang and Yao, Cong and Yang, Zhibo},
  booktitle={2024 IEEE/CVF Conference on Computer Vision and Pattern Recognition (CVPR)}, 
  title={OMNIPARSER: A Unified Framework for Text Spotting, Key Information Extraction and Table Recognition}, 
  year={2024},
  volume={},
  number={},
  pages={15641-15653},
  doi={10.1109/CVPR52733.2024.01481}}

@article{robustTableDetection_Ma_2022,
   title={Robust Table Detection and Structure Recognition from Heterogeneous Document Images},
   volume={133},
   ISSN={0031-3203},
   url={http://dx.doi.org/10.1016/j.patcog.2022.109006},
   DOI={10.1016/j.patcog.2022.109006},
   journal={Pattern Recognition},
   publisher={Elsevier BV},
   author={Ma, Chixiang and Lin, Weihong and Sun, Lei and Huo, Qiang},
   year={2023},
   month=jan, pages={109006} }

@INPROCEEDINGS{tableformer_2022,
  author={Nassar, Ahmed and Livathinos, Nikolaos and Lysak, Maksym and Staar, Peter},
  booktitle={2022 IEEE/CVF Conference on Computer Vision and Pattern Recognition (CVPR)}, 
  title={TableFormer: Table Structure Understanding with Transformers}, 
  year={2022},
  volume={},
  number={},
  pages={4604-4613},
  doi={10.1109/CVPR52688.2022.00457}}

@inproceedings{Qiao2021ICDAR_22,
    author = {Qiao, Liang and Li, Zaisheng and Cheng, Zhanzhan and Zhang, Peng and Pu, Shiliang and Niu, Yi and Ren, Wenqi and Tan, Wenming and Wu, Fei},
    title = {LGPMA: Complicated Table Structure Recognition with Local and Global Pyramid Mask Alignment},
    year = {2021},
    isbn = {978-3-030-86548-1},
    publisher = {Springer-Verlag},
    address = {Berlin, Heidelberg},
    url = {https://doi.org/10.1007/978-3-030-86549-8_7},
    doi = {10.1007/978-3-030-86549-8_7},
    booktitle = {Document Analysis and Recognition – ICDAR 2021: 16th International Conference, Lausanne, Switzerland, September 5–10, 2021, Proceedings, Part I},
    pages = {99–114},
    numpages = {16},
    location = {Lausanne, Switzerland}
}

@misc{innodata,
  title = {{Innodata}},
  howpublished = {https://solutions.innodata.com/document-intelligence/},
  note = {Accessed: 2025-05-15}
}

@misc{docugami,
  title = {{Docugami}},
  howpublished = {https://www.docugami.com/},
  note = {Accessed: 2025-05-15}
}

@inproceedings{Weir2020ACM_25,
    author = {Weir, Nathaniel and Utama, Prasetya and Galakatos, Alex and Crotty, Andrew and Ilkhechi, Amir and Ramaswamy, Shekar and Bhushan, Rohin and Geisler, Nadja and H\"{a}ttasch, Benjamin and Eger, Steffen and Cetintemel, Ugur and Binnig, Carsten},
    title = {DBPal: A Fully Pluggable NL2SQL Training Pipeline},
    year = {2020},
    isbn = {9781450367356},
    publisher = {Association for Computing Machinery},
    address = {New York, NY, USA},
    url = {https://doi.org/10.1145/3318464.3380589},
    doi = {10.1145/3318464.3380589},
    booktitle = {Proceedings of the 2020 ACM SIGMOD International Conference on Management of Data},
    pages = {2347–2361},
    numpages = {15},
    location = {Portland, OR, USA},
    series = {SIGMOD '20}
}

@inproceedings{Basik2018ACM_26,
author = {Basik, Fuat and H\"{a}ttasch, Benjamin and Ilkhechi, Amir and Usta, Arif and Ramaswamy, Shekar and Utama, Prasetya and Weir, Nathaniel and Binnig, Carsten and Cetintemel, Ugur},
title = {{DBPal: A Learned NL-Interface for Databases}},
year = {2018},
isbn = {9781450347037},
publisher = {Association for Computing Machinery},
address = {New York, NY, USA},
note = {\url{https://doi.org/10.1145/3183713.3193562}},
doi = {10.1145/3183713.3193562},
booktitle = {Proceedings of the 2018 International Conference on Management of Data},
pages = {1765–1768},
numpages = {4},
location = {Houston, TX, USA},
series = {SIGMOD '18}
}

@inproceedings{Wang2019_27,
  title={{RAT-SQL}: Relation-Aware Schema Encoding and Linking for {Text-to-SQL} Parsers},
  author={Bailin Wang and Richard Shin and Xiaodong Liu and Oleksandr Polozov and Matthew Richardson},
  booktitle={Annual Meeting of the Association for Computational Linguistics},
  year={2019},
  note = { \url{https://api.semanticscholar.org/CorpusID:207863446}}
}

@INPROCEEDINGS{Brunner2021ValueNetAN_28,
  author={Brunner, Ursin and Stockinger, Kurt},
  booktitle={2021 IEEE 37th International Conference on Data Engineering (ICDE)}, 
  title={{ValueNet: A Natural Language-to-SQL System that Learns from Database Information}}, 
  year={2021},
  volume={},
  number={},
  pages={2177-2182},
  doi={10.1109/ICDE51399.2021.00220}}

@article{sen2020VLDB_29,
author = {Sen, Jaydeep and Lei, Chuan and Quamar, Abdul and \"{O}zcan, Fatma and Efthymiou, Vasilis and Dalmia, Ayushi and Stager, Greg and Mittal, Ashish and Saha, Diptikalyan and Sankaranarayanan, Karthik},
title = {{ATHENA++: natural language querying for complex nested SQL queries}},
year = {2020},
issue_date = {August 2020},
publisher = {VLDB Endowment},
volume = {13},
number = {12},
issn = {2150-8097},
url = {https://doi.org/10.14778/3407790.3407858},
doi = {10.14778/3407790.3407858},
journal = {Proc. VLDB Endow.},
month = {jul},
pages = {2747–2759},
numpages = {13}
}

@article{AutoTables_57,
author = {Li, Peng and He, Yeye and Yan, Cong and Wang, Yue and Chaudhuri, Surajit},
title = {{Auto-Tables: Synthesizing Multi-Step Transformations to Relationalize Tables without Using Examples}},
year = {2023},
issue_date = {July 2023},
publisher = {VLDB Endowment},
volume = {16},
number = {11},
issn = {2150-8097},
url = {https://doi.org/10.14778/3611479.3611534},
doi = {10.14778/3611479.3611534},
journal = {Proc. VLDB Endow.},
month = {jul},
pages = {3391–3403},
numpages = {13}
}

@misc{tablevqa_2024,
      title={{TableVQA-Bench: A Visual Question Answering Benchmark on Multiple Table Domains}}, 
      author={Yoonsik Kim and Moonbin Yim and Ka Yeon Song},
      year={2024},
      eprint={2404.19205},
      archivePrefix={arXiv},
      primaryClass={cs.CV},
      url={https://arxiv.org/abs/2404.19205}, 
      note = {\url{https://arxiv.org/abs/2404.19205}}
}

@misc{openai2023gpt4-turbo,
  title = {{GPT-4 Turbo Model}},
  author = {OpenAI},
  year = {2023},
  note = {\url{https://platform.openai.com/docs/models/gpt-4-turbo}},
}

@misc{scitat_2024,
      title={{SCITAT: A Question Answering Benchmark for Scientific Tables and Text Covering Diverse Reasoning Types}}, 
      author={Xuanliang Zhang and Dingzirui Wang and Baoxin Wang and Longxu Dou and Xinyuan Lu and Keyan Xu and Dayong Wu and Qingfu Zhu and Wanxiang Che},
      year={2024},
      eprint={2412.11757},
      archivePrefix={arXiv},
      primaryClass={cs.CL},
      url = {https://arxiv.org/abs/2412.11757},
      note = {\url{https://arxiv.org/abs/2412.11757}}, 
}

@inproceedings{hybridqa_2021,
    title = "{HybridQA: A Dataset of Multi-Hop Question Answering over Tabular and Textual Data}",
    author = "Chen, Wenhu  and
      Zha, Hanwen  and
      Chen, Zhiyu  and
      Xiong, Wenhan  and
      Wang, Hong  and
      Wang, William Yang",
    editor = "Cohn, Trevor  and
      He, Yulan  and
      Liu, Yang",
    booktitle = "Findings of the Association for Computational Linguistics: EMNLP 2020",
    month = nov,
    year = "2020",
    address = "Online",
    publisher = "Association for Computational Linguistics",
    url = "https://aclanthology.org/2020.findings-emnlp.91/",
    doi = "10.18653/v1/2020.findings-emnlp.91",
    pages = "1026--1036",
}

@inproceedings{finqa_2022,
          title={{FinQA: A Dataset of Numerical Reasoning over Financial Data}},
          author={Chen, Zhiyu and Chen, Wenhu and Smiley, Charese and Shah, Sameena and Borova, Iana and Langdon, Dylan and Moussa, Reema and Beane, Matt and Huang, Ting-Hao and Routledge, Bryan R and others},
          booktitle={Proceedings of the 2021 Conference on Empirical Methods in Natural Language Processing},
          pages={3697--3711},
          year={2021}
        }

@misc{ottqa_2021,
  title = {{Open Question Answering over Tables and Text}},
  author = {Wenhu Chen and Ming-Wei Chang and Eva Schlinger and William Wang and William W. Cohen},
  year = {2021},
  eprint = {2010.10439},
  archivePrefix = {arXiv},
  primaryClass = {cs.CL},
  note = {\url{https://arxiv.org/abs/2010.10439}}
}

@misc{aitqa_2021,
      title={{AIT-QA: Question Answering Dataset over Complex Tables in the Airline Industry}}, 
      author={Yannis Katsis and Saneem Chemmengath and Vishwajeet Kumar and Samarth Bharadwaj and Mustafa Canim and Michael Glass and Alfio Gliozzo and Feifei Pan and Jaydeep Sen and Karthik Sankaranarayanan and Soumen Chakrabarti},
      year={2021},
      eprint={2106.12944},
      archivePrefix={arXiv},
      primaryClass={cs.CL},
      url={https://arxiv.org/abs/2106.12944}, 
note = {\url{https://arxiv.org/abs/2106.12944}}
}

@misc{tabfact_2020,
      title={{TabFact: A Large-scale Dataset for Table-based Fact Verification}}, 
      author={Wenhu Chen and Hongmin Wang and Jianshu Chen and Yunkai Zhang and Hong Wang and Shiyang Li and Xiyou Zhou and William Yang Wang},
      year={2020},
      eprint={1909.02164},
      archivePrefix={arXiv},
      primaryClass={cs.CL},
      note = {\url{https://arxiv.org/abs/1909.02164}},
}

@article{fetaqa_2021,
    title = "{FeTaQA: Free-form Table Question Answering}",
    author = "Nan, Linyong  and
      Hsieh, Chiachun  and
      Mao, Ziming  and
      Lin, Xi Victoria  and
      Verma, Neha  and
      Zhang, Rui  and
      Kry{\'s}ci{\'n}ski, Wojciech  and
      Schoelkopf, Hailey  and
      Kong, Riley  and
      Tang, Xiangru  and
      Mutuma, Mutethia  and
      Rosand, Ben  and
      Trindade, Isabel  and
      Bandaru, Renusree  and
      Cunningham, Jacob  and
      Xiong, Caiming  and
      Radev, Dragomir  and
      Radev, Dragomir",
    editor = "Roark, Brian  and
      Nenkova, Ani",
    journal = "Transactions of the Association for Computational Linguistics",
    volume = "10",
    year = "2022",
    address = "Cambridge, MA",
    publisher = "MIT Press",
    url = "https://aclanthology.org/2022.tacl-1.3/",
    doi = "10.1162/tacl_a_00446",
    pages = "35--49"
}

@inproceedings{wikitablequestions_2017,
    title = "{Search-based Neural Structured Learning for Sequential Question Answering}",
    author = "Iyyer, Mohit  and
      Yih, Wen-tau  and
      Chang, Ming-Wei",
    editor = "Barzilay, Regina  and
      Kan, Min-Yen",
    booktitle = "Proceedings of the 55th Annual Meeting of the Association for Computational Linguistics (Volume 1: Long Papers)",
    month = jul,
    year = "2017",
    address = "Vancouver, Canada",
    publisher = "Association for Computational Linguistics",
    url = "https://aclanthology.org/P17-1167/",
    doi = "10.18653/v1/P17-1167",
    pages = "1821--1831"
}

@article{tablebench_2025, 
title={{TableBench: A Comprehensive and Complex Benchmark for Table Question Answering}}, volume={39}, url={https://ojs.aaai.org/index.php/AAAI/article/view/34739}, DOI={10.1609/aaai.v39i24.34739},  number={24}, journal={Proceedings of the AAAI Conference on Artificial Intelligence}, author={Wu, Xianjie and Yang, Jian and Chai, Linzheng and Zhang, Ge and Liu, Jiaheng and Du, Xeron and Liang, Di and Shu, Daixin and Cheng, Xianfu and Sun, Tianzhen and Li, Tongliang and Li, Zhoujun and Niu, Guanglin}, year={2025}, month={Apr.}, pages={25497-25506} }

@inproceedings{tatqa2021,
    title = "{TAT-QA: A Question Answering Benchmark on a Hybrid of Tabular and Textual Content in Finance}",
    author = "Zhu, Fengbin  and
      Lei, Wenqiang  and
      Huang, Youcheng  and
      Wang, Chao  and
      Zhang, Shuo  and
      Lv, Jiancheng  and
      Feng, Fuli  and
      Chua, Tat-Seng",
    editor = "Zong, Chengqing  and
      Xia, Fei  and
      Li, Wenjie  and
      Navigli, Roberto",
    booktitle = "Proceedings of the 59th Annual Meeting of the Association for Computational Linguistics and the 11th International Joint Conference on Natural Language Processing (Volume 1: Long Papers)",
    month = aug,
    year = "2021",
    address = "Online",
    publisher = "Association for Computational Linguistics",
    url = "https://aclanthology.org/2021.acl-long.254",
    doi = "10.18653/v1/2021.acl-long.254",
    pages = "3277--3287"
}

@inproceedings{tatdqa2022,
    author = {Zhu, Fengbin and Lei, Wenqiang and Feng, Fuli and Wang, Chao and Zhang, Haozhou and Chua, Tat-Seng},
    title = {{Towards Complex Document Understanding By Discrete Reasoning}},
    year = {2022},
    isbn = {9781450392037},
    publisher = {Association for Computing Machinery},
    address = {New York, NY, USA},
    url = {https://doi.org/10.1145/3503161.3548422},
    doi = {10.1145/3503161.3548422},
    booktitle = {Proceedings of the 30th ACM International Conference on Multimedia},
    pages = {4857–4866},
    numpages = {10},
    location = {Lisboa, Portugal},
    series = {MM '22}
}

@article{NLtoSQLsurvey2024,
  author       = {Boyan Li and
                  Yuyu Luo and
                  Chengliang Chai and
                  Guoliang Li and
                  Nan Tang},
  title        = {{The Dawn of Natural Language to SQL: Are We Fully Ready? [Experiment,
                  Analysis {\&} Benchmark {]}}},
  journal      = {Proc. {VLDB} Endow.},
  volume       = {17},
  number       = {11},
  pages        = {3318--3331},
  year         = {2024},
  url          = {https://www.vldb.org/pvldb/vol17/p3318-luo.pdf},
  doi          = {10.14778/3681954.3682003}
}

@misc{pivottabler,
  title = {{R package pivottabler}},
  howpublished ={https://github.com/cbailiss/pivottabler/},
  note = {Accessed: 2025-05-15}
}

@book{pivotTable2005,
author = {Jelen, Bill and Alexander, Michael},
title = {{Pivot Table Data Crunching}},
year = {2005},
isbn = {0672327945},
publisher = {Que Corp.},
address = {USA}
}

@article{10.1431751.3231766,
author = {He, Yeye and Chu, Xu and Ganjam, Kris and Zheng, Yudian and Narasayya, Vivek and Chaudhuri, Surajit},
title = {{Transform-data-by-example (TDE): an extensible search engine for data transformations}},
year = {2018},
publisher = {VLDB Endowment},
volume = {11},
number = {10},
journal = {Proc. VLDB Endow.},
month = jun,
pages = {1165–1177},
numpages = {13}
}

@article{10.147407790.3407831,
author = {Jin, Zhongjun and He, Yeye and Chauduri, Surajit},
title = {{Auto-transform: learning-to-transform by patterns}},
year = {2020},
issue_date = {August 2020},
publisher = {VLDB Endowment},
volume = {13},
number = {12},
issn = {2150-8097},
journal = {Proc. VLDB Endow.},
month = jul,
pages = {2368–2381},
numpages = {14}
}

@misc{azure_document_intelligence,
  title        = {{Azure AI Document Intelligence}},
  author       = {Microsoft Corporation},
  year         = {2025},
  howpublished = {https://azure.microsoft.com/en-us/products/ai-services/ai-document-intelligence},
  note         = {Accessed: 2025-05-15}
}

@inproceedings{tapas-paper,
   title={{TaPas: Weakly Supervised Table Parsing via Pre-training}},
   url={http://dx.doi.org/10.18653/v1/2020.acl-main.398},
   DOI={10.18653/v1/2020.acl-main.398},
   booktitle={Proceedings of the 58th Annual Meeting of the Association for Computational Linguistics},
   publisher={Association for Computational Linguistics},
   author={Herzig, Jonathan and Nowak, Pawel Krzysztof and Müller, Thomas and Piccinno, Francesco and Eisenschlos, Julian},
   year={2020} }

@misc{tapex-paper,
      title={{TAPEX: Table Pre-training via Learning a Neural SQL Executor}}, 
      author={Qian Liu and Bei Chen and Jiaqi Guo and Morteza Ziyadi and Zeqi Lin and Weizhu Chen and Jian-Guang Lou},
      year={2022},
      eprint={2107.07653},
      archivePrefix={arXiv},
      primaryClass={cs.CL},
      url={https://arxiv.org/abs/2107.07653}, 
}

@misc{chemmengath2021topictransferabletablequestion,
      title={{Topic Transferable Table Question Answering}}, 
      author={Saneem Ahmed Chemmengath and Vishwajeet Kumar and Samarth Bharadwaj and Jaydeep Sen and Mustafa Canim and Soumen Chakrabarti and Alfio Gliozzo and Karthik Sankaranarayanan},
      year={2021},
      eprint={2109.07377},
      archivePrefix={arXiv},
      primaryClass={cs.CL},
      url={https://arxiv.org/abs/2109.07377}, 
}

@misc{liu2024surveynl2sqllargelanguage,
      title={{A Survey of NL2SQL with Large Language Models: Where are we, and where are we going?}}, 
      author={Xinyu Liu and Shuyu Shen and Boyan Li and Peixian Ma and Runzhi Jiang and Yuxin Zhang and Ju Fan and Guoliang Li and Nan Tang and Yuyu Luo},
      year={2024},
      eprint={2408.05109},
      archivePrefix={arXiv},
      primaryClass={cs.DB},
      url={https://arxiv.org/abs/2408.05109}, 
note = {\url{https://arxiv.org/abs/2408.05109}}
}

@inproceedings{MMLongbenchdoc_neurips2024,
  author       = {Yubo Ma and
                  Yuhang Zang and
                  Liangyu Chen and
                  Meiqi Chen and
                  Yizhu Jiao and
                  Xinze Li and
                  Xinyuan Lu and
                  Ziyu Liu and
                  Yan Ma and
                  Xiaoyi Dong and
                  Pan Zhang and
                  Liangming Pan and
                  Yu{-}Gang Jiang and
                  Jiaqi Wang and
                  Yixin Cao and
                  Aixin Sun},
  editor       = {Amir Globersons and
                  Lester Mackey and
                  Danielle Belgrave and
                  Angela Fan and
                  Ulrich Paquet and
                  Jakub M. Tomczak and
                  Cheng Zhang},
  title        = {{MMLONGBENCH-DOC: Benchmarking Long-context Document Understanding
                  with Visualizations}},
  booktitle    = {Advances in Neural Information Processing Systems 38: Annual Conference
                  on Neural Information Processing Systems 2024, NeurIPS 2024, Vancouver,
                  BC, Canada, December 10 - 15, 2024},
  year         = {2024},
  url          = {http://papers.nips.cc/paper\_files/paper/2024/hash/ae0e43289bffea0c1fa34633fc608e92-Abstract-Datasets\_and\_Benchmarks\_Track.html},
}

@inproceedings{spiqa_neurips2024,
  author       = {Shraman Pramanick and
                  Rama Chellappa and
                  Subhashini Venugopalan},
  editor       = {Amir Globersons and
                  Lester Mackey and
                  Danielle Belgrave and
                  Angela Fan and
                  Ulrich Paquet and
                  Jakub M. Tomczak and
                  Cheng Zhang},
  title        = {{SPIQA: A Dataset for Multimodal Question Answering on Scientific
                  Papers}},
  booktitle    = {Advances in Neural Information Processing Systems 38: Annual Conference
                  on Neural Information Processing Systems 2024, NeurIPS 2024, Vancouver,
                  BC, Canada, December 10 - 15, 2024},
  year         = {2024},
  url          = {http://papers.nips.cc/paper\_files/paper/2024/hash/d74033a247989e8f6f3bf9e0c9629fb5-Abstract-Datasets\_and\_Benchmarks\_Track.html},
 }

@article{zhu2024autotqa,
  title={{Autotqa: Towards autonomous tabular question answering through multi-agent large language models}},
  author={Zhu, Jun-Peng and Cai, Peng and Xu, Kai and Li, Li and Sun, Yishen and Zhou, Shuai and Su, Haihuang and Tang, Liu and Liu, Qi},
  journal={Proceedings of the VLDB Endowment},
  volume={17},
  number={12},
  pages={3920--3933},
  year={2024},
  publisher={VLDB Endowment}
}

@article{urban2023caesura,
  title={{CAESURA: language models as multi-modal query planners}},
  author={Urban, Matthias and Binnig, Carsten},
  journal={arXiv preprint arXiv:2308.03424},
  year={2023}
}

@article{zhang2024reactable,
  title={{ReAcTable: enhancing ReAct for table question answering}},
  author={Zhang, Yunjia and Henkel, Jordan and Floratou, Avrilia and Cahoon, Joyce and Deep, Shaleen and Patel, Jignesh M},
  journal={Proceedings of the VLDB Endowment},
  volume={17},
  number={8},
  pages={1981--1994},
  year={2024},
  publisher={VLDB Endowment}
}

@article{burdick2020table,
  title={{Table extraction and understanding for scientific and enterprise applications}},
  author={Burdick, Douglas and Danilevsky, Marina and Evfimievski, Alexandre V and Katsis, Yannis and Wang, Nancy},
  journal={Proceedings of the VLDB Endowment},
  volume={13},
  number={12},
  pages={3433--3436},
  year={2020},
  publisher={VLDB Endowment}
}

@article{kim2020natural,
  title={{Natural language to SQL: Where are we today?}},
  author={Kim, Hyeonji and So, Byeong-Hoon and Han, Wook-Shin and Lee, Hongrae},
  journal={Proceedings of the VLDB Endowment},
  volume={13},
  number={10},
  pages={1737--1750},
  year={2020},
  publisher={VLDB Endowment}
}

@inproceedings{mitsopoulou2025analysis,
  title={{Analysis of text-to-SQL benchmarks: limitations, challenges and opportunities}},
  author={Mitsopoulou, Anna and Koutrika, Georgia},
  booktitle={Proceedings 28th International Conference on Extending Database Technology, EDBT 2025},
  pages={199--212},
  year={2025},
  organization={OpenProceedings. org}
}

@inproceedings{wang2020natural,
  title={{A natural language interface for database: Achieving transfer-learnability using adversarial method for question understanding}},
  author={Wang, Wenlu and Tian, Yingtao and Wang, Haixun and Ku, Wei-Shinn},
  booktitle={2020 IEEE 36th International conference on data engineering (ICDE)},
  pages={97--108},
  year={2020},
  organization={IEEE}
}

@inproceedings{song2025fevisqa,
  title={{Fevisqa: Free-form question answering over data visualizations}},
  author={Song, Yuanfeng and Lu, Jinwei and Song, Yuanwei and Cao, Caleb Chen and Wong, Raymond Chi-Wing and Zhang, Haodi},
  booktitle={2025 IEEE 41st International Conference on Data Engineering (ICDE)},
  pages={2726--2739},
  year={2025},
  organization={IEEE}
}

@inproceedings{acharya2022question,
  title={{Question answering system using NLP and BERT}},
  author={Acharya, Shreya and Sornalakshmi, K and Paul, Bidisha and Singh, Anshul},
  booktitle={2022 3rd International Conference on Smart Electronics and Communication (ICOSEC)},
  pages={925--929},
  year={2022},
  organization={IEEE}
}

@article{peng2024live,
  title={{Live: Learnable in-context vector for visual question answering}},
  author={Peng, Yingzhe and Hu, Xinting and Peng, Jiawei and Geng, Xin and Yang, Xu and others},
  journal={Advances in Neural Information Processing Systems},
  volume={37},
  pages={9773--9800},
  year={2024}
}

@inproceedings{shen2024ambiguous,
  title={{Ambiguous Entity Oriented Targeted Document Detection}},
  author={Shen, Wei and Wen, Haixu},
  booktitle={2024 IEEE 40th International Conference on Data Engineering (ICDE)},
  pages={874--886},
  year={2024},
  organization={IEEE}
}

@inproceedings{lin2025querying,
  title={{Querying templatized document collections with large language models}},
  author={Lin, Yiming and Hulsebos, Madelon and Ma, Ruiying and Shankar, Shreya and Zeighami, Sepanta and Parameswaran, Aditya G and Wu, Eugene},
  booktitle={2025 IEEE 41st International Conference on Data Engineering (ICDE)},
  pages={2422--2435},
  year={2025},
  organization={IEEE}
}

@inproceedings{lu2012dataset,
  title={{A dataset search engine for the research document corpus}},
  author={Lu, Meiyu and Bangalore, Srinivas and Cormode, Graham and Hadjieleftheriou, Marios and Srivastava, Divesh},
  booktitle={2012 IEEE 28th International Conference on Data Engineering},
  pages={1237--1240},
  year={2012},
  organization={IEEE}
}

@inproceedings{li2023toward,
  title={{Toward a unified framework for unsupervised complex tabular reasoning}},
  author={Li, Zhenyu and Li, Xiuxing and Duan, Zhichao and Dong, Bowen and Liu, Ning and Wang, Jianyong},
  booktitle={2023 IEEE 39th International Conference on Data Engineering (ICDE)},
  pages={1691--1704},
  year={2023},
  organization={IEEE}
}
}}

\clearpage
\newpage

 \appendix





\section{HCT properties}
\label{appendix:table_properties}

\subsection{HCT Property definitions and examples}
\label{appendix:table_properties_defs_and_examples}

Definitions of HCT specific table properties are given in Table~\ref{tab:hct_property_definitions} and expanded upon with examples in this section.

\begin{figure}
    \centering
    \includegraphics[width=\columnwidth]{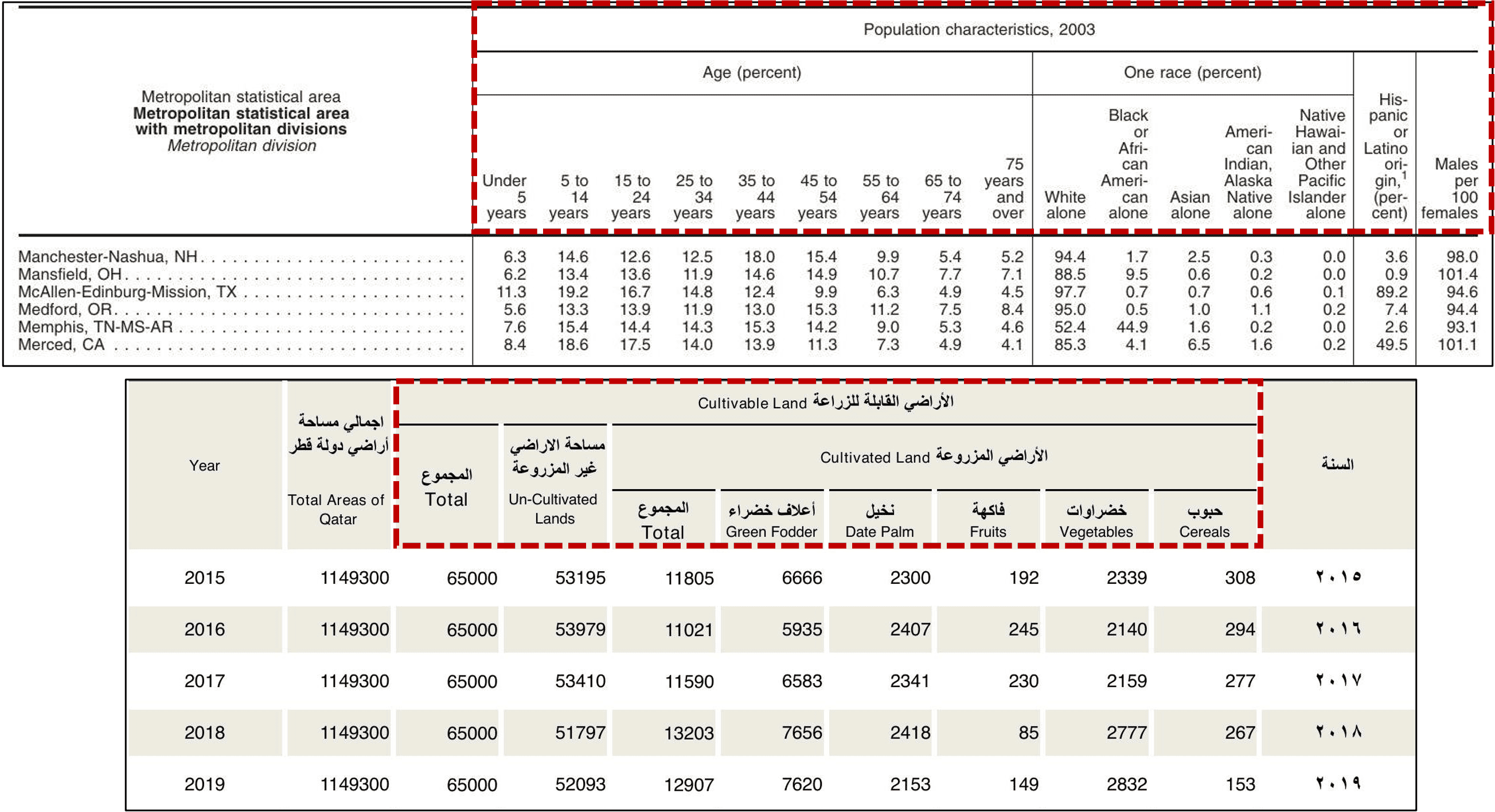}
    \caption{Example of Balanced Column Nesting}
    \label{fig:balanced_column_nesting_example}
\end{figure}

\noindent\textbf{Balanced Column Nesting:} Column nesting can be likened to a tree-like data structure. Similar to the definition of a balanced tree, we can see that each branch of column names (from the top most parent till each leaf) has the same depth. In case of all the table shown in the Figure~\ref{fig:balanced_column_nesting_example}, the depth is 2.

\begin{figure}
    \centering
    \includegraphics[width=\columnwidth]{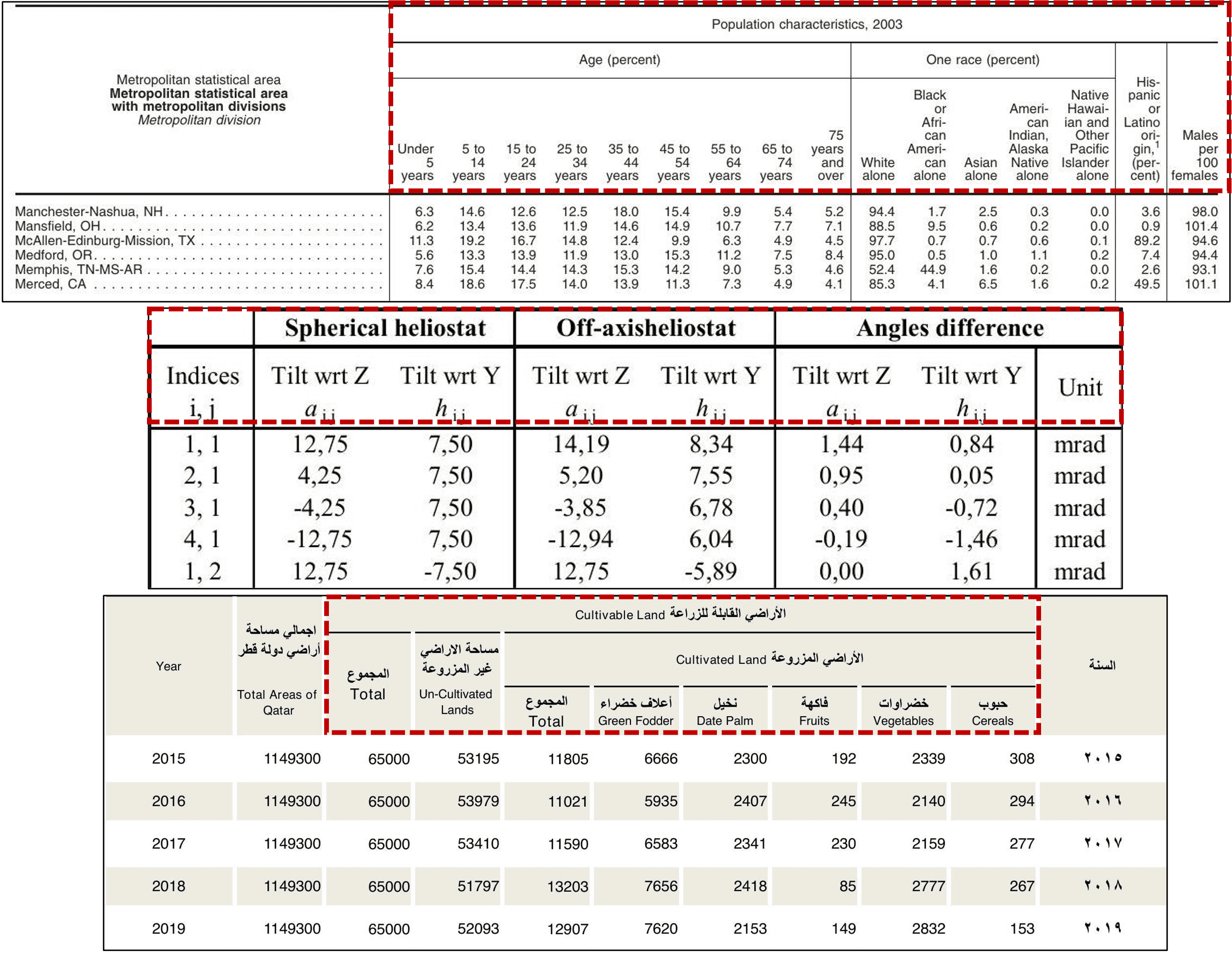}
    \caption{Example of Unbalanced Column Nesting}
    \label{fig:unbalanced_column_nesting_example}
\end{figure}

\noindent\textbf{Unbalanced Column Nesting:} In this the depth of all branches in the nested column tree do not have the same depth. In Figure~\ref{fig:unbalanced_column_nesting_example}, the first table has "Population characteristics, 2003" as the top most parent, and has branches with depth of 2; the second table has the parent "Cultivable Land" with branches of depth 2 and 3 as well.

\begin{figure}
    \centering
    \includegraphics[width=\columnwidth]{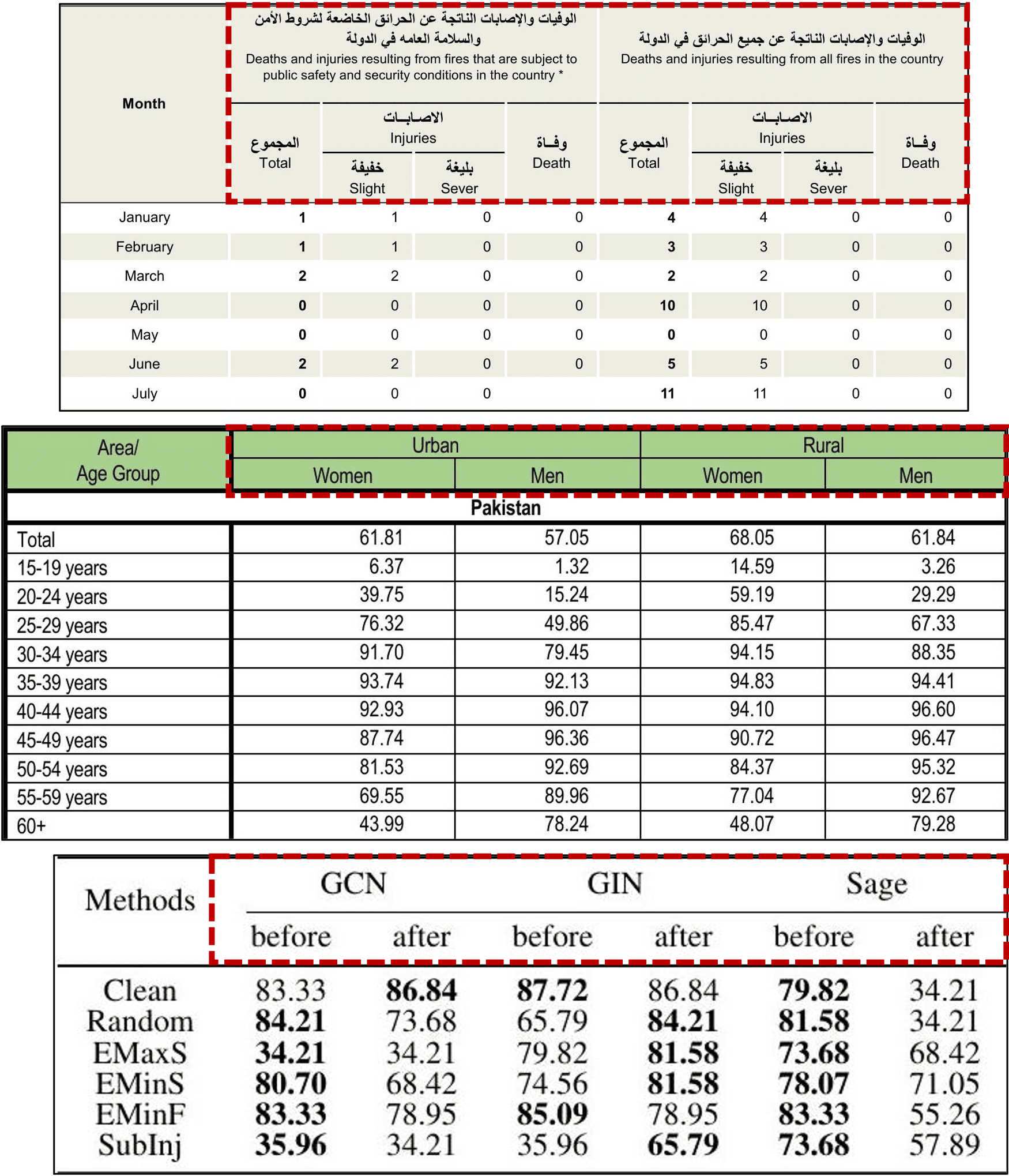}
    \caption{Example of Symmetric Column Nesting}
    \label{fig:symmetric_column_nesting_example}
\end{figure}

\noindent\textbf{Symmetric Column Nesting:} Column nesting is labeled as symmetric when the children of all the cells in highest level of nesting that has more than 1 cell are all identical. In the case of the first table in Figure~\ref{fig:symmetric_column_nesting_example}, the top two cells have identical child branches (Total, Injuries-Slight, Injuries-Sever, Death). Whereas the second table has the parent cells "Urban" and "Rural" both with identical children values ("Women" and "Men"). We do not consider column values such as "Month" (in the first table) and "Area/Age Group" in the second table as they do not exhibit any nesting. The first table also shows an example of where column nesting is unbalanced but is still symmetrical. 

\vspace{1em}

\begin{figure}
    \centering
    \includegraphics[width=\columnwidth]{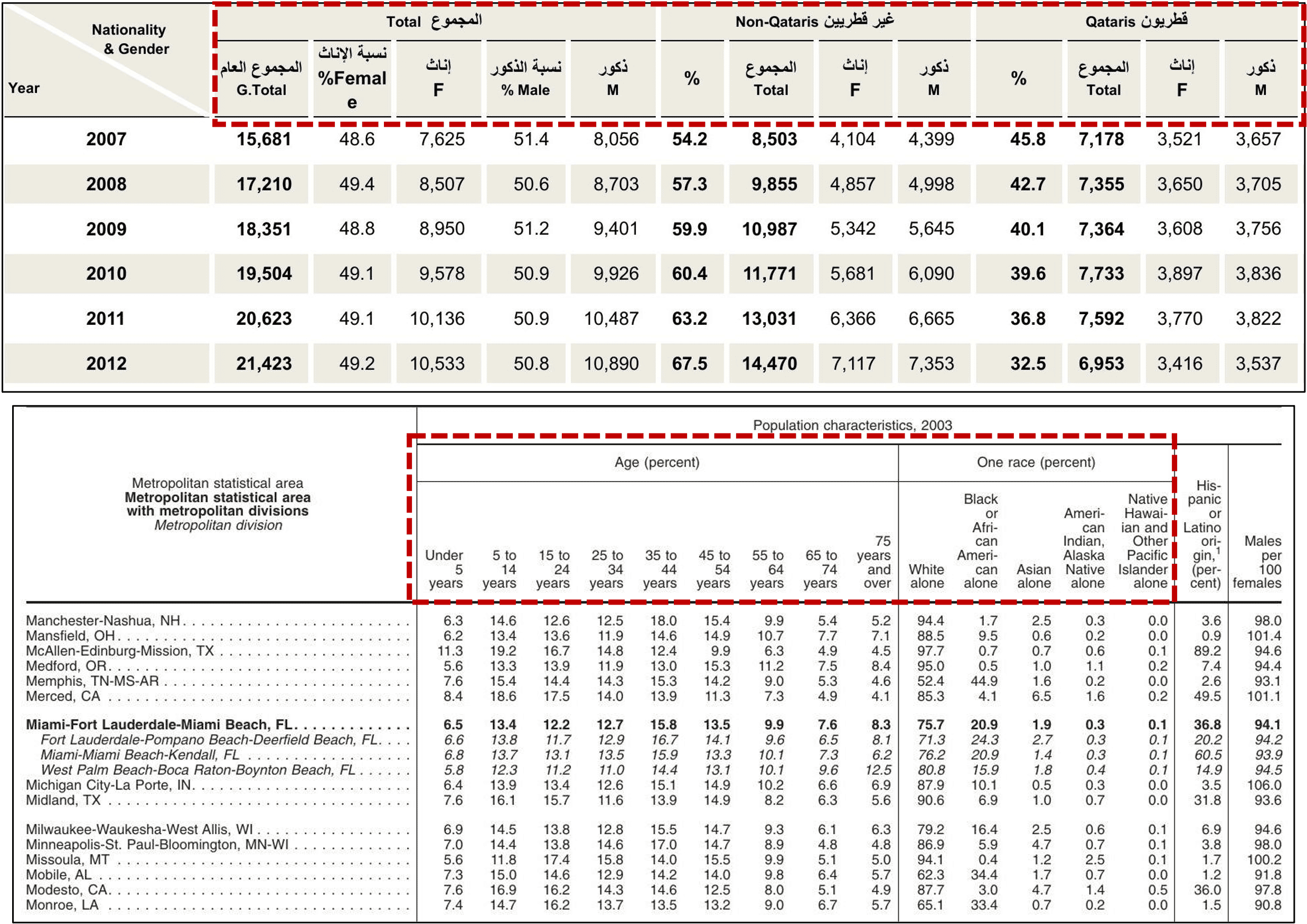}
    \caption{Example of Asymmetric Column Nesting}
    \label{fig:asymmetric_column_nesting_example}
\end{figure}

\noindent\textbf{Asymmetric Column Nesting:} Asymmetric column nesting is when the conditions stated above from symetric column nesting are not met. This entails parent cells having different children cells (branches). In the first table in Figure~\ref{fig:asymmetric_column_nesting_example} we see the the parent cell "Total" has the children ("G.Total", "\%Female", "F", "\%Male", "M") whereas the other 2 parent cells ("Non-Qataris" and "Qataris" have a different set of children ("\%", "Total", "F", "M"). Similarly, in second table, we see that the parent cells "Age (percent)" and "One race (percent)" have different children cells, thus making this column nesting assymetric. 

\vspace{1em}

\begin{figure}
    \centering
    \includegraphics[width=\columnwidth]{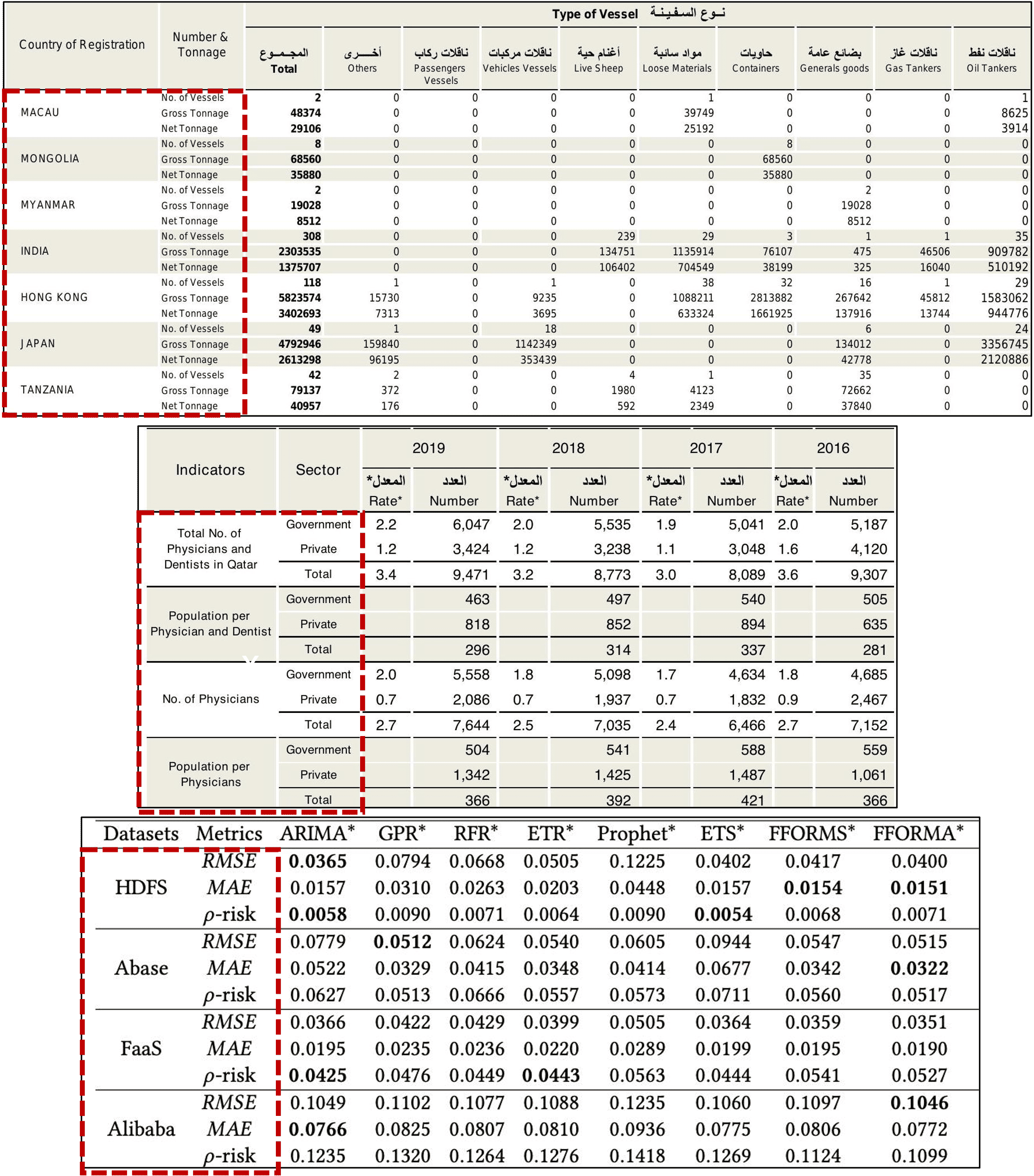}
    \caption{Example of Balanced Row Nesting}
    \label{fig:balanced_row_nesting_example}
\end{figure}

\noindent\textbf{Balanced Row Nesting:} Very similar to column nesting, balanced row nesting entails that all row nesting branches have the same depth (in this case depth is measured left-to-right rather than top-down as with columns). The 3 tables in Figure~\ref{fig:balanced_row_nesting_example} show this property as the children values (the second column from the left in each table) are identical across their parent cells (left most column in each table). 

\vspace{1em}

\vspace{1em}

\vspace{1em}

\begin{figure}
    \centering
    \includegraphics[width=\columnwidth]{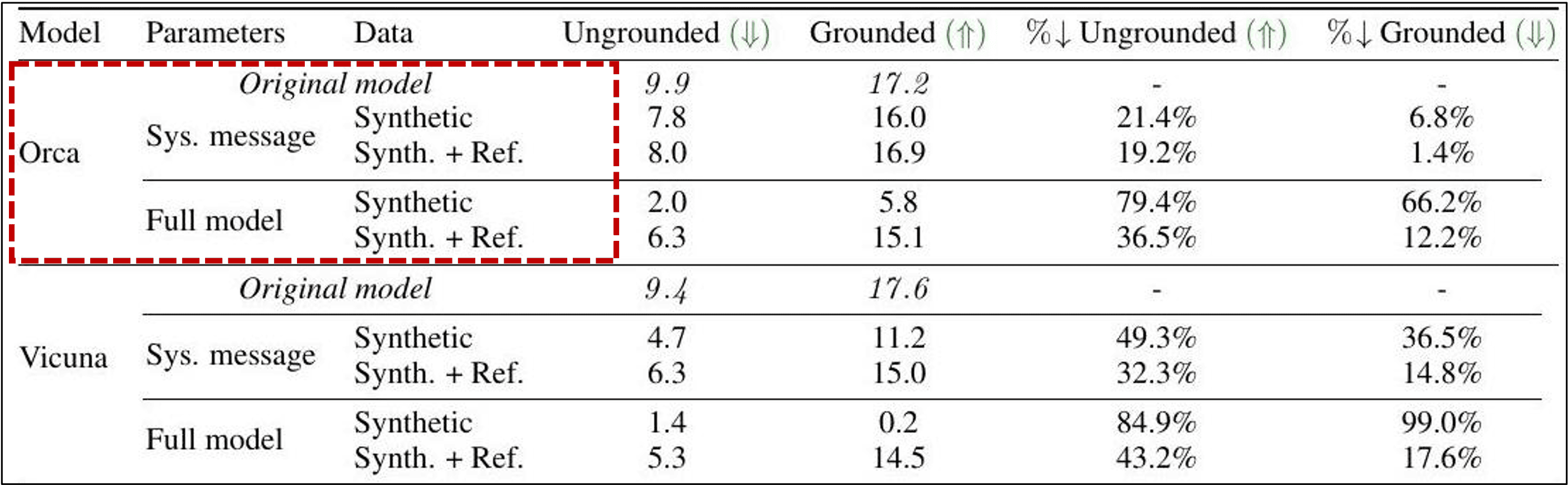}
    \caption{Example of Unbalanced Row Nesting}
    \label{fig:unbalanced_row_nesting_example}
\end{figure}

\noindent\textbf{Unbalanced Row Nesting:} This is when the conditions for balanced row nesting are unmet and branches of nesting in rows have different depths. In the table shown in Figure~\ref{fig:unbalanced_row_nesting_example} we see that the branches "Orca - Sys. message - Synthetic" (depth of 3) and "Orca - Original model" (depth of 2) have different depths, thus they are unbalanced.

\vspace{1em}

\begin{figure}
    \centering
    \includegraphics[width=\columnwidth]{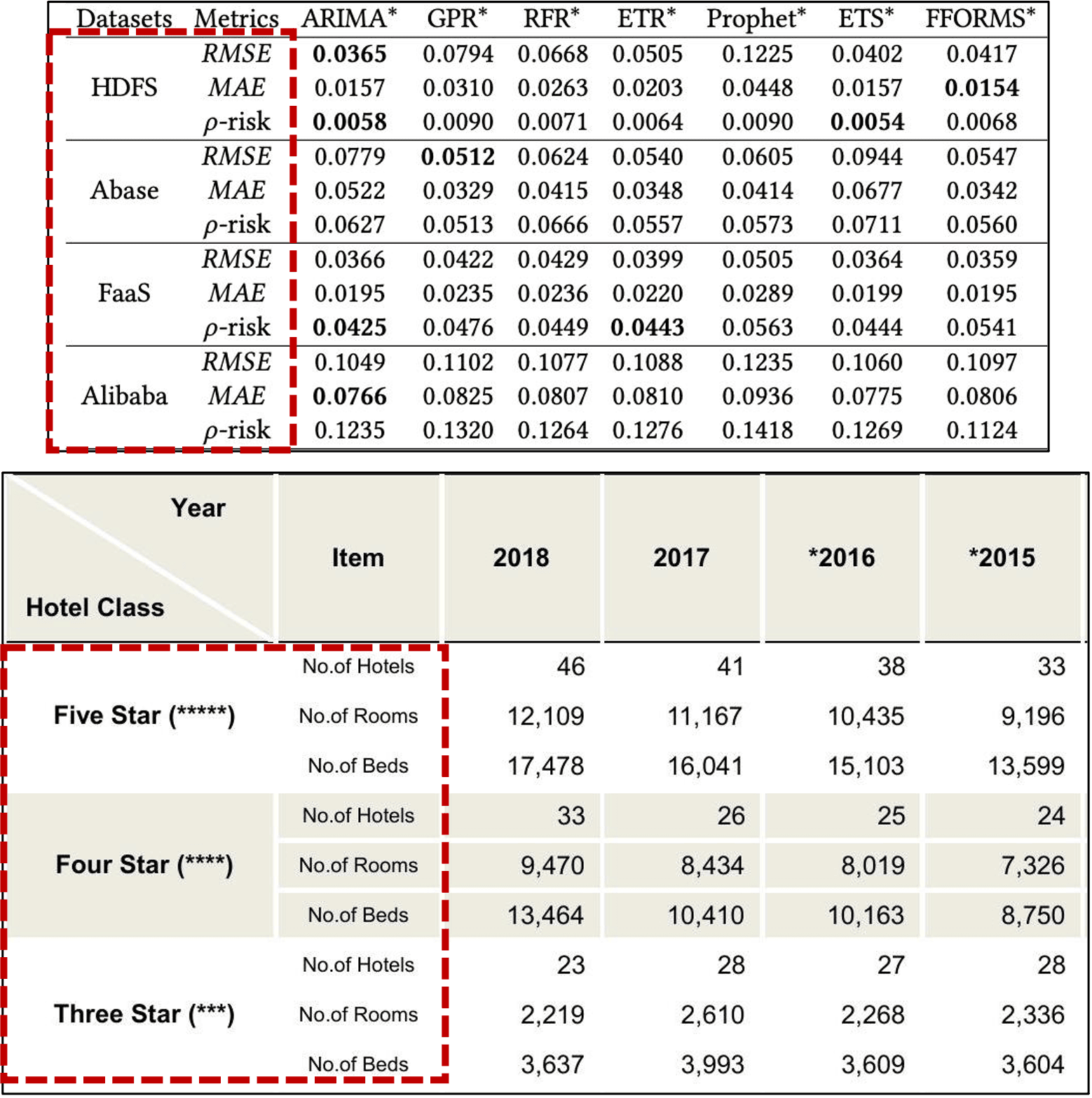}
    \caption{Example of Symmetric Row Nesting}
    \label{fig:symmetric_row_nesting_example}
\end{figure}

\noindent\textbf{Symmetric Row Nesting:} Similar to symmetric column nesting, this property entails each parent cell in the first column (left most) that exhibits nesting properties having identical children cells (branches). We can see in the first table in Figure~\ref{fig:symmetric_row_nesting_example} that all the parent cells ("HDFS", "Abase", "Faas", "Alibaba") have the same children cells ("RMSE", "MAE", "p-risk"). Similarly, all the parent cells in the second table have the same children cells ("No.of Hotels", "No.of Rooms", "No.of Beds"). 

\vspace{1em}

\begin{figure}
    \centering
    \includegraphics[width=\columnwidth]{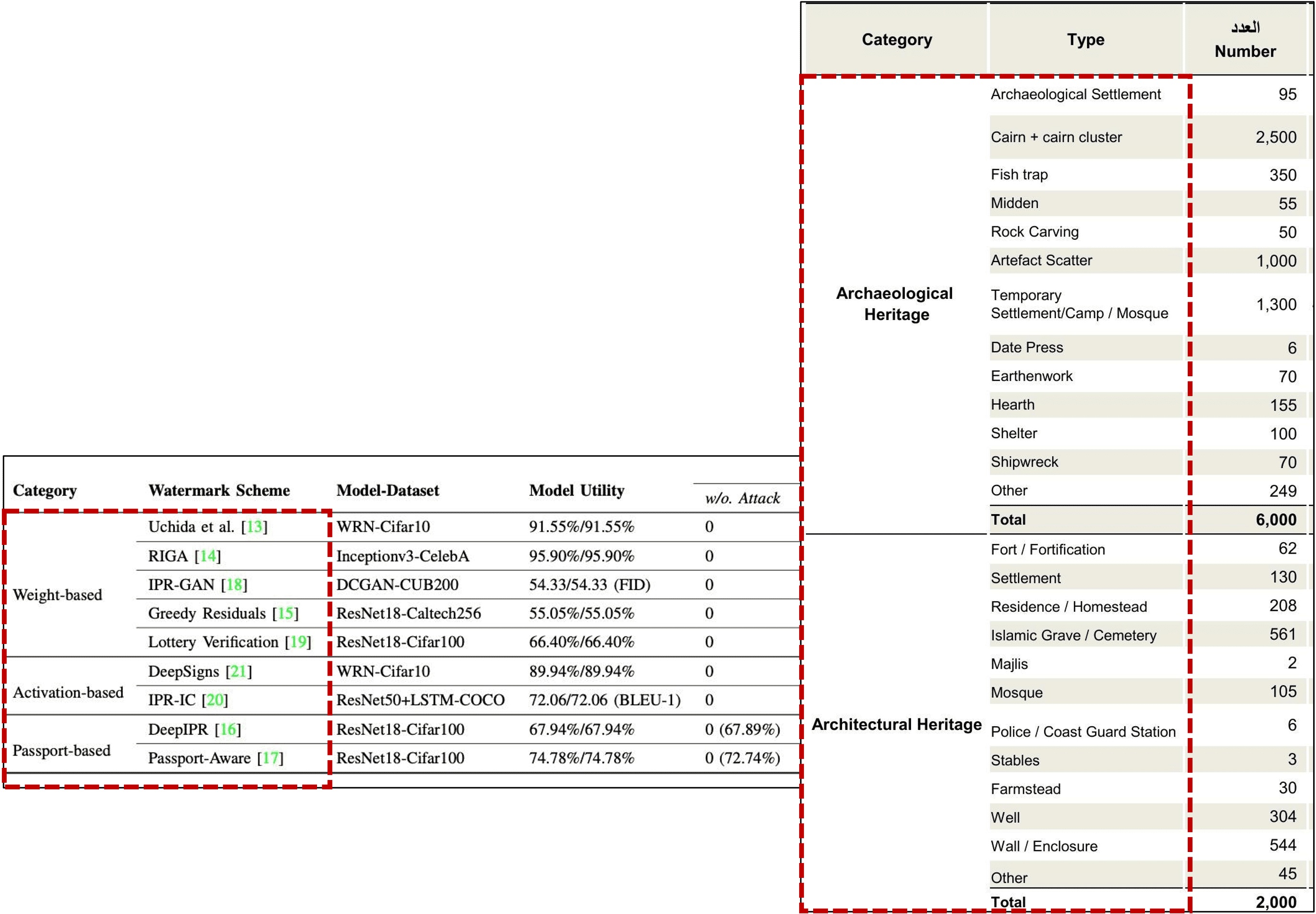}
    \caption{Example of Asymmetric Row Nesting}
    \label{fig:asymmetric_row_nesting_example}
\end{figure}

\noindent\textbf{Asymmetric Row Nesting:} Antithetical to symetric row nesting, this property entails the parent cells having different children cells. The table on the left in Figure~\ref{fig:asymmetric_row_nesting_example} shows different children values for each of the parent cells ("Weight-based" with 5 children cells, "Activation-based" with 2 children cells, "Passport-based" with 2 children cells that have different values from all the others). The same can be seen in the table on the right as the children values are different for both parent cells.

\vspace{1em}

\begin{figure}
    \centering
    \includegraphics[width=\columnwidth]{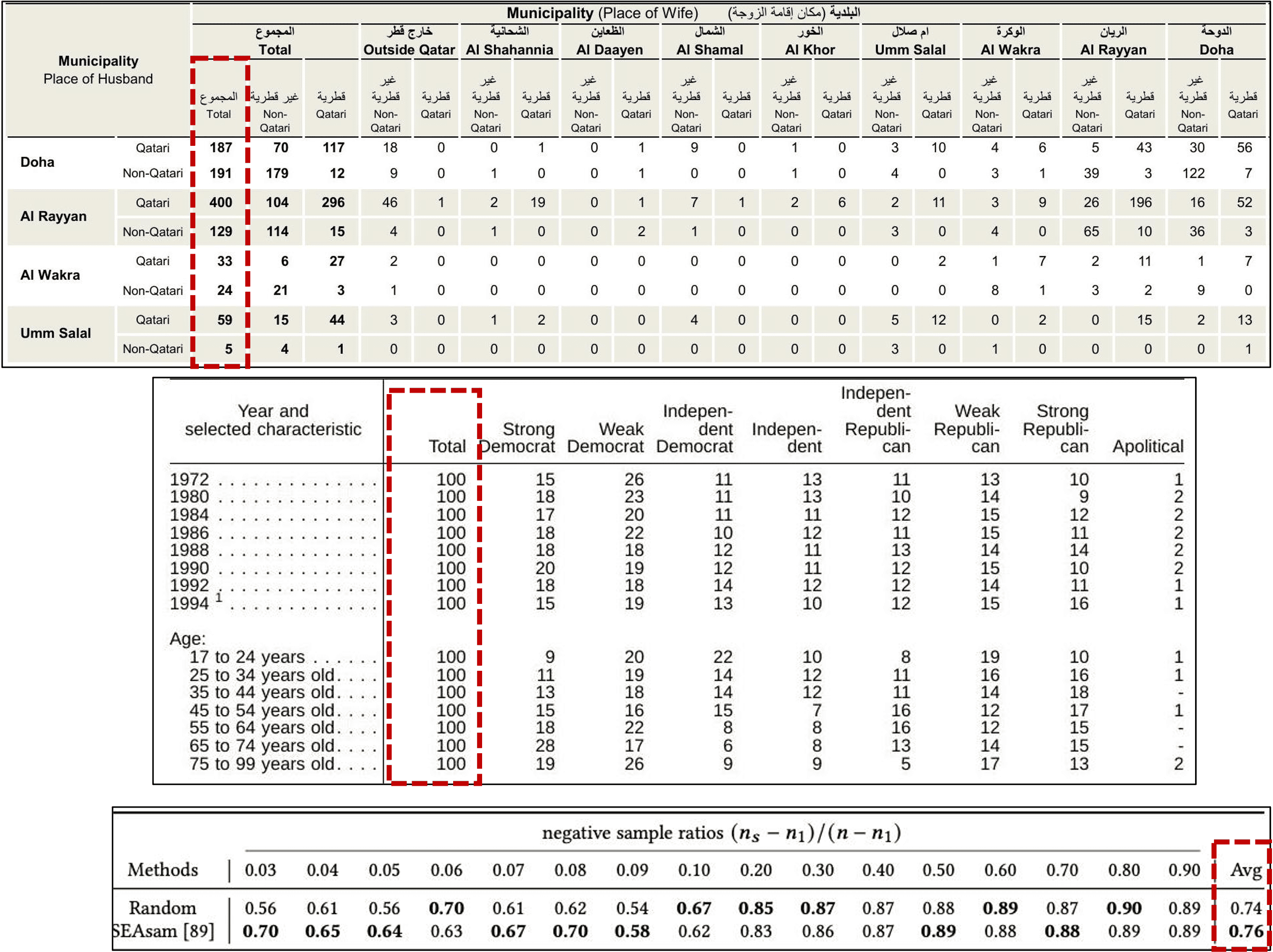}
    \caption{Example of Global Column Aggregation}
    \label{fig:global_column_aggregation_example}
\end{figure}

\noindent\textbf{Global Column Aggregation:} This property entails that there is a column in the table that meets one of the following criteria: (1) is the aggregate of all columns in the table including or excluding local column-group aggregate columns or (2) is the aggregate of all local column-group aggregate columns present in the table. In this case an aggregation can mean any operation or function (e.g sum, max, min, average, different, or any other equation applied consistently across the column that uses values from other columns). For the first table in Figure~\ref{fig:global_column_aggregation_example}, the column "Total - Total" is a global aggregate as it meets both the conditions: it is the sum of all columns excluding "Total - NonQatari" and "Total - Qatari" which are local-group aggregate columns, and it is the sum of the aforementioned local-group aggregate columns (which are the only such columns in the table). Similarly, the "Total" column in the second table and the "Avg" column in the third table perform the sum and arithmetic mean (average) operations on all the other columns in their tables respectively. 

\vspace{1em}

\begin{figure}
    \centering
    \includegraphics[width=\columnwidth]{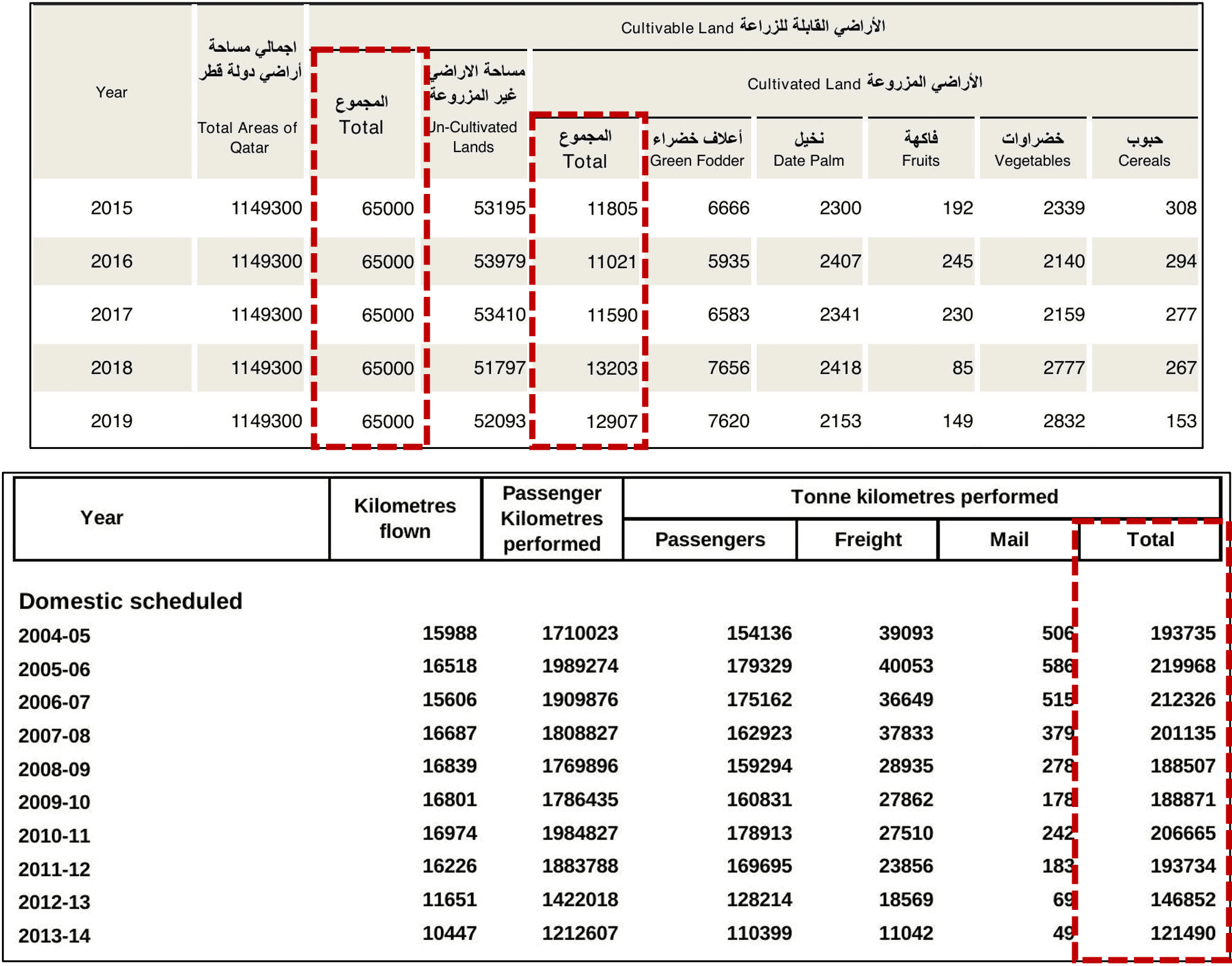}
    \caption{Example of Local Column-Group Aggregation}
    \label{fig:local_column_group_aggregation_example}
\end{figure}

\noindent\textbf{Local Column-Group Aggregation:} A column that aggregates only a subset of numerical columns (including other local column-group aggregation columns) within the table. The first table in Figure~\ref{fig:local_column_group_aggregation_example} has two such columns. "Cultivable Land - Cultivated Land - Total" is the sum of all the other children columns under the parent cell "Cultivable Land - Cultivated Land". "Cultivable Land - Total" is the sum of the aforementioned local-group aggregate and the "Cultivable Land - UnCultivated Lands" column. Similarly, in second table the "Tonne kilometres performed - Total" column sums all the children columns of the parent cell "Tonne kilometres performed" and thus is a local column-group aggregate column.

\vspace{1em}

\begin{figure}
    \centering
    \includegraphics[width=\columnwidth]{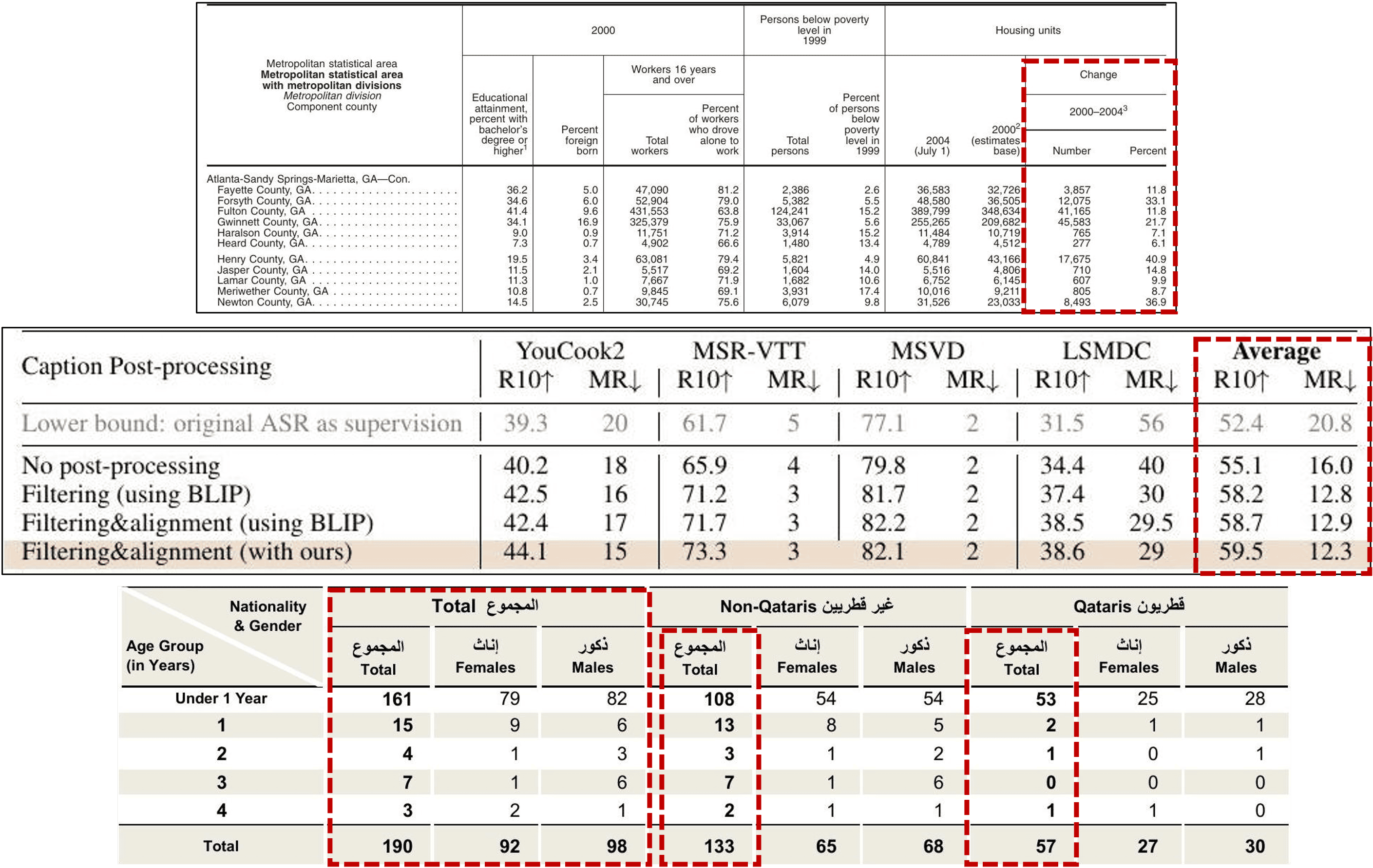}
    \caption{Example of Explicit Column Aggregation Terms}
    \label{fig:explicit_column_aggregation_terms_example}
\end{figure}

\noindent\textbf{Explicit Column Aggregation Terms:} A column where the cell value clearly indicates aggregation (e.g., “Total”, "Percentage", "Sum"). For nested columns, if even one of the cells in the branch has an explicit cell value then the aggregation is considered as explicit. Figure~\ref{fig:explicit_column_aggregation_terms_example} shows three tables that show columns having values like "Change", "Average", and "Total". 

\vspace{1em}

\begin{figure}
    \centering
    \includegraphics[width=\columnwidth]{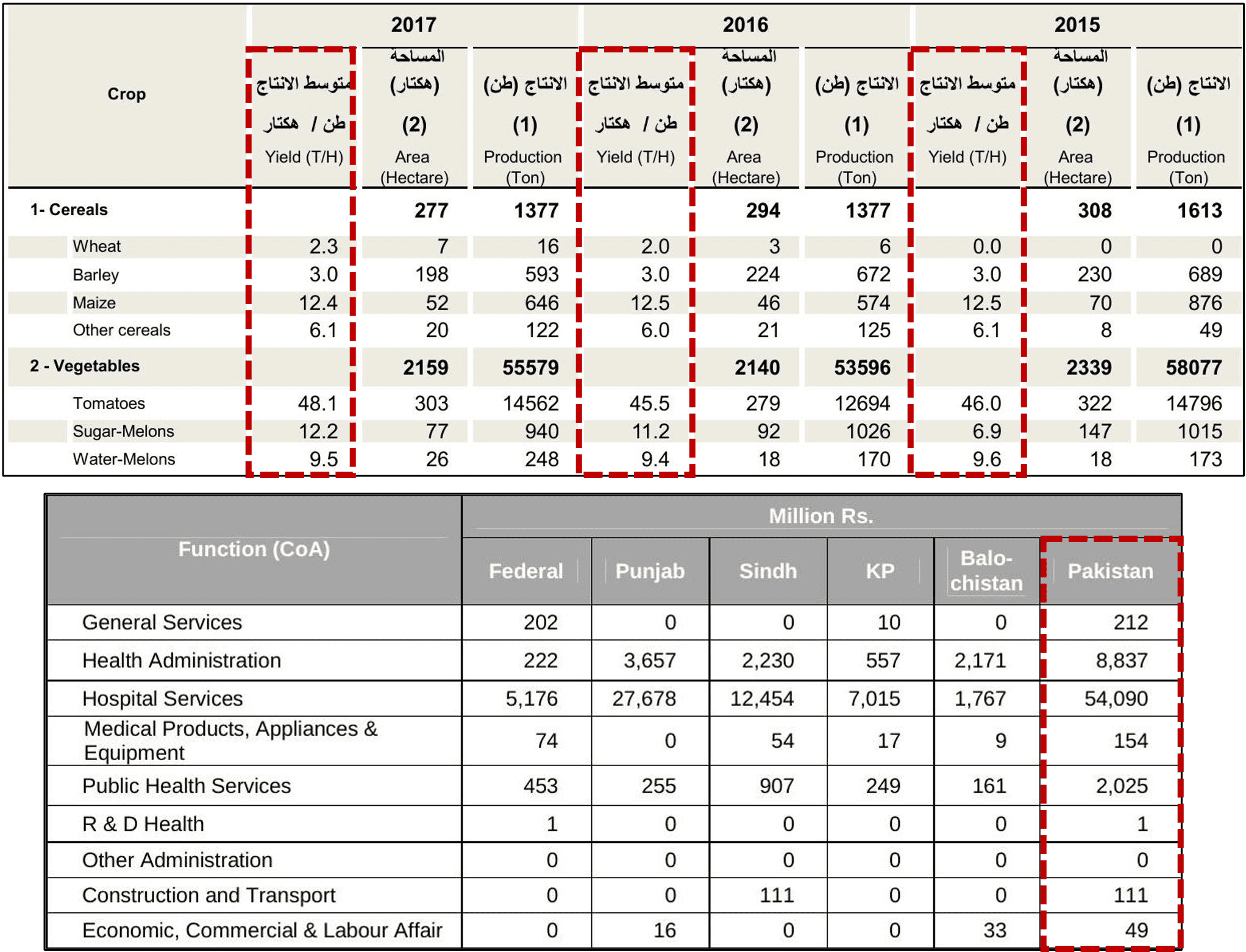}
    \caption{Example of Implicit Column Aggregation Terms}
    \label{fig:implicit_column_aggregation_terms_example}
\end{figure}

\noindent\textbf{Implicit Column Aggregation Terms:} A column that aggregates data, but the column name does not make this explicit. Figure~\ref{fig:implicit_column_aggregation_terms_example} illustrates this ambiguity as the column "Yield" in the first table is the ratio of the other two columns. Although arguable that the name "yield" is indicative of a aggregation, we do not consider such terms that require some degree of domain familiarity to be understood. Thus, it is considered as an implicit aggregation term. The second table presents a more definitive example of this property as the column "Pakistan" is the summation of all the other columns (which represent different parts of Pakistan). Since it is difficult to understand that this is an aggregation without having prior knowledge about Pakistani geography, we consider this an implicit aggregation term. 

\vspace{1em}

\begin{figure}
    \centering
    \includegraphics[width=\columnwidth]{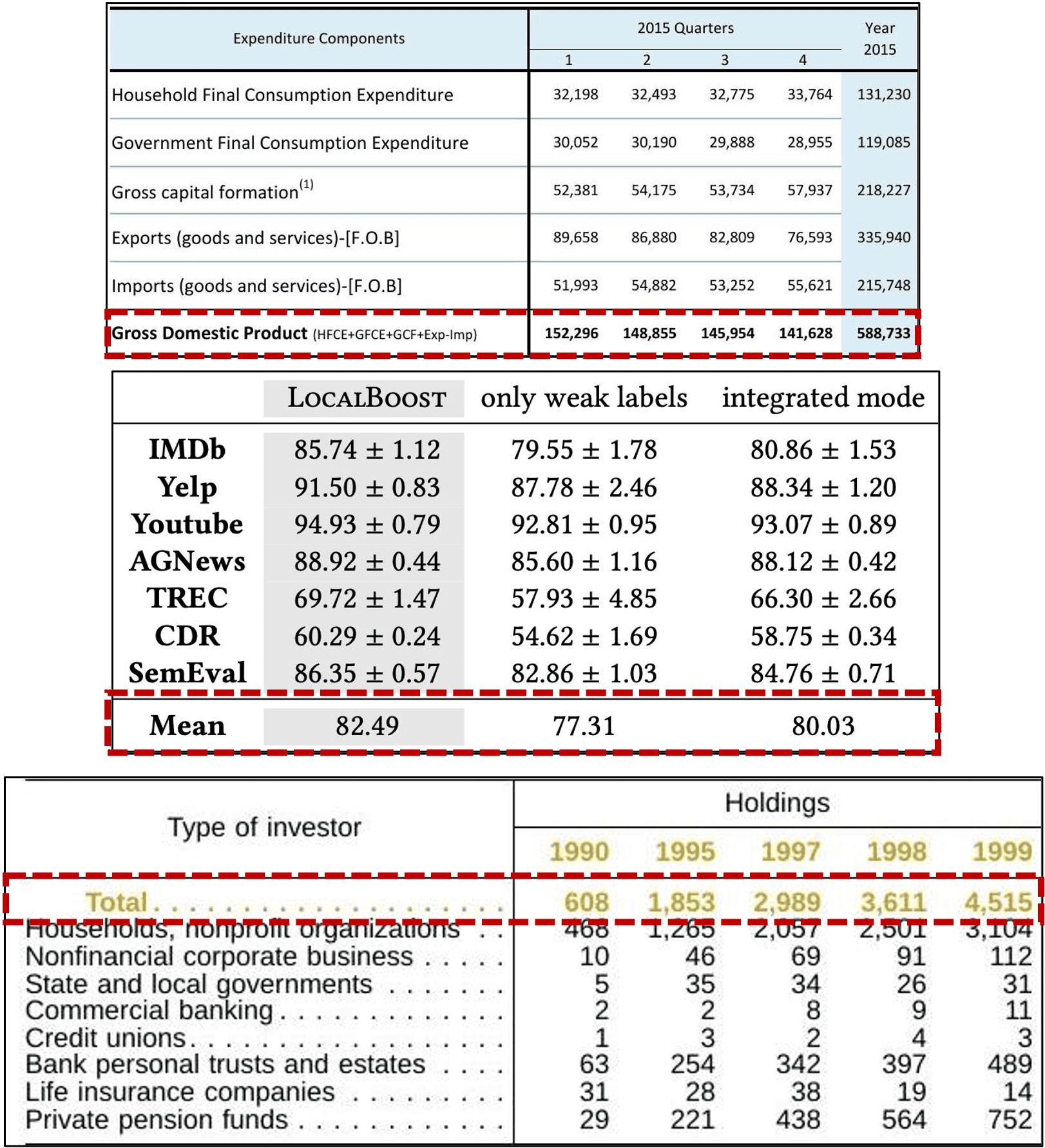}
    \caption{Example of Global Row Aggregation}
    \label{fig:global_row_aggregation_example}
\end{figure}

\noindent\textbf{Global Row Aggregation:} A row that aggregates (using any function) all other rows in the table. Figure~\ref{fig:global_row_aggregation_example} presents examples of such aggregate rows that use the sum and arithmetic mean aggregations. 

\vspace{1em}

\begin{figure}
    \centering
    \includegraphics[width=\columnwidth]{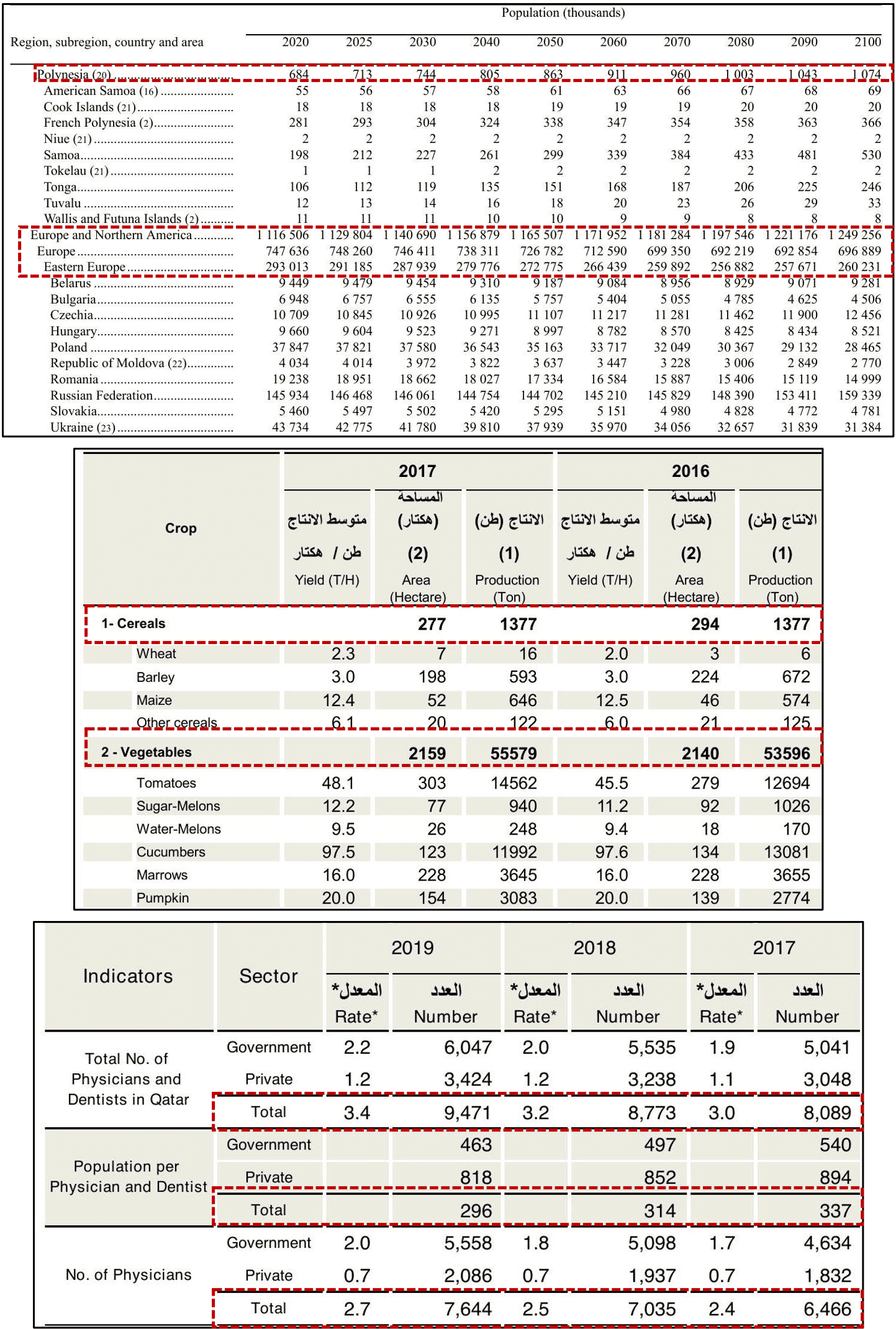}
    \caption{Example of Local Row-Group Aggregation}
    \label{fig:local_row_group_aggregation_example}
\end{figure}

\noindent\textbf{Local Row-Group Aggregation:} A row that aggregates a specific subset of rows (including other row-group aggregate rows) but not the entire table. Examples are shown in Figure~\ref{fig:local_row_group_aggregation_example} where the highlighted rows aggregate (sum in this case) subsets of rows. This property in particular has various visual manifestations: indentation in the first table, numbered rows in the second, explicit "total" cell in third.

\vspace{1em}

\begin{figure}
    \centering
    \includegraphics[width=\columnwidth]{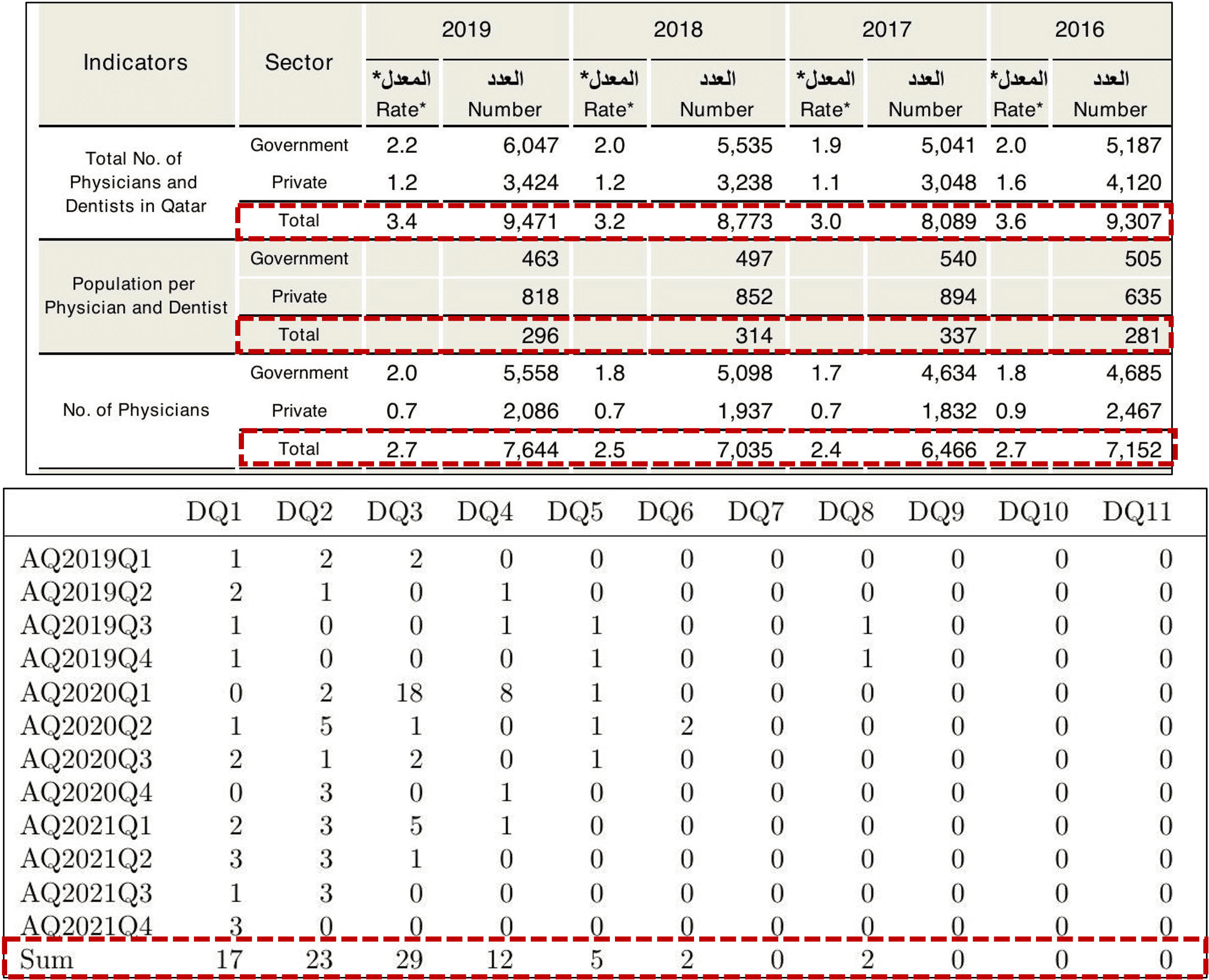}
    \caption{Example of Explicit Row Aggregation Terms}
    \label{fig:explicit_row_aggregation_terms_example}
\end{figure}

\noindent\textbf{Explicit Row Aggregation Terms:} A row where at least one cell explicitly indicates aggregation (e.g., “Total”, “Average”, “Sum”). This is clearly shown in Figure~\ref{fig:explicit_row_aggregation_terms_example}.

\vspace{1em}

\begin{figure}
    \centering
    \includegraphics[width=\columnwidth]{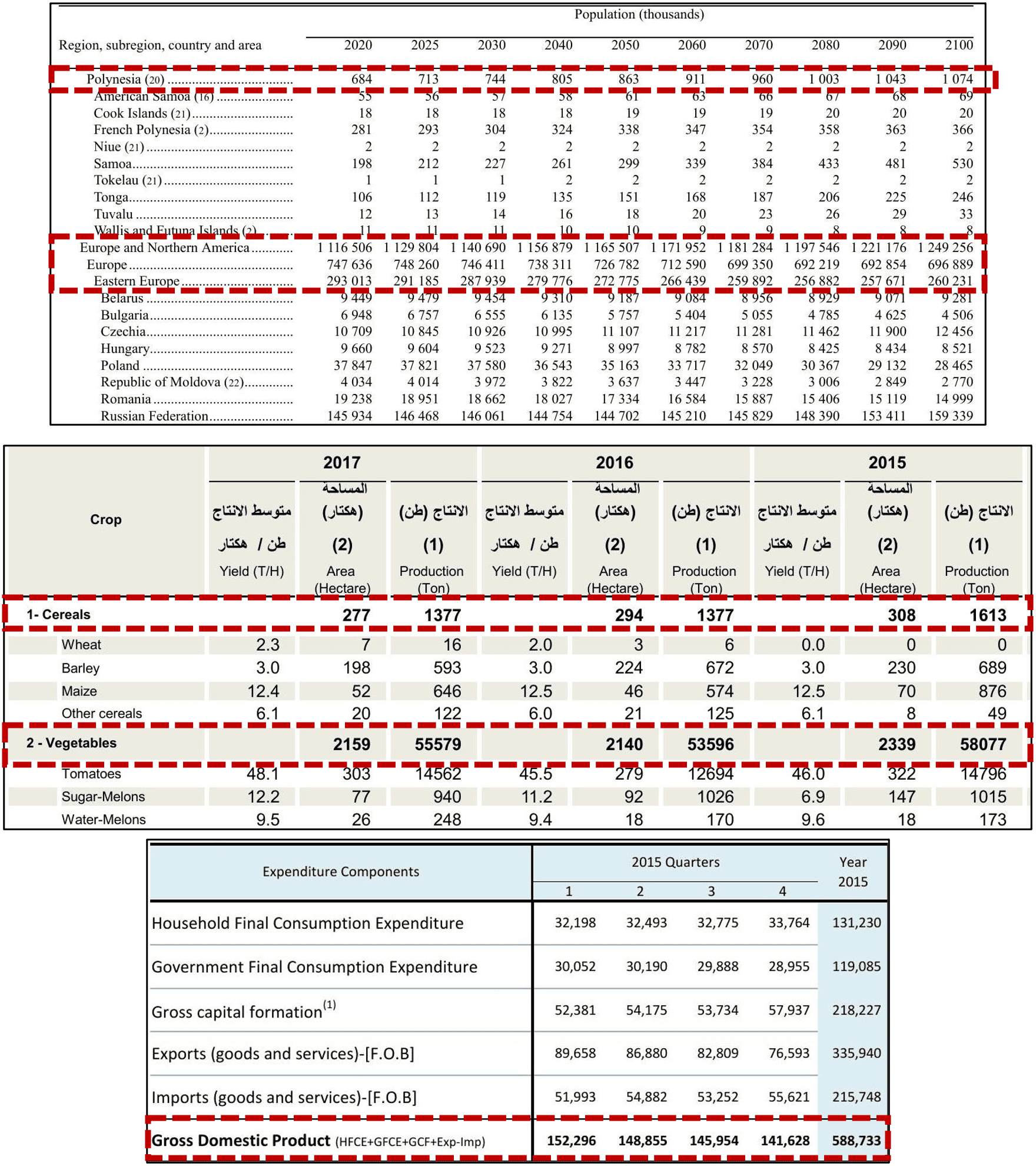}
    \caption{Example of Implicit Row Aggregation Terms}
    \label{fig:implicit_row_aggregation_terms_example}
\end{figure}

\noindent\textbf{Implicit Row Aggregation Terms:} A row that likely aggregates data, but none of its textual cells explicitly say so. Figure~\ref{fig:implicit_row_aggregation_terms_example} illustrates this. As stated earlier, this property is often shown using numbered cells, indentation, domain specific terms (such as "Gross Domestic Product", font stylization, and other visual characteristics.

\vspace{1em}

\begin{figure}
    \centering
    \includegraphics[width=\columnwidth]{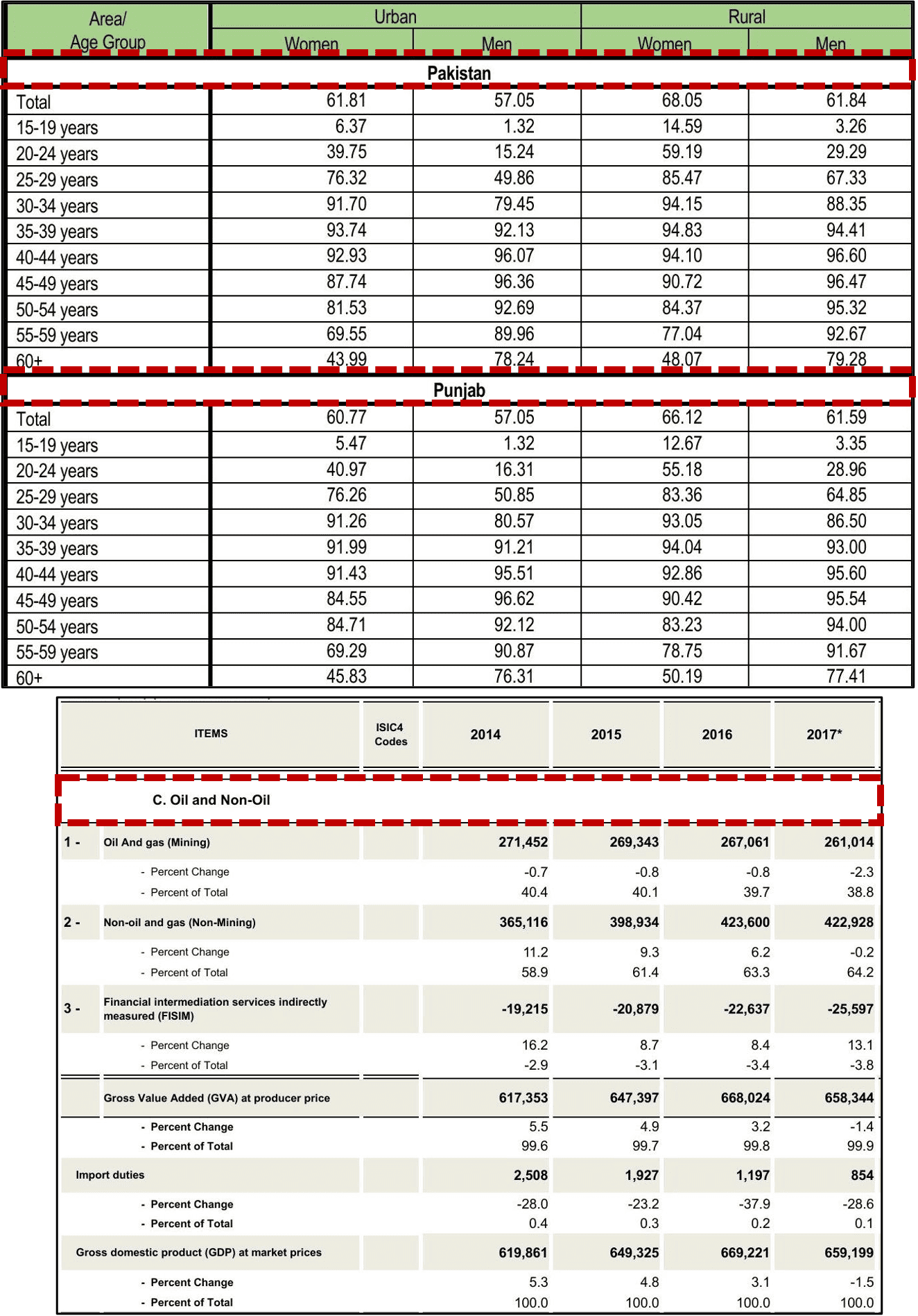}
    \caption{Example of Row Group Label}
    \label{fig:row_group_label_example}
\end{figure}

\noindent\textbf{Row Group Label:} A row that acts as a heading or title for a set of rows above or below it. See Figure~\ref{fig:row_group_label_example} for examples. The first table has the row group labels "Pakistan" and "Punjab" respectively. The second table has "C. Oil and Non-Oil". 

\vspace{1em}

\begin{figure}
    \centering
    \includegraphics[width=\columnwidth]{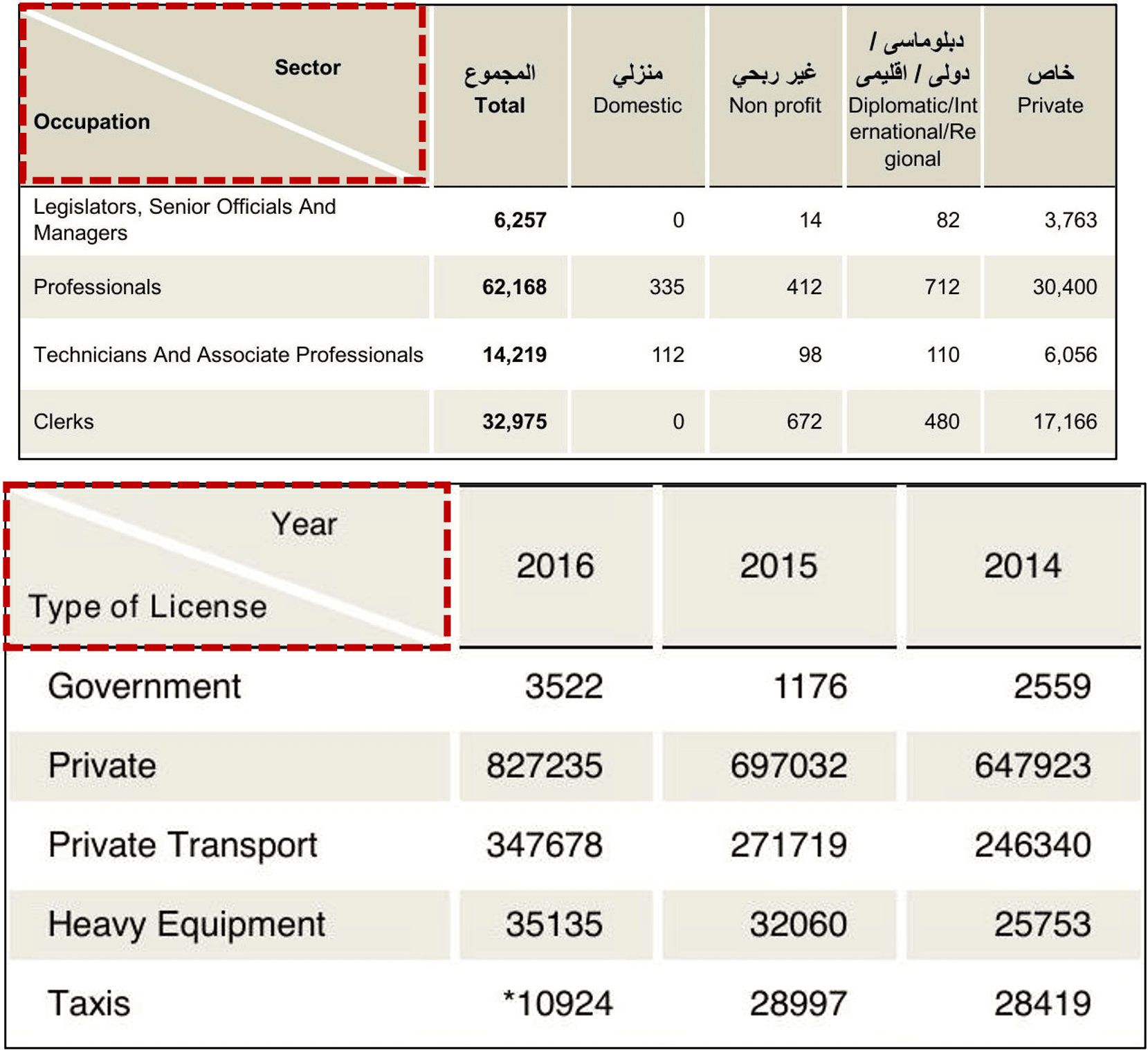}
    \caption{Example of Split Header Cell}
    \label{fig:split_header_cell_example}
\end{figure}

\noindent\textbf{Split Header Cell:} A single header cell split diagonally and containing two distinct values. This visual property is shown in Figure~\ref{fig:split_header_cell_example}. In the first table the term "Sector" describes the column names and "Occupation" the values in the column below it. Similarly, in the second table, "Year" describes the other column names, and "Type of License" is the column name for that particular column. 

\vspace{1em}

\begin{figure}
    \centering
    \includegraphics[width=\columnwidth]{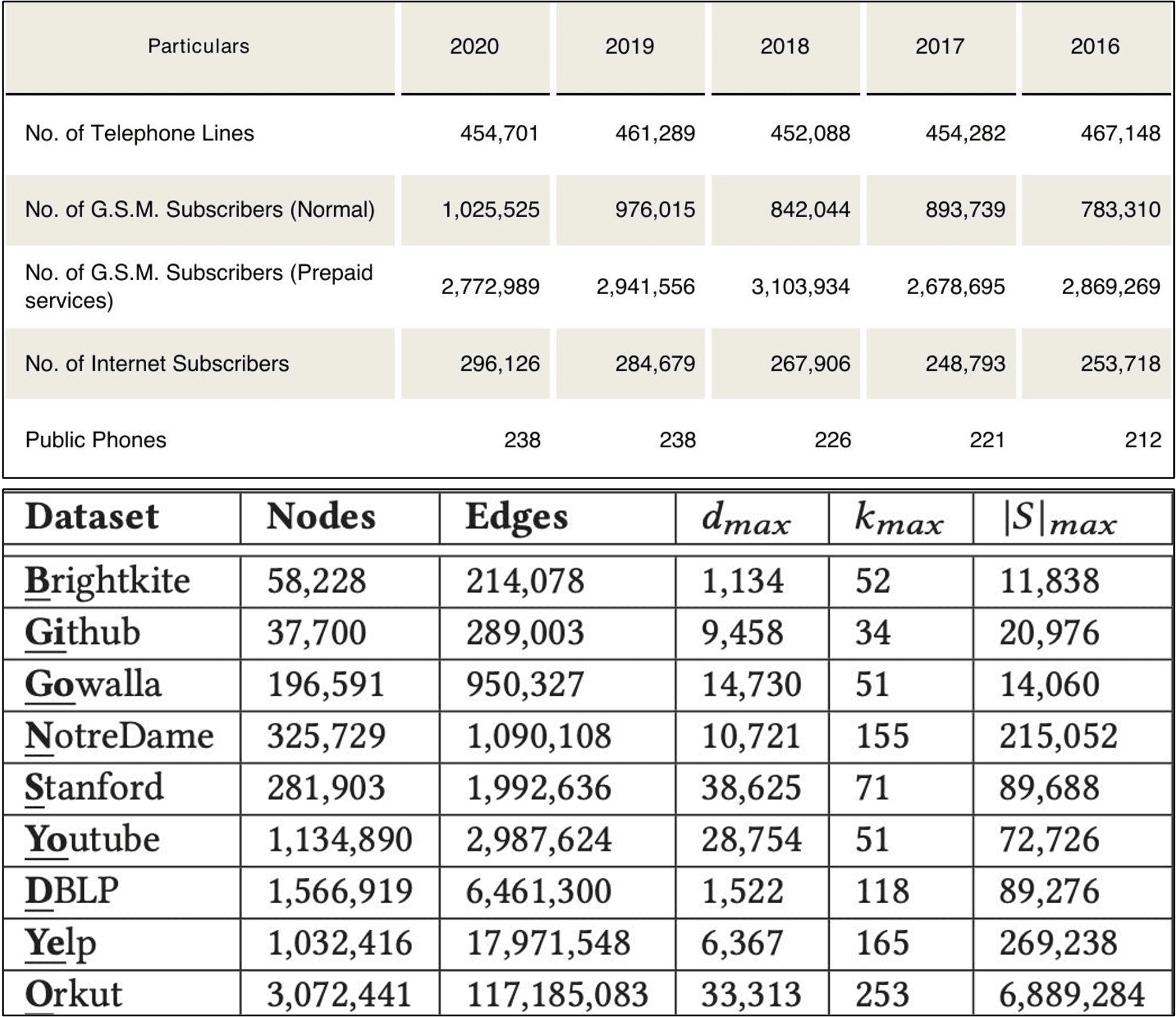}
    \caption{Example of Standard Relational Table}
    \label{fig:standard_relational_table_example}
\end{figure}

\noindent\textbf{Standard Relational Table:} A table that does not exhibit any of the defined HCT properties — no nesting, no aggregation, no labels. Figure~\ref{fig:standard_relational_table_example} shows examples of this simple table form.

\vspace{1em}


\subsection{Table Annotation Tool}
\label{appendix:table_properties_annotation_tool}

\begin{figure}
    \centering
    \includegraphics[width=\columnwidth]{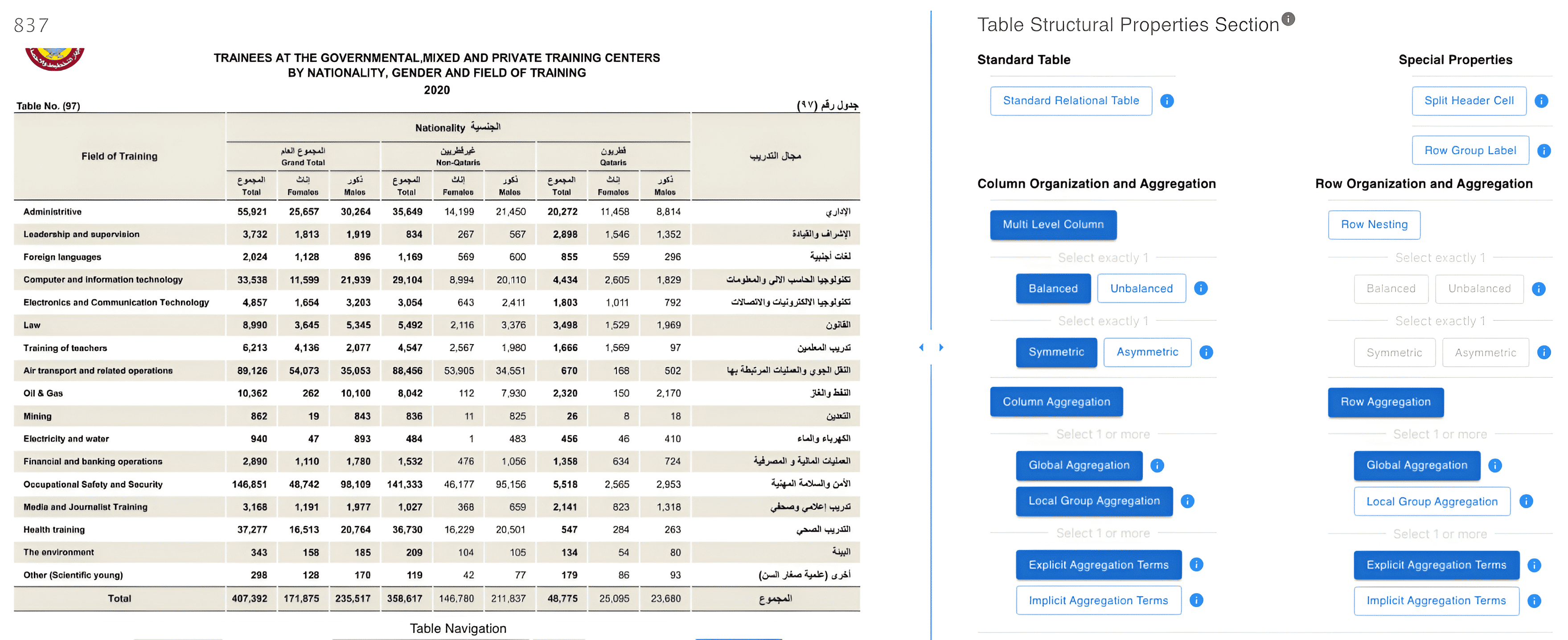}
    \caption{A user-friendly tool for annotating/viewing HCTs and their properties.}
    \label{fig:TableProperties_tool_demo}
\end{figure}
\vspace{1em}


We developed a user-friendly tool for annotating tables and setting their properties as illustrated in Figure~\ref{fig:TableProperties_tool_demo}. 
The left side visualizes one table at a time, while the right side lists the possible allowed properties. 

As an example, the illustrated table in Figure~\ref{fig:TableProperties_tool_demo} contains ``Column Nesting''. And since the child column names (e.g., \{``Grand Total'', ``Females'', ``Males''\}) across all groups are at the same level (same depth from the top parent cell), it is considered a ``Balanced Column Nesting''. Similarly,  since all branches of the parents have identical structure and cell values, the column nesting is considered ``Symmetric''.
Similar rules apply to the ``Row Nesting" property. For example, the table in Figure~\ref{fig:TableProperties_tool_demo} does not containing any row nesting, thus that property is not selected. The table also contains column aggregation at both levels,  local (e.g., ``Total'' under ``Non-Qataris'', ``Males'' under ``Grand Total''), and global (e.g., ``Total'' under ``Grand Total''). At the row level, this table contains only global aggregation (the last row). 
We also keep track of whether the aggregation columns carry explicit aggregate-meaningful labels (e.g., SUM, TOTAL, AVERAGE, MIN, MAX, etc.) as in the table illustrated in Figure~\ref{fig:TableProperties_tool_demo}. 

Certainly, given the compound complexity among the different properties, we could have captured the properties at a more coarse or grained granularity. Nevertheless, we opt for this level of details because it sufficiently captures the underlying complexity of the HCT objects,  and would provide good insights on how these properties correlate to QA performance without being dragged into too many features which requires high dimensional analysis. the annotation tool was developed in-house and we plan to release it on GitHub\footnote{\url{https://github.com/shahmeer99/HCT-QA-Benchmark/}} page soon.

\subsection{Understanding the distribution of table properties}
\label{appendix:table_properties_prop_distribution}

\begin{figure}[h!]
    \centering
    \begin{tabular}{ccc}
    \includegraphics[width=0.45\columnwidth]{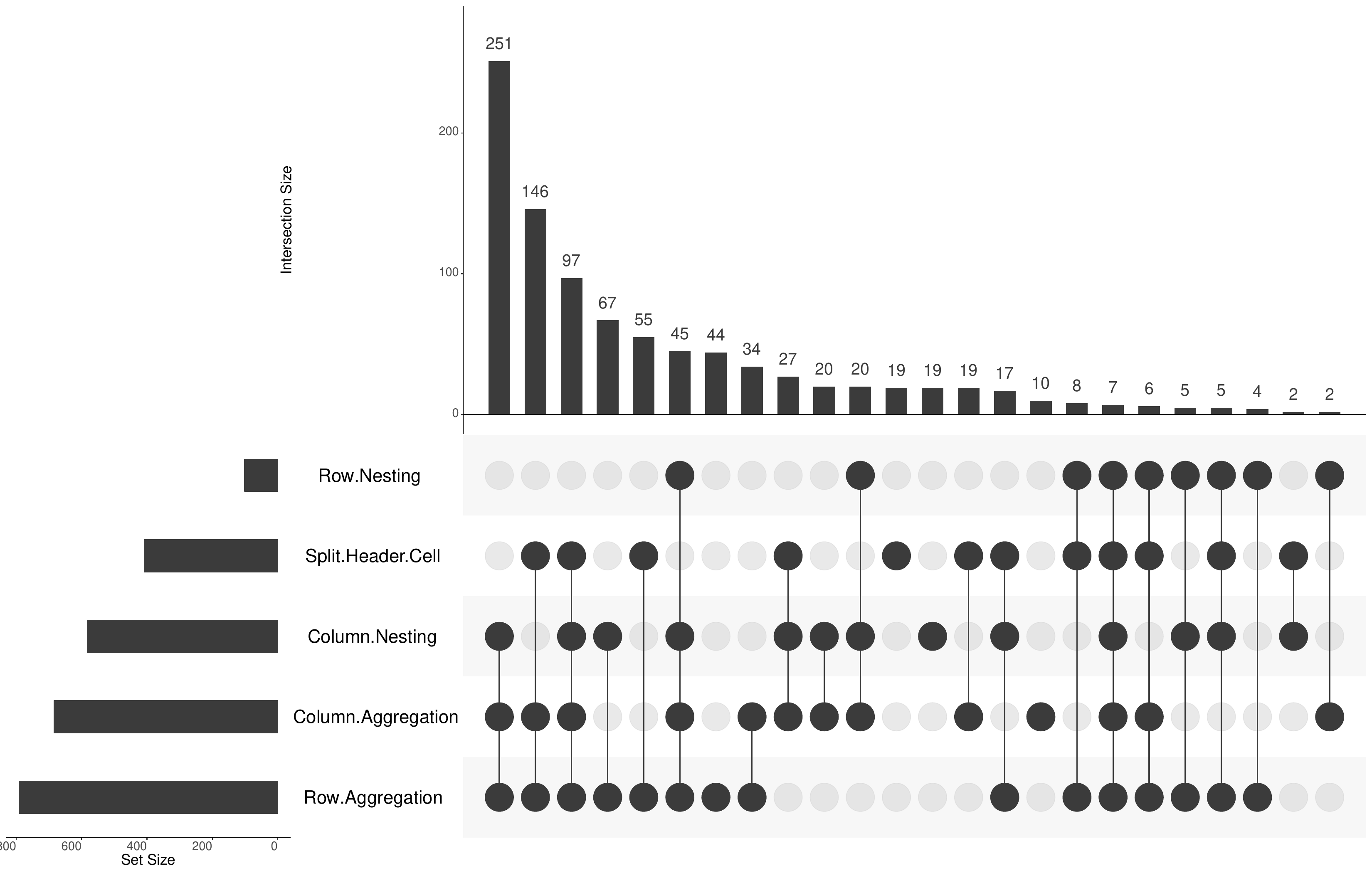}&&
    \includegraphics[width=0.45\columnwidth]{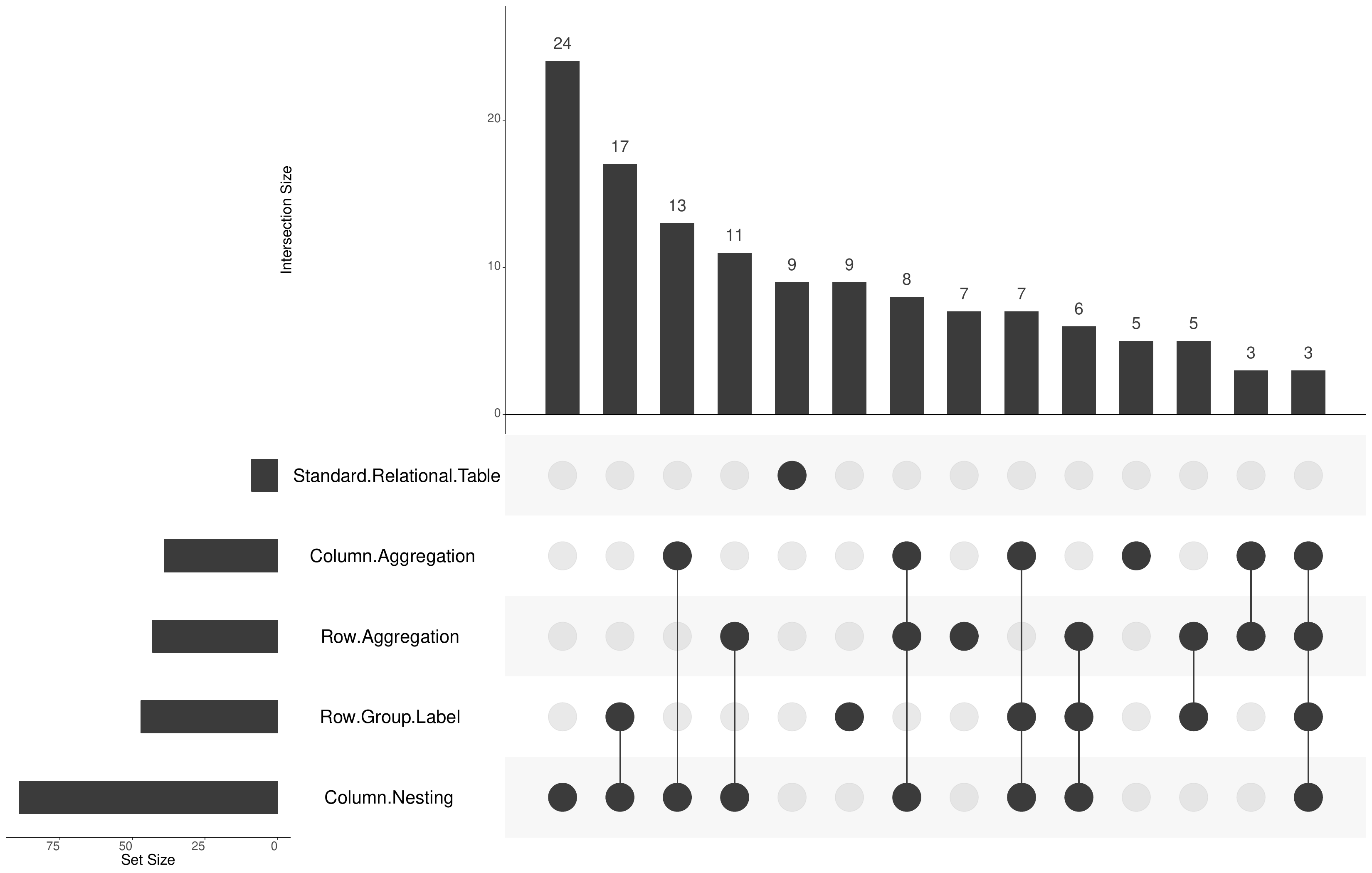}\\
    (a) \textbf{QNPC} &&(b) \textbf{Pak. C}\\
    \includegraphics[width=0.45\columnwidth]{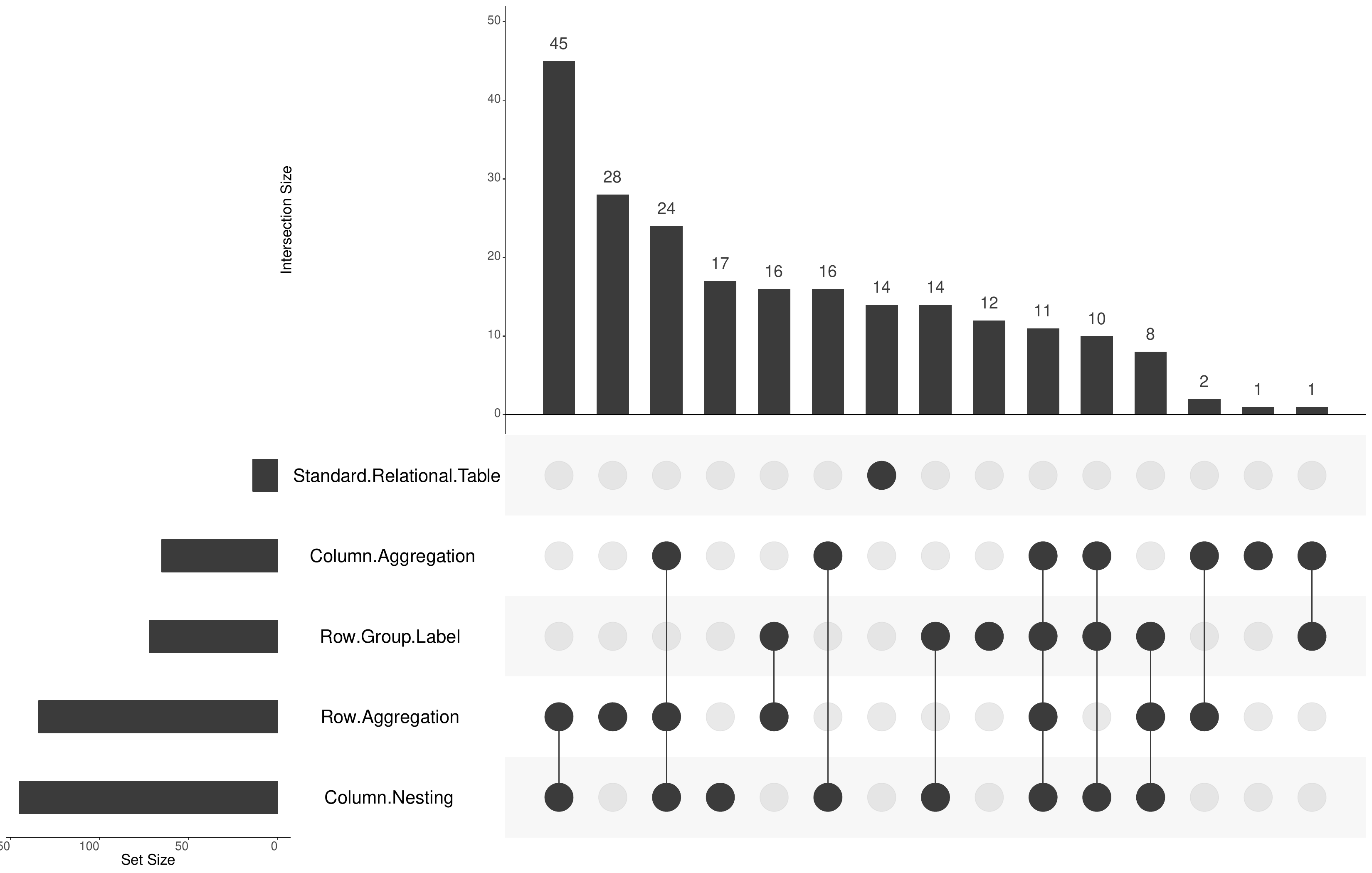}&&
    \includegraphics[width=0.45\columnwidth]{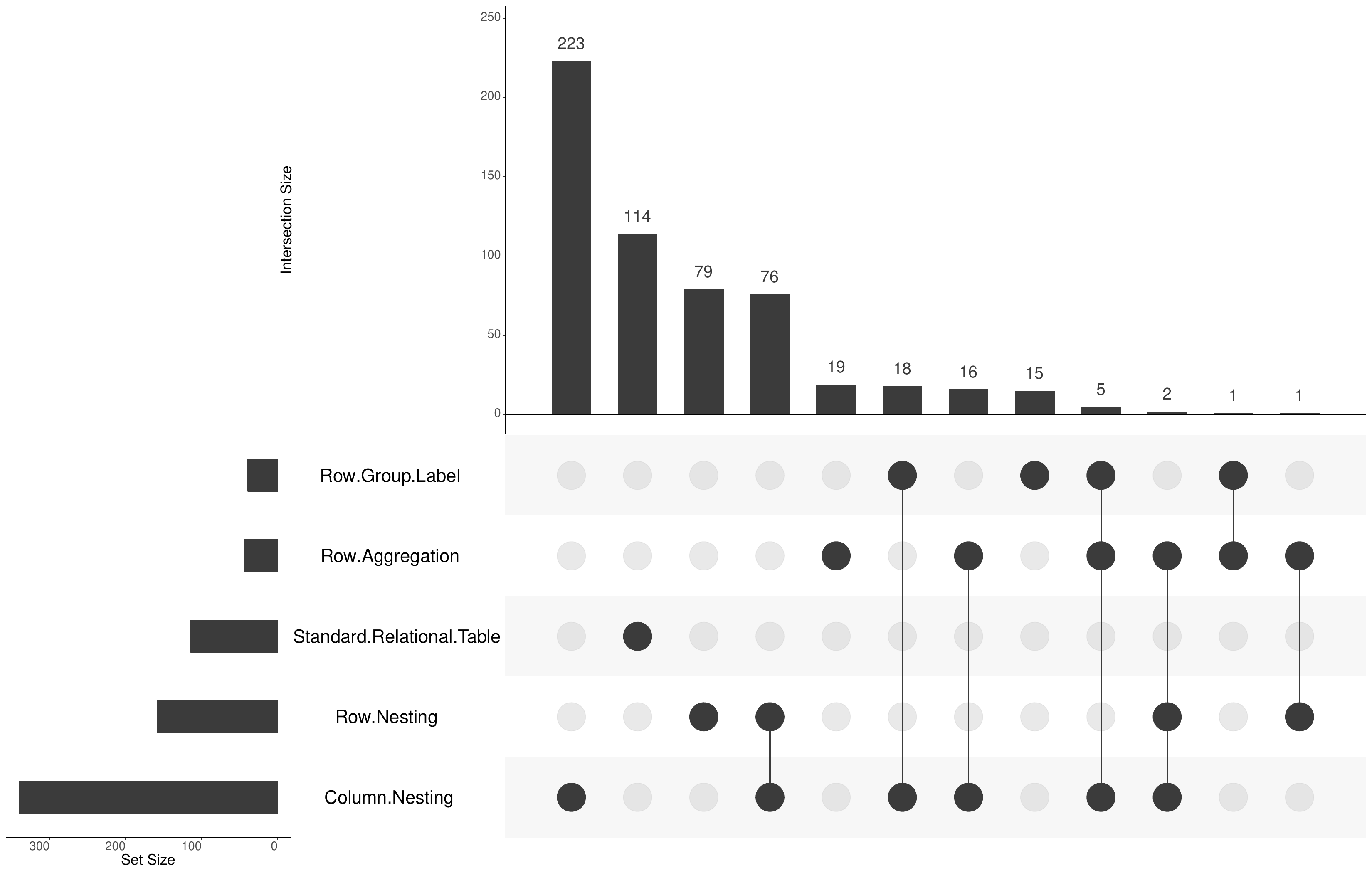}\\
    (c) \textbf{US C.} &&(d) \textbf{ArXiv}\\
    \end{tabular}
    \caption{Distribution of main table properties for each real dataset.\label{fig:upset_table_main}}

\end{figure}

\begin{figure}[h!]
    \centering
    \begin{tabular}{ccc}
    \includegraphics[width=0.45\columnwidth]{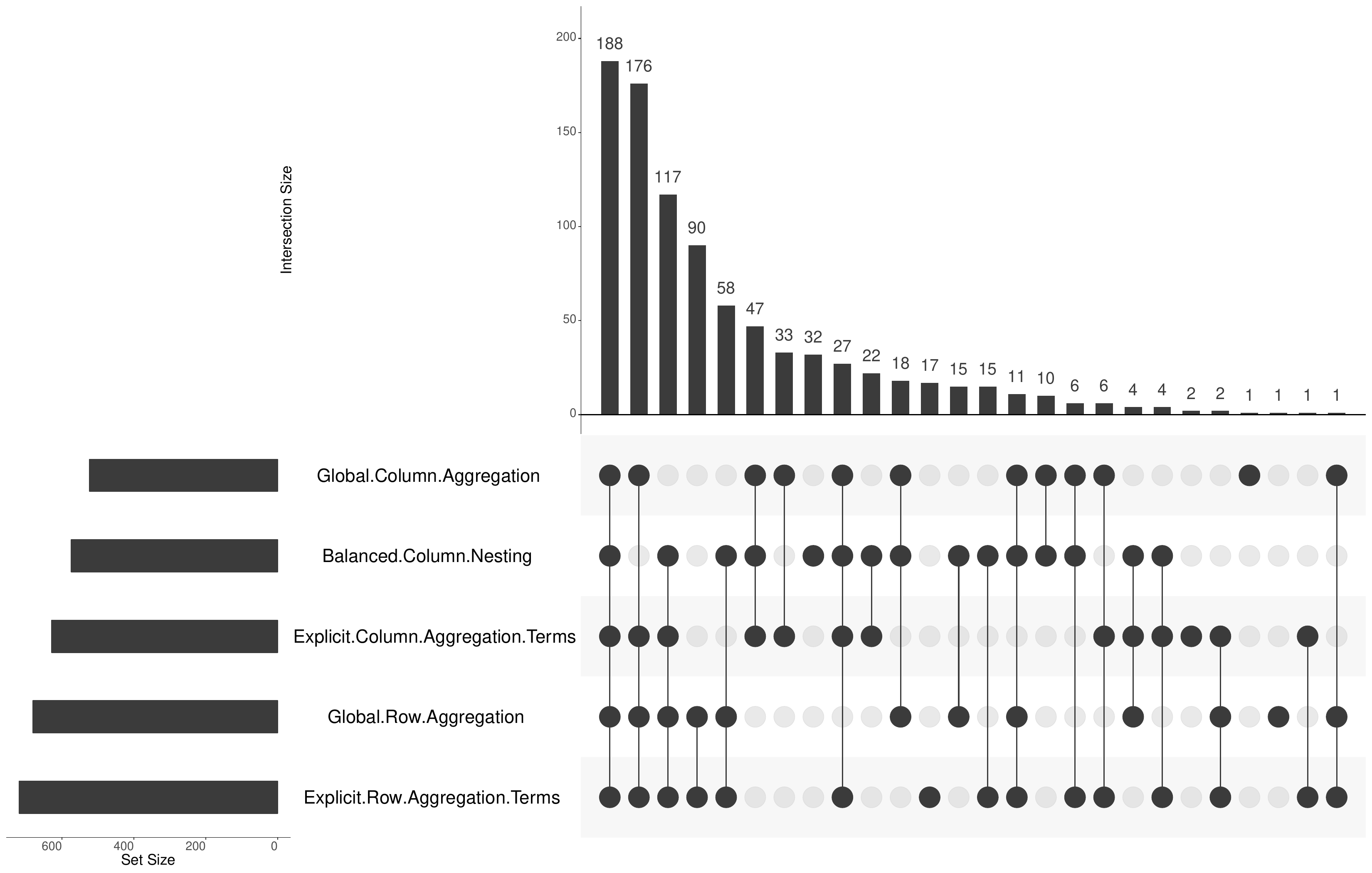}&&
    \includegraphics[width=0.45\columnwidth]{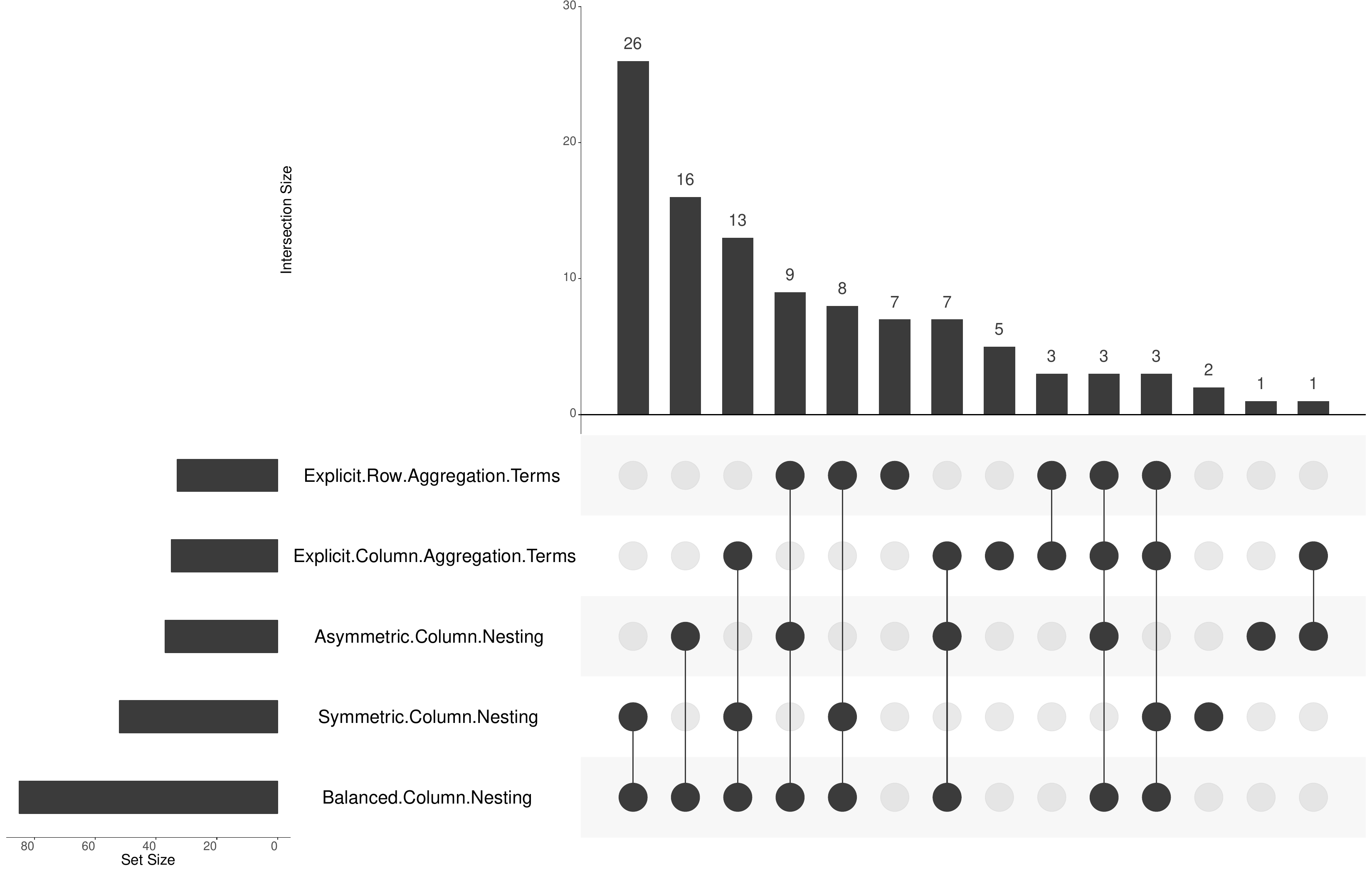}\\
    (a) \textbf{QNPC} &&(b) \textbf{Pak. C}\\
    \includegraphics[width=0.45\columnwidth]{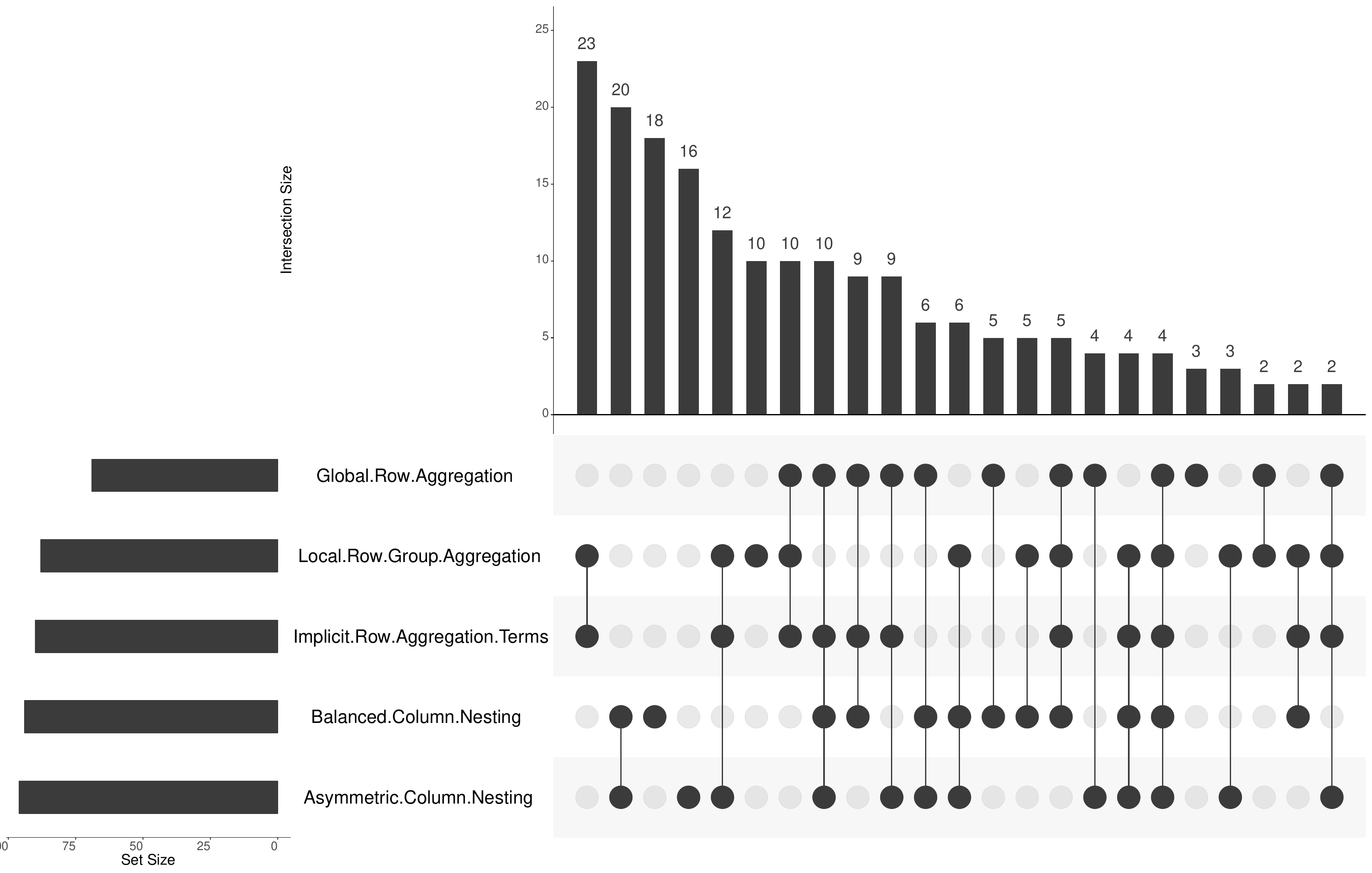}&&
    \includegraphics[width=0.45\columnwidth]{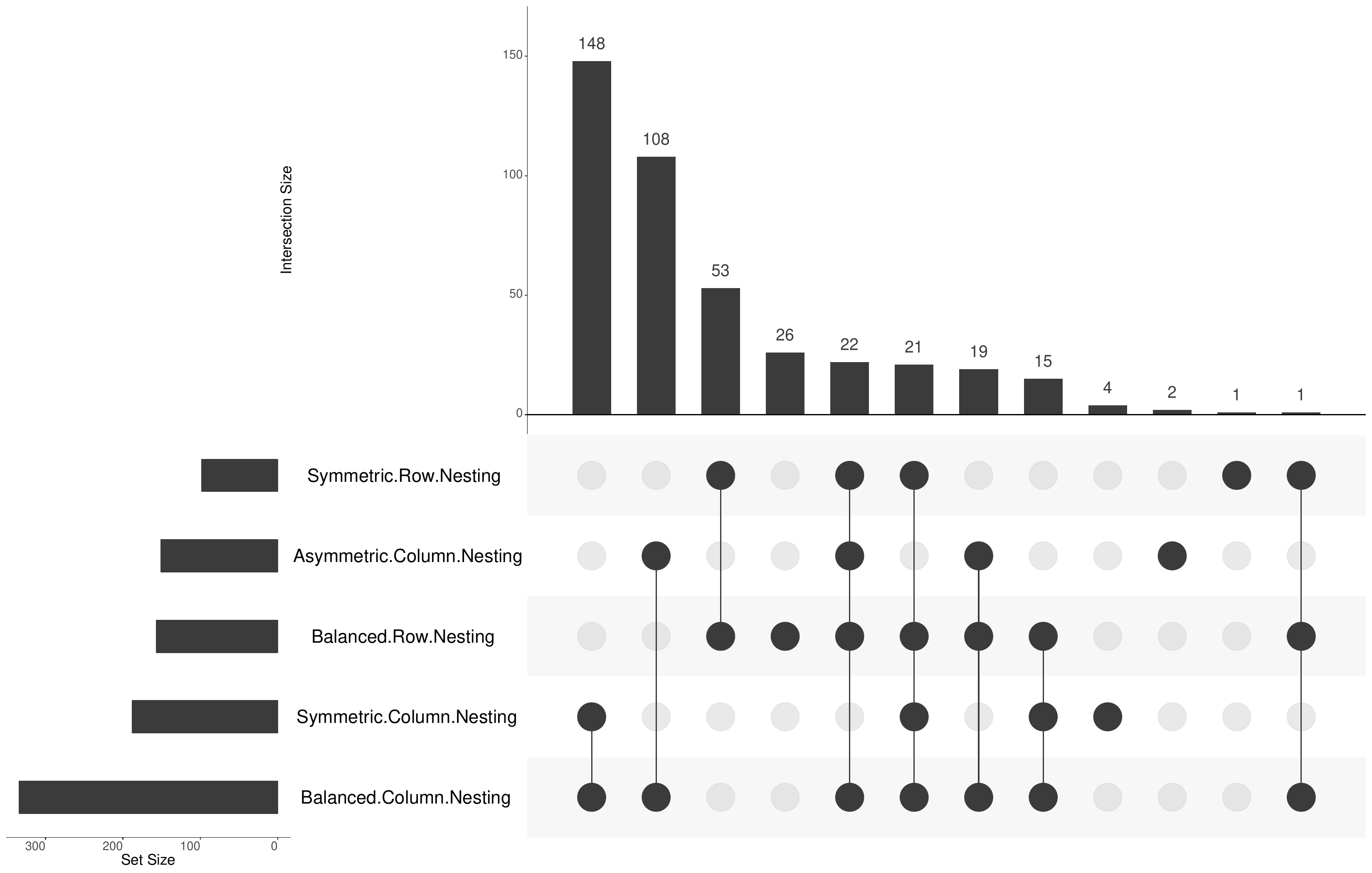}\\
    (c) \textbf{US C.}&&(d) \textbf{ArXiv}\\
    \end{tabular}
    \caption{Distribution of detailed table properties for each real dataset.\label{fig:upset_table_sub}}

\end{figure}

The Figure~\ref{fig:upset_table_main} and  Figure~\ref{fig:upset_table_sub} show the distribution of the top-$5$ main and detailed real table properties per dataset. The property distribution is very different between datasets. For instance, column and row nesting are present in \textbf{QNPC} and \textbf{ArXiv} datasets  (Figure~\ref{fig:upset_table_main}(a) and (d)) but no row nesting appears in \textbf{US C.} or \textbf{Pak C.} datasets. Column and row aggregation appear in all datasets except in \textbf{ArXiv} with only a small proportion of row aggregation. \textbf{QNPC} datasets have the most various and complex combinations of main properties, while \textbf{ArXiv} have the least various.

Regarding detailed table properties in Figure~\ref{fig:upset_table_sub}, \textbf{QNPC} and \textbf{US C.} show quite complex and various combinations compared to \textbf{Pak. C} and \textbf{ArXiv}. While balanced column nesting is a typical property for all the tables, it is the major one in \textbf{Pak. C} and \textbf{ArXiv} while it is more balanced in the other two datasets.

\section{Synthetic Data Generation}
\label{appendix:synthetic_table_generation}

\subsection{Synthetic Generator}
\label{appendix:synthetic_gen_process}

In the following subsections, we detail the configuration templates we used for setting the domain vocabulary (Section~\ref{appendix:synthetic_config_semantics}) common to both \hcts, relational tables, SQL queries, and natural language questions.
Then we detail the templates used for
generating the synthetic \hcts (Section~\ref{appendix:synthetic_config_tables}), and those used for generating the SQL queries and the Natural Language questions (Section~\ref{appendix:synthetic_config_questions}).

The codes of the synthetic generator and the instructions for installing and running them are available at \url{https://github.com/qcri/HCTQA-Benchmark/blob/main/synthetic_data_generator/README_SYNTHETIC_GENERATOR.md}.

\subsection{Domain vocabulary templates}
\label{appendix:synthetic_config_semantics}

\begin{figure*}[h!]
    \centering
    \includegraphics[width=0.95\textwidth]{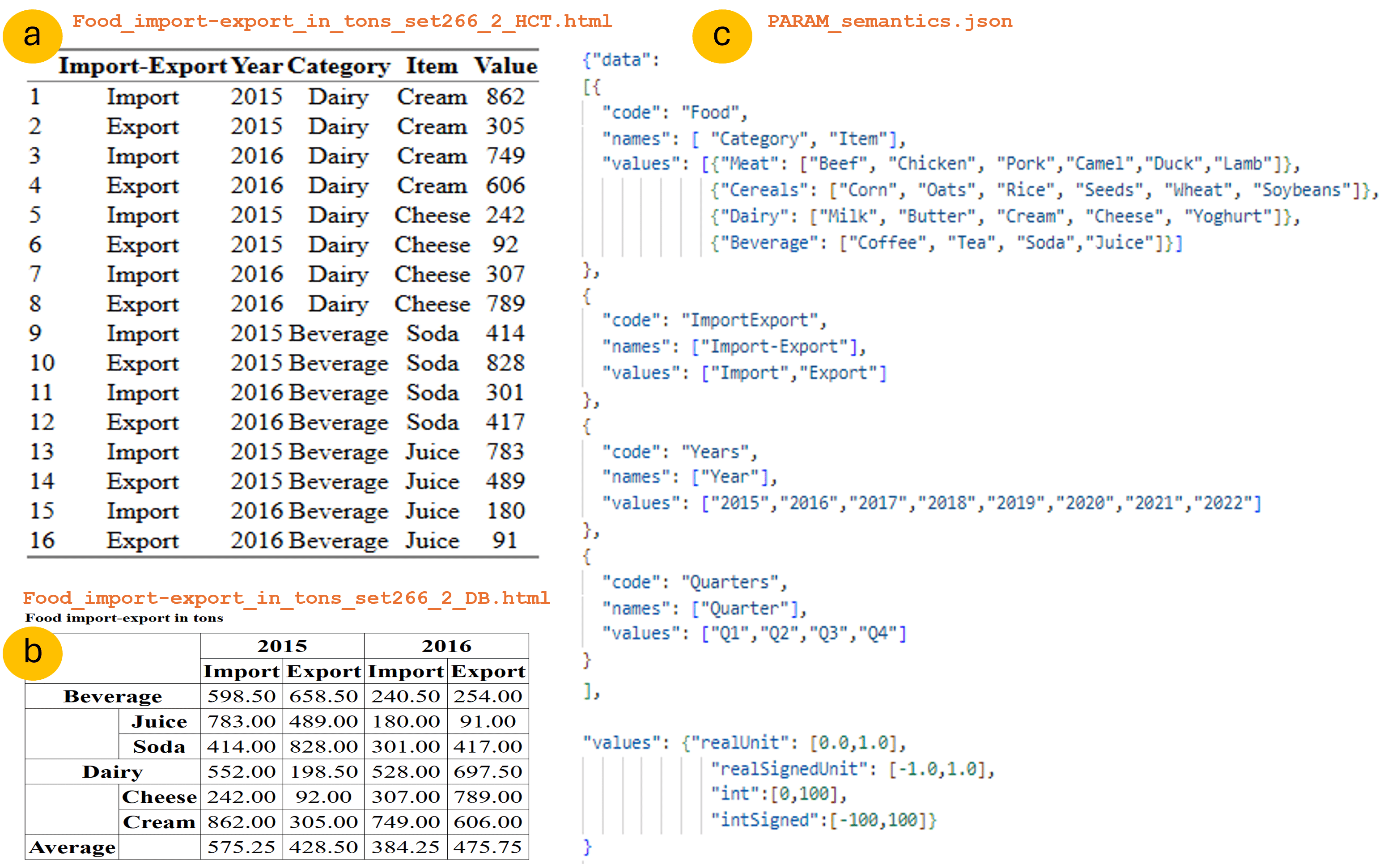} 
    \caption{Domain vocabulary for generating synthetic tables and questions. Examples of $T_{REL}$ (a) and a corresponding $T_{HCT}$ (b) related to the amount of food import-export in some place. (c) JSON format to describe Food Import-Export domain vocabulary used to generate these tables (see Section~\ref{appendix:synthetic_config_tables}) and the related questions (see Section~\ref{appendix:synthetic_config_questions}). Not all vocabulary is used in $T_{REL}$ and $T_{HCT}$. 
    \label{fig:synthetic_semantics_template}}
\end{figure*}

In this section, all \ttt{terms} refer to Figure~\ref{fig:synthetic_semantics_template}.

Our synthetic data relies on domain vocabulary naming the dataset's attributes and values. 
A synthetic $T_{HCT}$ (Figure~\ref{fig:synthetic_semantics_template}b) is obtained by pivoting a relational table $T_{REL}$ (Figure~\ref{fig:synthetic_semantics_template}a) composed of at least two nominal attributes and a single numerical attribute always named \ttt{Value}.

There are two types of nominal attributes: 
\begin{itemize}
\item \textbf{Independent attributes} like \ttt{Import-Export}, \ttt{Years}, or \ttt{Quarters} can be organized as row or column headers at any nesting level in $T_{HCT}$. 
\item \textbf{Hierarchical attributes} like food \ttt{Category} (\textit{e.g.} \ttt{Meat, Cereals, Dairy, Beverage}) and food \ttt{Item} (\eg \ttt{Cream, Soda, Milk}) are dependent on each other. If both attributes appear in the \hct, the food \ttt{Item} must be displayed at a lower level than the food \ttt{Category}, \textit{i.e.} closer to the inner part of the \hct, and all must appear together either in row headers or column headers.
\end{itemize}
 
Domain vocabulary is encoded in a JSON file (Figure~\ref{fig:synthetic_semantics_template}c) with two main keys: \ttt{data} and \ttt{values}. 

\begin{itemize}
\item \ttt{data} refers to a list of attributes (\eg \ttt{ImportExport}, \ttt{Years}, \ttt{Quarters}) or groups of attributes (\eg \ttt{Food}), each described with three keys:
\begin{itemize}
    \item \ttt{code} is the primary key used in table and question templates to identify the independent attributes (\eg \ttt{ImportExport} or \ttt{Years}) and the hierarchical groups of attributes (\eg \ttt{Food}). 
    \item \ttt{names} is the list of attribute names displayed in $T_{REL}$ column headers but typically hidden (implicit) in $T_{HCT}$ and $Q_{NL}$. It is a list with a single name for independent attributes, or with as many names as levels for hierarchical groups of attributes (\eg \ttt{Food}), ordered from top (outer) to bottom (inner) levels (\eg \ttt{Category}, \ttt{Item}). 
    \item \ttt{values} is the list of values (\eg \ttt{2015, 2016...} for the attribute \ttt{Year}). It can also be the hierarchy of values of a group of attributes. For instance, in \ttt{Food} group, \ttt{Meat} is a value of the attribute \ttt{Category}, and \ttt{Beef}, \ttt{Chicken}..., are values of the attribute \ttt{Item}, whose parents are \ttt{Meat} in the \ttt{Food} hierarchy. Attribute values are always displayed as row or column headers in $T_{HCT}$.
\end{itemize}

\item \ttt{values} refers to the numerical values to be generated in both the core part of $T_{HCT}$ and the \ttt{Value} column of $T_{REL}$. This object contains shortcut keys (\eg \ttt{realUnit} or \ttt{intSigned}) to generate specific ranges and types of numbers when called in table templates. The value associated with such shortcuts must be a list of two numbers $[m, M]$ representing the minimum $m$ and maximum $M$ possible values. If any of these min and max is a decimal number, the generated value will be a real number with two decimal digits drawn uniformly in this interval. If both min and max are integers, the generated value will be an integer drawn uniformly in this interval. The generator will ensure no replica in the generated values, reals or integers, possibly overflowing the max value if asked to generate more numeric values than the range allows.

\end{itemize}

\noindent\textbf{Customization of domain vocabulary} All \ttt{terms} in domain vocabulary are customizable except the JSON keys \ttt{data}, \ttt{code}, \ttt{names}, and \ttt{values}, following the rules above. The number of decimals of generated real numbers is specified by \ttt{NUM\_DECIMAL\_DIGITS\_REAL\_FORMAT} in \ttt{config.R}. 
The domain vocabulary is stored in the file \ttt{PARAM\_semantics.json}.

\subsection{Table templates}
\label{appendix:synthetic_config_tables}


The generic table templates described in the file \ttt{PARAM\_tableTemplate.json} are used to generate a (large) set of $T_{REL}$ and $T_{HCT}$ tables from the domain vocabulary (See Section~\ref{appendix:synthetic_config_semantics}). 

This template contains the following keys:
\begin{itemize}
    
    \item \ttt{replica} is an integer setting how many \textit{random}  variants of each initial table template from the list \ttt{tables} will be generated.
    
    \item \ttt{shuffle} is a list of column and row shuffling commands that will be applied to all the initial table templates listed in \ttt{tables}. It is used to vary the structure of the initial table. \ttt{none} is the default; \ttt{row} mixes only row attributes; \ttt{cols} mixes only column attributes; \ttt{rowscols} mixes rows and columns, keeping them as rows or columns, respectively; \ttt{all} mixes across rows and columns (some columns can become rows and vice-versa, except for hierarchical attributes which stay grouped in rows or columns). The shuffling process is a random permutation of the rows or columns to be mixed.  

    \item \ttt{col\_row\_levels} is a list of authorized depths for nested rows and columns. It must be a list of strings "$dc$\_$dr$" indicating the depth of the column header ($dc$), and of the row header ($dr$), with $dc$ and $dr$ in $\{1,2,3\}$.
    
    \item \ttt{col\_row\_name\_pos} is a list of authorized displays for row and column attribute names. It must be a list of strings "C\_R" indicating where the column (C) or row (R) attribute shall be displayed. R and C can be \ttt{none} to hide the attribute name, \ttt{left} to show it on the left side of the header, or \ttt{top} to show it at the top of the header. (\textit{Place-holder for a future version, ignored by the generator}).
    
    \item \ttt{col\_row\_agg\_pos} is a list of authorized positions for aggregate rows or columns. It must be a list of strings "C\_R" indicating on which side of the header the aggregate column (C) or row (R) shall be displayed: R can be set to \ttt{top} or \ttt{bottom}, C can be set to \ttt{left} or \ttt{right}, and both can be set to \ttt{none} (Default) to hide the corresponding aggregate.
    
    \item \ttt{row\_format} can be set to \ttt{new} to use one column for each level of row header, or \ttt{indent} to create an empty row for all level $L$ and indent all rows at level $L+1$. The header of the empty row is the attribute value for that row. An aggregate can fill this row if \ttt{col\_row\_agg\_pos} is set to \ttt{*\_top}.  
    
    \item \ttt{tables} is a list of table content specifications (values, headers, and aggregates) with coherent domain vocabulary. Each initial table template in the list has the following keys: 
    
    \begin{itemize}
        
        \item \ttt{valueName} is the title displayed at the top of the $T_{HCT}$ table (\eg "Food import-export in tons")
        
        \item \ttt{values} is either a shortcut name referring to one of the keys (\eg \ttt{realUnit}) stored in \ttt{values} in the domain vocabulary (see Figure~\ref{fig:synthetic_semantics_template}), or directly an interval of the form $[m, M]$ following the description of the main key \ttt{values} in Section~\ref{appendix:synthetic_config_semantics} 
        
        \item \ttt{valueUnit} (\textit{Place-holder for a future version, not yet implemented}) is a character string to be added to give units to values in the core part of the table (\eg $\%$ sign after each value in the core part of $T_{REL}$).
        
        \item \ttt{rowCodes} is a list of one or more of the \ttt{data} \ttt{code} keys (\eg \ttt{Food} or \ttt{Quarters}) listed in the domain vocabulary. The generator will pick among these codes to generate the rows of $T_{HCT}$, and one part of the nominal attributes of $T_{REL}$.
        
        \item \ttt{rowSamples} is a list containing as many intervals $[m,M]$ as there are codes listed in \ttt{rowCodes}, in the same order. Each interval specifies the minimum $m$ and maximum $M$ number $n$ of values to select among the list $L$ of values of the specified attribute. The generator first picks $n$ at random in $[m,M]$, then selects $n$ consecutive values in $L$ starting at a random position in $L$ but ensuring there are always $n$ values in the result if the list $L$ is long enough, or return $L$ itself otherwise. For instance, let's consider an independent attribute \ttt{Years} with a list of values: $L = [2001, 2002, 2003, 2004, 2005,2006]$, if we set \ttt{rowSamples} to $[1,5]$, and the generator randomly picks $n=3$, it then picks a random index $i$ between $1$ and $4$, \eg $i=2$, to generate the final subset $[2002, 2003, 2004]$ of years to use in the table. For hierarchical attributes, the same process applies recursively to all levels of the hierarchy, starting from the top level. If the interval is set to $[0]$ instead, all the values of the corresponding independent attribute are generated, or all the values at all hierarchy levels are generated for hierarchical attributes. Note that the \ttt{col\_row\_levels} parameter limits the depth of row and column headers, so the attributes finally displayed in $T_{HCT}$.
        
        \item \ttt{colCodes} is the same as \ttt{rowCodes} but for columns. The generator will pick among these codes to generate the columns of $T_{HCT}$, and the other part of the nominal attributes of $T_{REL}$.
        
        \item \ttt{colSamples} is the same as \ttt{rowSamples} but for columns.
        
        \item \ttt{agg\_name1} is the name of the aggregate to display in the table.
        
        \item \ttt{agg\_fun1} is the aggregate function to use among \ttt{sum} (default), \ttt{avg} (average), \ttt{min}, \ttt{max}. 
    \end{itemize}
\end{itemize}

These generic table templates \ttt{PARAM\_tableTemplate.json} are used with the domain vocabulary \ttt{PARAM\_semantics.json} to generate individual templates using the script \ttt{1\_geneAllTablesJSONfromPatterns.R}. The individual templates generated and described below are stored in the file \ttt{PARAM\_tablesToGenerate.json}.

\noindent\textbf{Customization of generic table templates} These generic table templates are customizable except for all the JSON keys and following the rules above.

\begin{figure*}[ht!]
    \centering
    \begin{tabular}{c}
    \includegraphics[width=0.8\textwidth]{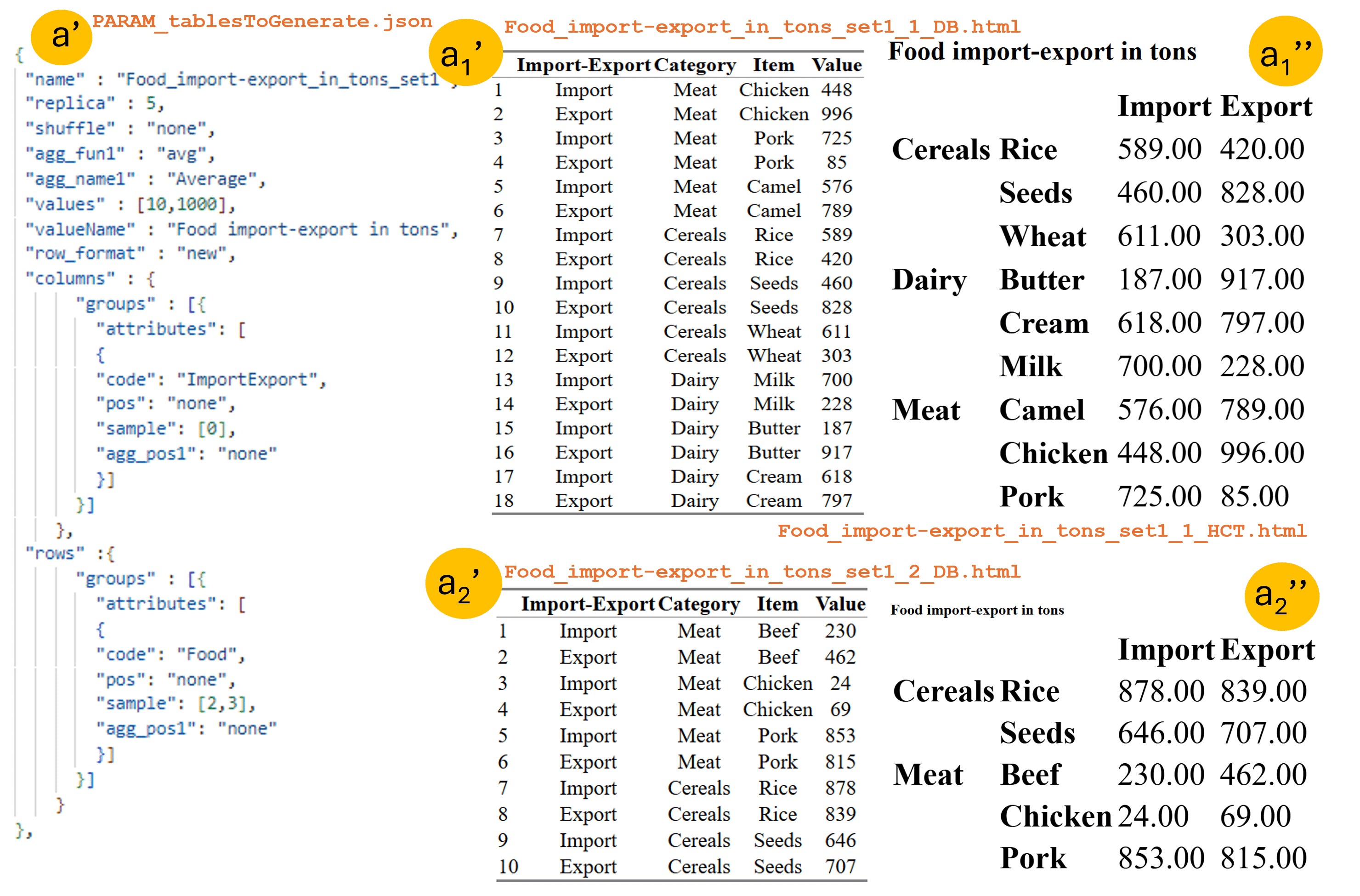}\\
    \includegraphics[width=0.82\textwidth]{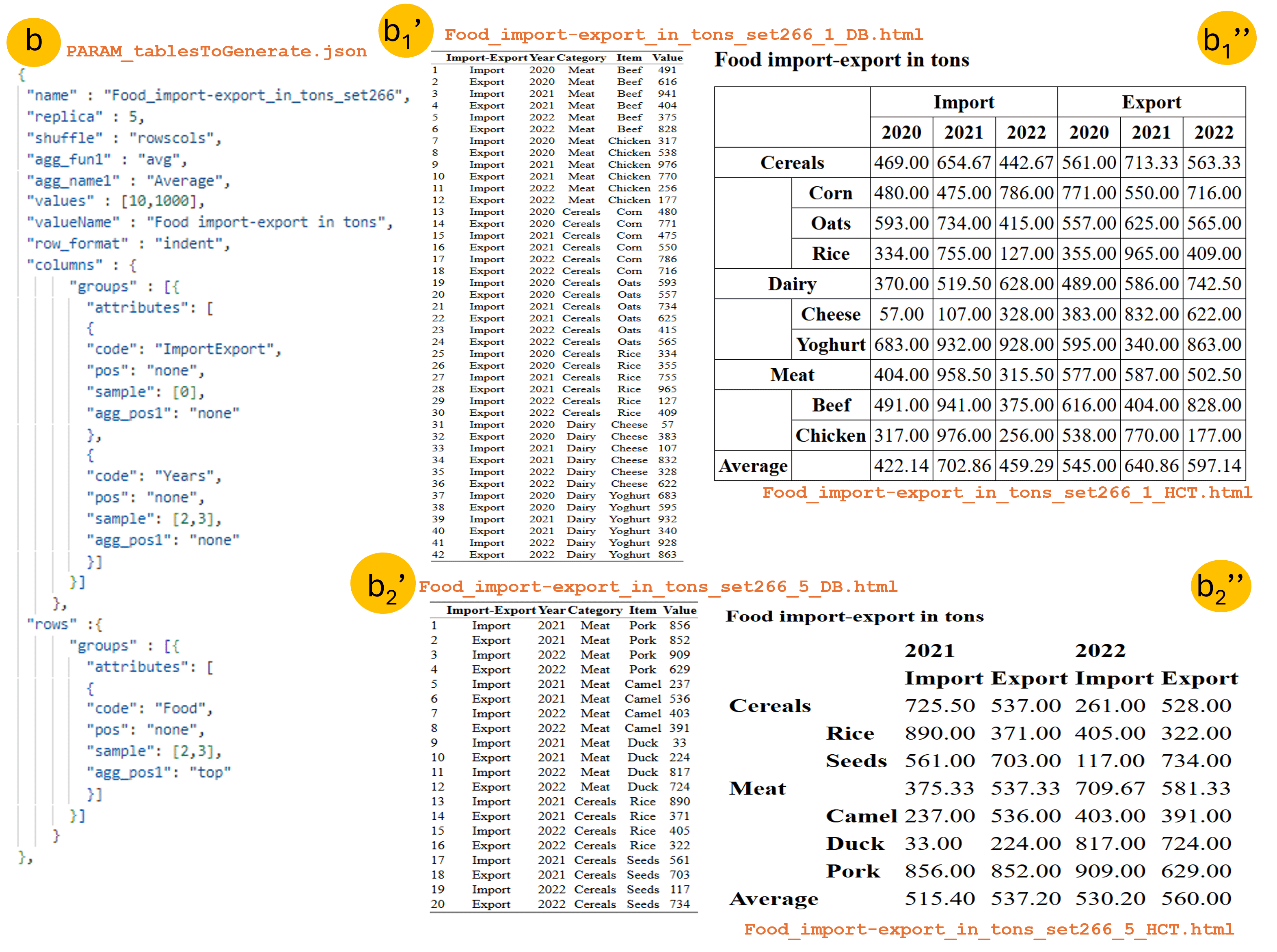}
    \end{tabular}
    \caption{Two examples of individual table templates (a,b), each with two random samples (1, 2) of $T_{REL}$ (a', b') and $T_{HCT}$ (a", b"). Templates a and b differ by the depth of column headers (\ttt{col\_row\_levels}), row indentation (\ttt{row\_format}), presence and position of aggregates (\ttt{agg\_pos1}). The sampled $T_{HCT}$ (1, 2) of a given template (a or b), further differs by the randomized order of the attributes (\ttt{shuffle}), number and names of the attribute values (\ttt{sample}), core numerical values (\ttt{values}), and presence of table cell borders (random).   
    \label{fig:synthetic_table_template_individual}}
\end{figure*}

An individual table template is a JSON object instantiating one combination of the parameters \ttt{shuffle}, \ttt{col\_row\_levels}, \ttt{col\_row\_name\_pos}, \ttt{col\_row\_agg\_pos}, and \ttt{row\_format} common to all initial table templates, and set in the generic table template. It contains the following keys:

\begin{itemize}
\item \ttt{name} is the identifier of the family of $T_{HCT}$ and $T_{REL}$ tables specified by this individual table template. 
\item \ttt{replica} is identical to \ttt{replica} set in the generic table template.
\item \ttt{agg\_fun1} is identical to \ttt{agg\_fun1} set in the generic table template.
\item \ttt{agg\_name1} is identical to \ttt{agg\_name1} set in the generic table template.
\item \ttt{values} is identical to \ttt{values} set in the generic table template.
\item \ttt{valueName} is identical to \ttt{valueName} set in the generic table template.
\item \ttt{shuffle} is one of the shuffling commands listed in \ttt{shuffle} of the generic table template.
\item \ttt{row\_format} is one of the row format commands listed in \ttt{row\_format} of the generic table template.
\item \ttt{columns} and \ttt{rows} are lists of attribute \ttt{codes} obtained after filtering the first $dr$ row attributes and $dc$ column attributes and setting the aggregate position using the \ttt{col\_row\_levels} and \ttt{agg\_pos1} parameters from the parameter combination of this instance. \ttt{sample} are identical to \ttt{colSamples} and \ttt{rowSamples} of the corresponding attribute set in the generic table template. 
\end{itemize}

Finally, \ttt{replica} instances of $T_{HCT}$ are generated using the script \ttt{2\_geneHCTfromJSONtables.R} which takes as input the files \ttt{PARAM\_semantics.json}, \ttt{PARAM\_tableTemplate.json}, and \ttt{PARAM\_tablesToGenerate.json}. It returns HTML files named \ttt{[name]\_n\_DB.html} ($T_{REL}$) and \ttt{[name]\_n\_HCT.html} ($T_{HCT}$) with $n$ the replica number. 

This process draws samples from each individual template, randomizing all the remaining parameters: the \ttt{shuffle} (stochastic) command is applied to rows and columns attributes, then the values of each attribute are sampled using the \ttt{sample} parameter, and the values in the core of the table are sampled from the \ttt{values} parameter. The absence or presence of cell borders in the $T_{HCT}$ layout is randomized for each final sample. 

Two examples of individual templates with two of the five sampled $T_{REL}$ and corresponding $T_{HCT}$ are shown in Figure~\ref{fig:synthetic_table_template_individual}.

\noindent\textbf{Customization of individual table templates} There is no possible customization of these individual table templates automatically generated from the generic table templates.

\subsection{Natural Language question templates}
\label{appendix:synthetic_config_questions}

\begin{figure*}[ht!]
    \centering
    \begin{tabular}{l}
    \includegraphics[width=0.9\textwidth]{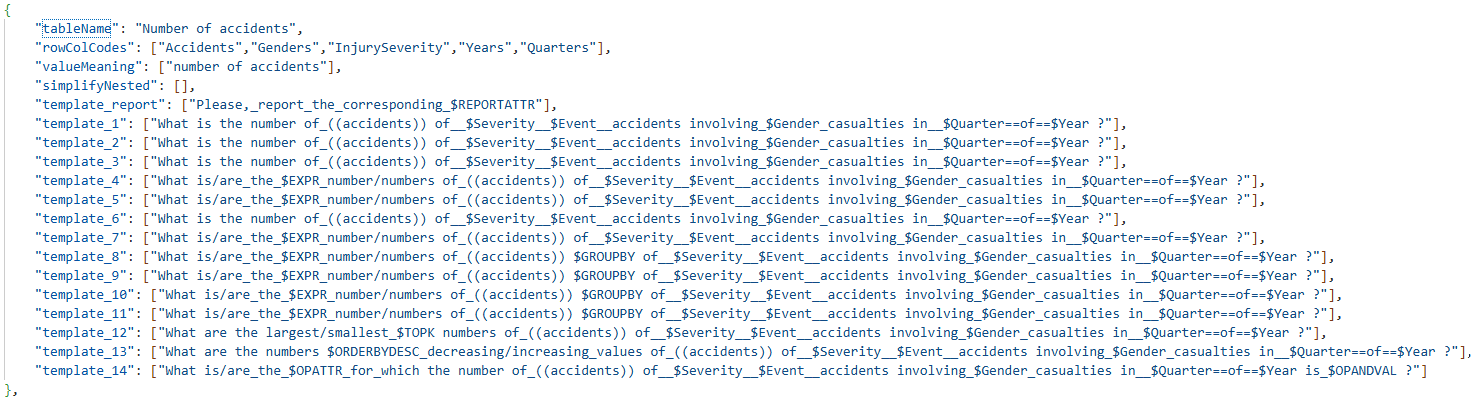}
    \\ 
    \multicolumn{1}{c}{(a)}\\
    \includegraphics[width=\textwidth]{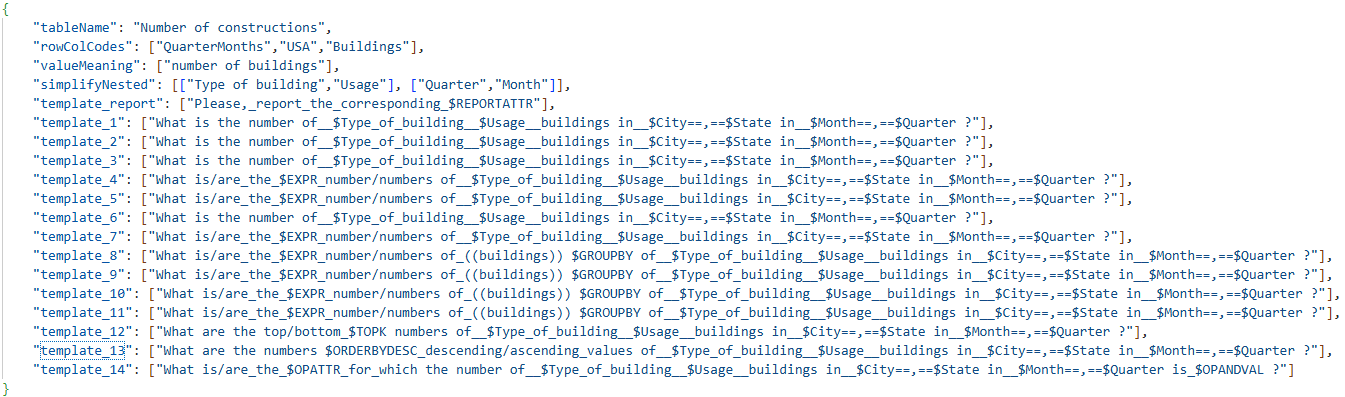}\\
    \multicolumn{1}{c}{(b)}\\
    \includegraphics[width=\textwidth]{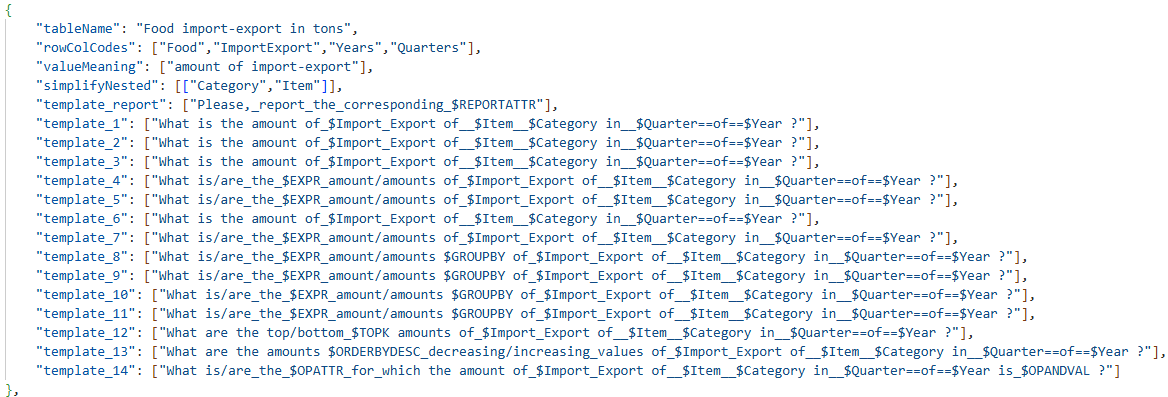}\\
    \multicolumn{1}{c}{(c)}\\
    \includegraphics[width=\textwidth]{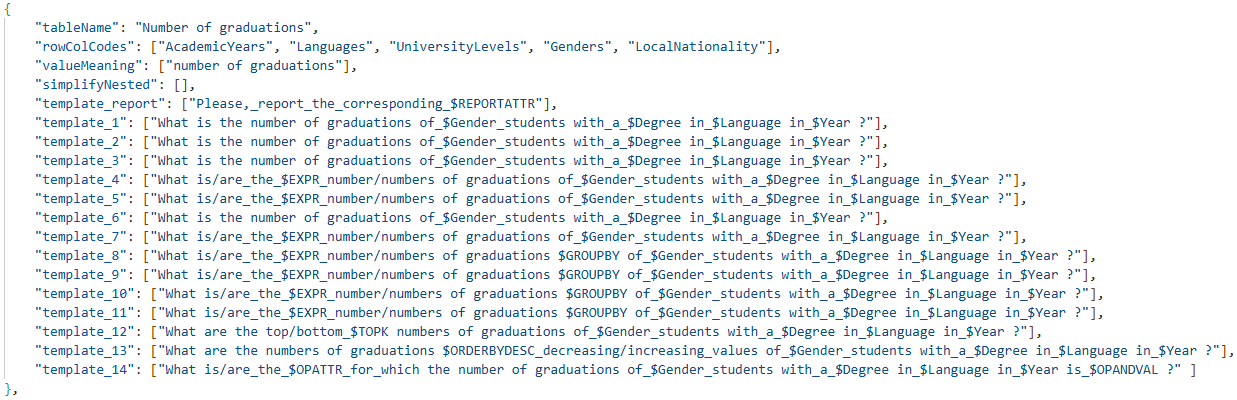}\\
    \multicolumn{1}{c}{(d)}
    \end{tabular}
    \caption{Synthetic natural question templates for the Accidents (a), Constructions (b), Food (c), and Graduations (d) domains. 
    \label{fig:synthetic_question_template}}
\end{figure*}

\begin{figure*}[ht!]
    \centering
    \begin{tabular}{l}
    \includegraphics[width=0.9\textwidth]{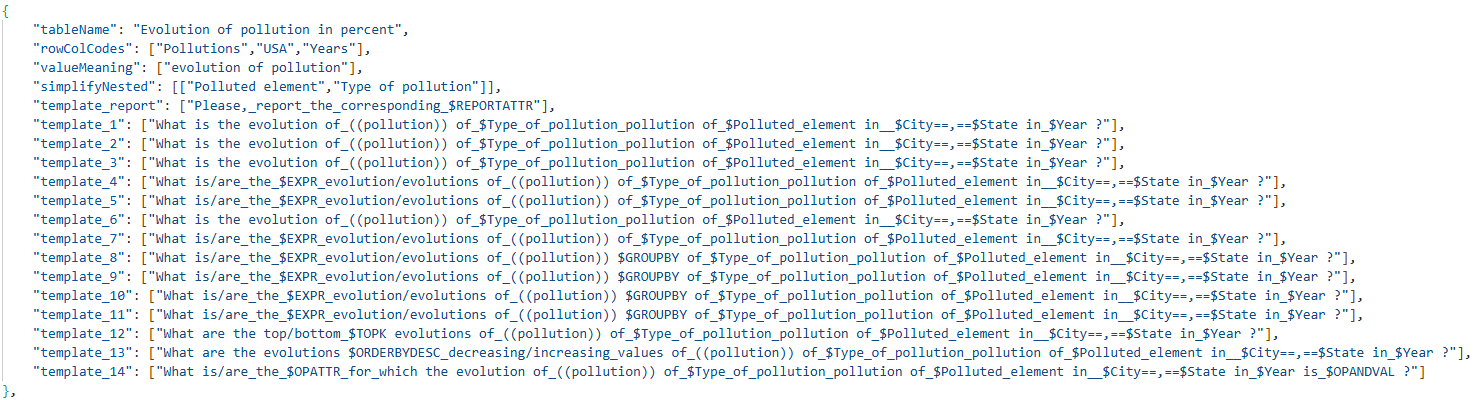}
    \\ 
    \multicolumn{1}{c}{(a)}\\
    \includegraphics[width=\textwidth]{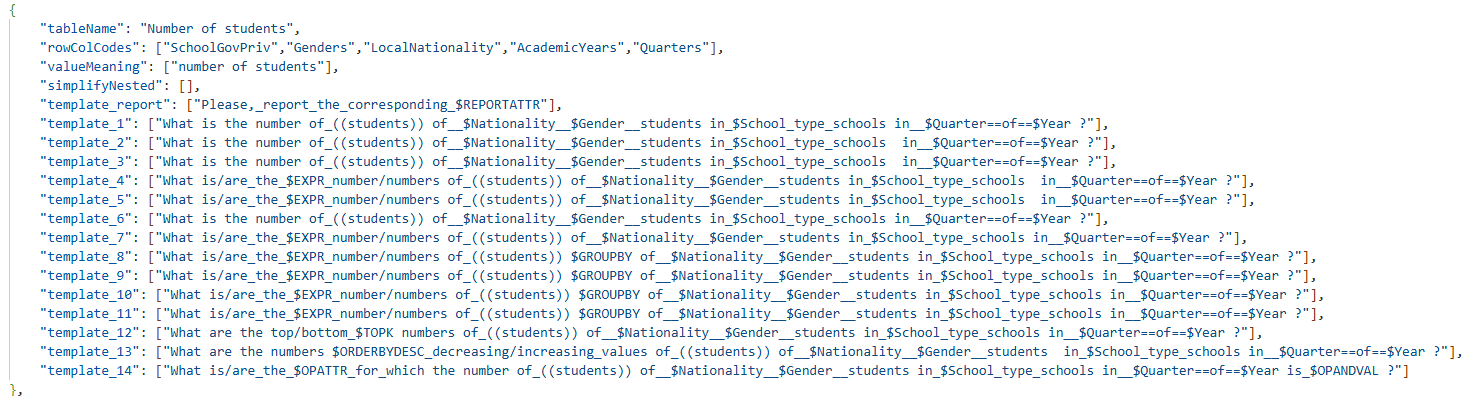}\\
    \multicolumn{1}{c}{(b)}\\
    \includegraphics[width=\textwidth]{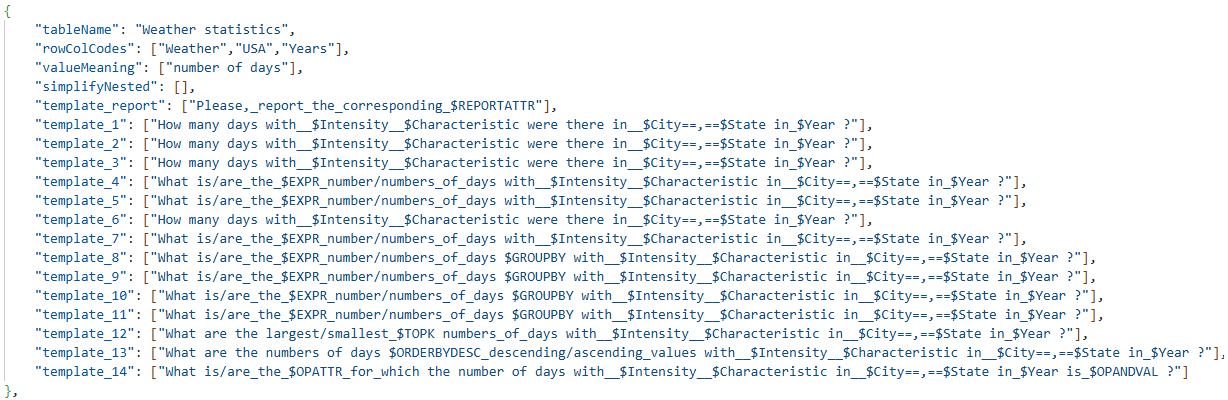}\\
    \multicolumn{1}{c}{(c)}
    \end{tabular}
    \caption{Synthetic natural question templates for the Pollution (a), Students (b), and Weather (c) domains. 
    \label{fig:synthetic_question_template_2}}
\end{figure*}

We defined $15$ SQL templates in Table~\ref{sql_templates} and transcribes them into the \hct domain using $15$ domain-specific  Natural Language templates (Figures~\ref{fig:synthetic_question_template} and \ref{fig:synthetic_question_template_2} show all the natural language question templates in JSON format related to the seven generated \hct domains). 

\begin{figure*}[t]
    \centering
    \includegraphics[width=0.9\textwidth]{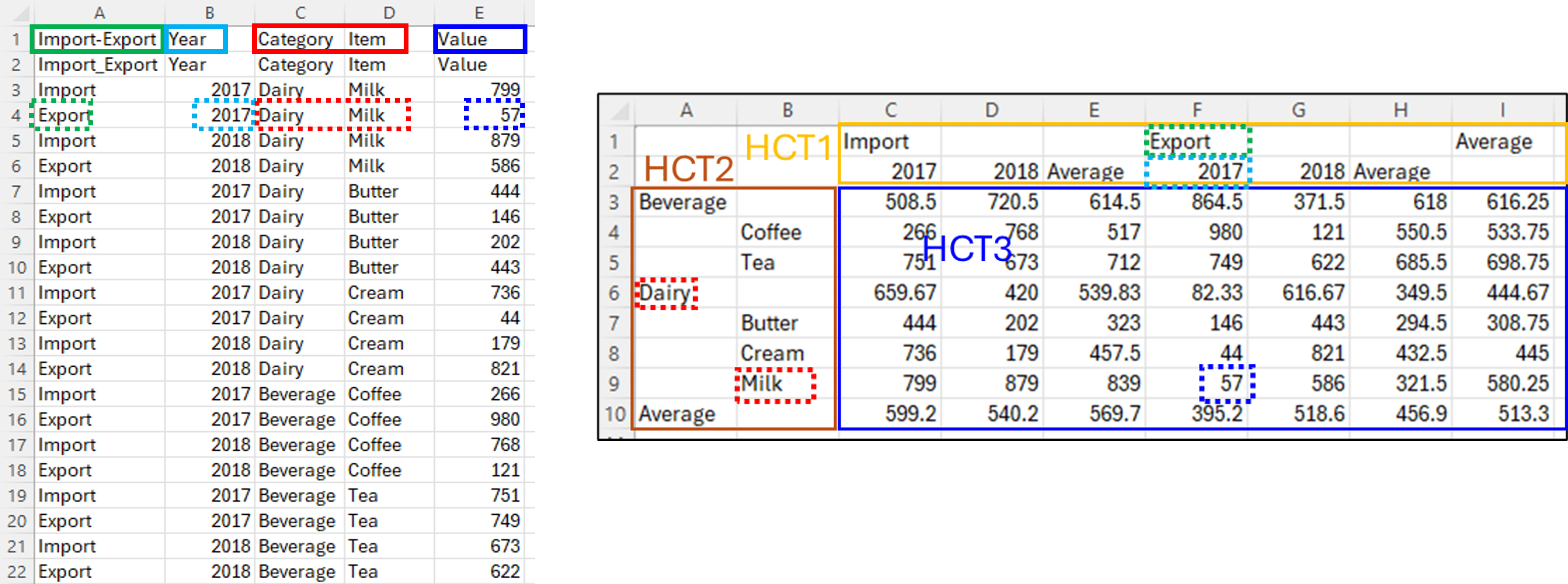}
    \caption{Example of a food import-export domain $T_{REL}$ (left) and $T_{HCT}$(right) to explain $Q_{NL}$ templates. All the values (dashed color frames) of columns A, B, C, and D of the relational table are factorized as row and column headers in the HCTs (HCT1, HCT2). In contrast, the whole value column  E (blue frame) is reproduced in the core area of the \hct (HCT3). In this instance, the ground-truth answer $A$ to the question \ttt{"What is the amount of Export of Milk in 2017?"} ($Q_{NL}$) related to $T_{HCT}$ is \ttt{"57"} (Dashed blue frame). This answer $A$ is obtained by running the SQL query $Q_{SQL}$ paired with $Q_{NL}$ onto $T_{REL}$.}
    \label{fig:synth_tables_process_appendix}
\end{figure*}

For instance, the $Q_{NL}1$ template below transcribes the $SQL_{1}$ template (Table~\ref{sql_templates}) for the Food Import-Export domain  \hct displayed in Figure \ref{fig:synth_tables_process_appendix}c:

\begin{quote}
\small\ttfamily
What is the amount of\_ \$Import\_Export of\_\_ \$Item\_\_ \$Category in\_\_ \$Quarter==of==\$Year?
\end{quote}

This template is instanciated with detailed row and column attribute values of the \hct, to generate the following natural language question:

\begin{quote}
\ttfamily
What is the amount of Export of Milk in 2017?
\end{quote}

Notice that not all the table attributes may be displayed in the \hct, and the values vary due to random selection of the \hct instance.

This question will be used as a prompt for the evaluated LLM along with the $T_{HCT}$ instance. 
The corresponding SQL query instance is:

\begin{quote}
\small\ttfamily
SELECT Value FROM DBdata \\
WHERE ((Import\_Export = 'Export' AND Year = '2017')) \\
AND ((Category = 'Dairy' AND Item = 'Milk'));
\end{quote}

and it will be run on the $T_{REL}$ instance paired with $T_{HCT}$ to get the ground-truth answer $A = "57"$.

Here are the $15$ types of natural language questions $Q_{NL}$ generated for the table $T_{HCT}$ in Figure~\ref{fig:synth_tables_process_appendix} instantiated from the template of Figure~\ref{fig:synthetic_question_template}a  ("," and ";" in answers, are column and row separators respectively):
\begin{itemize}
    \item $\mathbf{Q_{NL}1}$: \textit{one column, one row selection}: "What is the amount of Export of Milk in 2017?" $\rightarrow$ "57"
    \item $\mathbf{Q_{NL}2}$: \textit{one column, many rows selection}: "What is the amount of Export of Butter, Coffee, Milk, or Tea in 2017?" $\rightarrow$ "57; 146 ; 980; 749"
    \item $\mathbf{Q_{NL}3}$: \textit{many columns, one row selection}: "What is the amount of Export of Tea in 2017 or 2018?" $\rightarrow$ "749; 622"
    \item $\mathbf{Q_{NL}4}$: \textit{$Q_{NL}3$ + expression not in the table}: "What are the total and minimum amounts of Export or Import of Milk in 2017 or of Export of Milk in 2018?" $\rightarrow$ "1442,57"
    \item $\mathbf{Q_{NL}5}$ (opt): \textit{$Q_{NL}4$ + expression from the table}: (only if an aggregate exists in the \gls{refNameSingleton}) "What is the average amount of Export or Import of Coffee?" $\rightarrow$ "533.75"
    \item $\mathbf{Q_{NL}6}$: \textit{many rows, many columns selection}: "What is the amount of Export or Import of Butter, Cream, or Tea in 2017 or of Export of Butter, Cream, or Tea in 2018?"  $\rightarrow$ "444; 146; 443; 736; 44; 821; 751; 749; 622"
    \item $\mathbf{Q_{NL}7}$: \textit{$Q_{NL}2$ + aggregation not in the table}: "What are the total and minimum amounts of Import of Butter, Coffee, Cream, Milk, or Tea in 2017?" $\rightarrow$ "2996,266"
    \item $\mathbf{Q_{NL}8}$: \textit{$Q_{NL}6$ + grouping per column + reporting}: "What are the minimum and maximum amounts for each Import-Export and Year, of Export or Import of Butter, Coffee, Cream, Milk, or Tea in 2017 or 2018? Please, report the corresponding Import-Export and Year." $\rightarrow$ "Export,2017,44,980; Export,2018,121,821; Import,2017,266,799; Import,2018,179,879"
    \item $\mathbf{Q_{NL}9}$: \textit{$Q_{NL}2$ + grouping per row}:  "What is the minimum amount for each Category, of Import of Beverage or Dairy in 2018?" $\rightarrow$ "673;179"
    \item $\mathbf{Q_{NL}10}$: \textit{$Q_{NL}6$ + grouping per rows + reporting}: "What is the minimum amount for each Category, of Import of Beverage or Dairy in 2018? Please, report the corresponding Category." $\rightarrow$ "Beverage,673; Dairy,179"
    \item $\mathbf{Q_{NL}11}$: \textit{$Q_{NL}6$ + grouping per rows and columns + reporting}: "What is the minimum amount for each Category, Import-Export, and Year, of Import of Beverage or Dairy in 2017 or of Export or Import of Beverage or Dairy in 2018? Please, report the corresponding Category, Import-Export, and Year." $\rightarrow$ "Beverage,Export,2018,121; 
    Beverage,Import,2017,266;
    Beverage,Import,2018,673; 
    Dairy,Export,2018,443;
    Dairy,Import,2017,444; 
    Dairy,Import,2018,179"
    \item $\mathbf{Q_{NL}12}$: \textit{$Q_{NL}2$ + top-K + row filter}: "What are the bottom 5 amounts of Export of Beverage or Dairy in 2018?" $\rightarrow$ "121; 443; 586; 622; 821"
    \item $\mathbf{Q_{NL}13}$: \textit{$Q_{NL}2$ + row filter + ordering}: "What are the amounts ordered by increasing values of Import of Beverage or Dairy in 2017?" $\rightarrow$ "266; 444; 736; 751; 799"
    \item $\mathbf{Q_{NL}14}$: \textit{$Q_{NL}2$ + operation on column}: "What are the Category and Item for which the amount of Import in 2018 is greater than 513.3?" $\rightarrow$ "Dairy,Milk; Beverage,Coffee; Beverage,Tea"
    \item $\mathbf{Q_{NL}15}$: \textit{$Q_{NL}14$ with condition + reporting}: "What is the amount of Export in 2018 of Category and of Item for which the amount of Import in 2018 is greater than 513.3? Please, report the corresponding Category, Item, and amount of import-export." $\rightarrow$ "Dairy,Milk,586; Beverage,Coffee,121; Beverage,Tea,622"
\end{itemize}

Figures~\ref{fig:synthetic_question_template} and \ref{fig:synthetic_question_template_2} show the natural language question templates related to the seven generated \hct domains, as JSON objects with the following keys: 

\begin{itemize}
    \item \ttt{tableName} is identical to the \ttt{valueName} key in the generic table template (see Section~\ref{appendix:synthetic_config_tables}).
    \item \ttt{rowColCodes} gather the \ttt{rowCodes} and \ttt{colCodes} lists of attribute codes used in rows and columns of the generic table template. 
    \item \ttt{valueMeaning} is a text characterizing a value in the table continuing the sentence: "A value in this table represents the..." (\eg "amount of import-export").
    \item \ttt{simplifyNested} lists all the \ttt{names} group of hierarchical attributes used in the generic table template.
    \item \ttt{template\_report} is a sentence to report a specific list of attributes \ttt{\$REPORTATTR} selected for the query result in templates $8, 10, 11, 15$.  
    \item \ttt{template\_N} a list of templates for questions of type N (See the list of $15$ $Q_{NL}$ templates above). 
    \begin{itemize}
        \item A template is a natural language sentence with some place-holder variables to be filled by the generator depending on the table instance content and the type of question. All the templates must show all the possible attributes set in the generic table template, but the question generator will randomly pick the ones it will use, depending on the template type; the others will be absent.
        
        \item  We use \ttt{\$attribute\_name} to specify variables that will be automatically replaced by the generator with values from that attribute present in the table instance (\eg \ttt{\$Category} could be replaced by \ttt{Meat}, \ttt{Dairy}, \ttt{Cereals}, or \ttt{Beverage}). These \ttt{attribute\_name} must appear in the \ttt{rowColCodes} list for independent attributes, and in the \ttt{simplifyNested} list for groups of hierarchical attributes.

        \item All words in capital letters \ttt{\$RESERVED} and their connected words should not be modified, and must appear as-is in their respective template.

        \item In templates $4, 5, 7, 8, 9, 10, 11$, an expression \ttt{\$EXPR} will be replaced by one of the aggregation names \ttt{agg\_name1} (\ttt{min, max...}) in the generic table template. Some question can ask several expressions at once, like the min and the max, calling for a plural version of the sentence. We use $/$ to separate the singular and plural versions of a word that the generator will pick automatically depending on the number of expression terms generated. For instance, the template $7$: \ttt{What is/are\_the\_\$EXPR\_amount/amounts of\_\$Import\_Export of...} can generate \textit{"What is the minimum amount of Export of..."} or \textit{"What are the minimum and maximum amounts of Export of..."}

        \item In templates $8, 9, 10, 11$, \ttt{\$GROUPBY} will be replaced by \ttt{for each} followed by a list of attribute names. For instance, the template $9$: \ttt{What is/are\_the\_\$EXPR\_amount/amounts \$GROUPBY of\_\$Import\_Export of\_\_\$Item\_\_\$Category in\_\_\$Quarter==of==\$Year ?} can generate the question
        \textit{"What is the minimum amount for each Category, of Import of Beverage or Dairy in 2018?"}

        \item In template $12$, \ttt{top/bottom\_\$TOPK} will be replaced by an integer and either top or bottom automatically selected, \eg \textit{"What are the top 5 amounts of Import of...}

        \item In template $13$, the group \ttt{\$ORDERBYDESC\_decreasing/increasing\_values} must be used as-is and can generate the question \textit{"What are the amounts ordered by increasing values of Import of Beverage or Dairy in 2017?"} or \textit{"What are the amounts ordered by decreasing values of..."}

        \item In template $14$, the word \ttt{\$OPPATTR} will be replaced by a list of attribute names, and the word \ttt{\$OPANDVAL} by either \ttt{greater than} or \ttt{lower than} and a certain value, leading to a question like: \textit{"What are the Category and Item for which the amount of Import in 2018 is greater than 513.3?"}
        
        \item We use \ttt{\_} to link words like \ttt{of}, \ttt{in} to their attribute. If an attribute is not present in an instance of a question, all the words connected to it with \ttt{\_} will be absent too (\eg \ttt{of\_\$Import\_Export}). 

        \item We use \ttt{\_\_} to link several attributes together. For instance, we use $\_\_$ in \ttt{of\_\_\$Item\_\_\$Category} so if \ttt{\$Item} is absent, the \ttt{of} word stays connected to \ttt{\$Category} like in \ttt{of Meat}. This pattern can generate \ttt{of Beef} and \ttt{of Meat}, but will not generate \ttt{of Beef Meat} because these attributes are part of the same hierarchical group.  

        \item We use \ttt{==} to link several attributes separated by a word that only exist if both surrounding attributes are present. For instance: we can say \ttt{in Q2 of 2022} or \ttt{in 2022} or \ttt{in Q2}. Thus, we tell that \ttt{of} is only present if both \ttt{\$Quarter} and \ttt{\$Year} are present together, by coding \ttt{in\_\_\$Quarter==of==\$Year}.

        \item We use \ttt{((not\_an\_attribute))} to ensure the presence of a word which is not an attribute. For instance, we want to enable the generation of \ttt{the number of English male students} but also \ttt{the number of students} if the \ttt{\$Nationality} and \ttt{\$Gender} attributes are absent from the generated question. As \ttt{students} is not an attribute value, we use the template \ttt{of\_((students)) of\_\_\$Nationality\_\_\$Gender\_\_students} to enable this option. The generator will only produce the first part \ttt{of\_((students))} if the word \ttt{students} does not appear in the rest of the generated sentence (See Figure~\ref{fig:synthetic_question_template}b).
 
        \item The order of the attributes used in the template is always preserved. 
        
        \item Additional rules, a template shall never start with a \ttt{\$} and a space must be set before the final question mark. Attribute \ttt{names} composed of several words separated by \ttt{-} should replace it by \ttt{\_}. 
    \end{itemize}
\end{itemize}

All $14$ templates are repeated with variations for the $7$ domains in the \ttt{PARAM\_NLquestionTemplates.json} file, guiding the creation of the same templates for new domains. \textbf{Note that the question $Q_{NL}15$ is automatically generated from templates $13$ and $14$.}

The script \ttt{3\_geneSQLandNLQfromJSONtemplates.R} reads the question template file \ttt{PARAM\_NLquestionTemplates.json} and the other parameter files, to generate the natural language questions $Q_{NL}$ and corresponding SQL queries $Q_{SQL}$ for each question template as well as their ground truth answers $A$ from the tables generated from \ttt{PARAM\_tablesToGenerate.json}.

\noindent\textbf{Customization of question templates} These question templates are customizable except for all the JSON keys and following the rules above.




\section{Additional Experiments and Details}
\label{appendix:additional_experiments}

\clearpage  

\subsection{Additional Trend Analysis \& Observations}
\label{appendix:additional_experiments_trend_analysis}

\begin{figure}[t]
    \centering
    \includegraphics[width=\columnwidth]{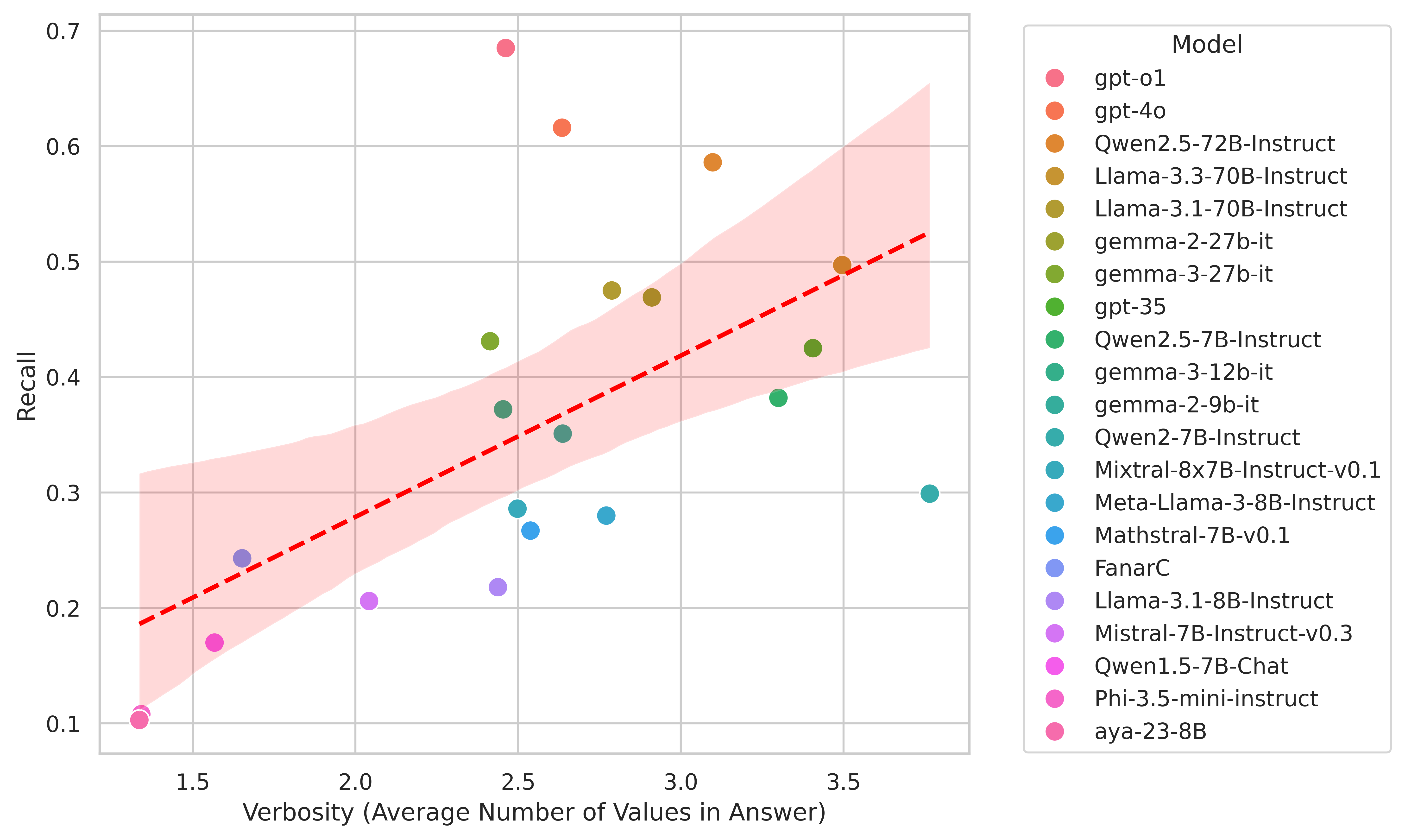}
    \caption{Recall Score vs. Verbosity for Text-Only Models on Real World HCTs.}
    \label{fig:verbosity_vs_recall}
\end{figure}

\noindent\textbf{(RQ8) How does model verbosity affect the output correctness? } Our initial hypothesis was that if models simply take the kitchen-sink approach and output more values in their answers then they would have a higher probability of matching values in the ground truth and thereby inflate their recall scores. As shown in Figure~\ref{fig:verbosity_vs_recall}, there is no clear correlation between answer length and recall. Most models, including the best performing ones, seem to be clustered around the middle on the verbosity scale. The average number of values in the \textit{ground-truth} answers for all queries is ~4, yet most models tend to provide between 2-3 values in their answers. This suggests that models may in fact be conservative in their answer verbosity as which is antithetical to the overly verbose nature of these generative models.

\begin{figure}[ht!]
    \centering
    \includegraphics[width=\columnwidth]{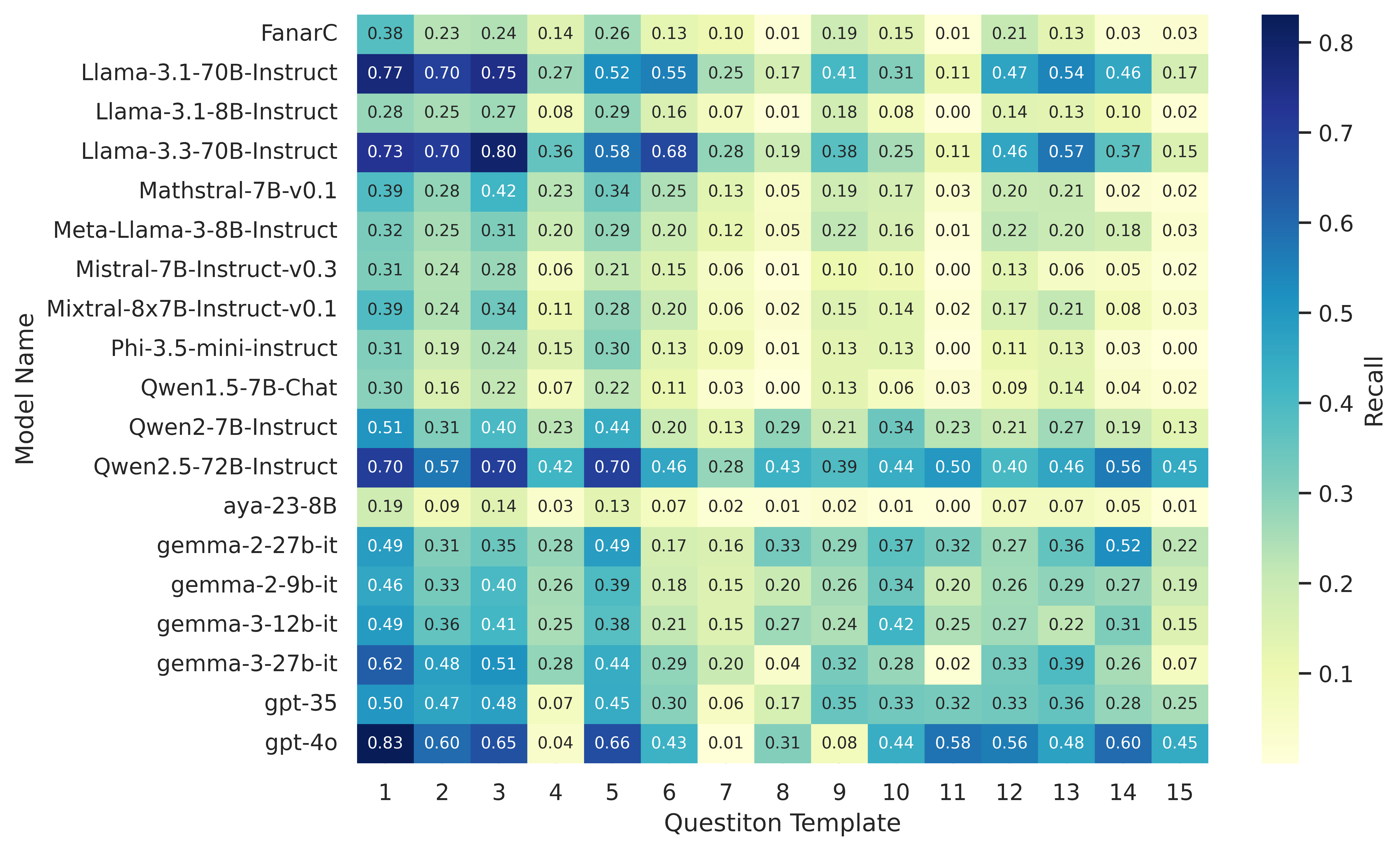}
    \caption{Recall Score for Text-Only Models on all Synthetic HCTs per Question Template.}
    \label{fig:heamtap_per_template_performance}
\end{figure}

\noindent\textbf{(RQ9) How do model perform on different synthetic QA templates?} 
The synthetic queries we generated have 15 types including various types of selection, projection, expression, and aggregation.
The model performance per template type is shown in Figure~\ref{fig:heamtap_per_template_performance}. 

As expected, models tend to do best on selection queries be it single-cell or multi-cell selection. This is clear as templates 1, 2, and 3 have the highest average recall across all models. 
Models tend to do better on queries involving expressions on columns (templates 5, 14) than queries involving aggregations (template 7). In fact, models struggle the most with template 7 which requires doing an actual aggregation (rather than retrieving an existing one from the \hct). Models like LLaMA-3.3-70B significantly outperform even models like GPT-4o and GPT-35 on this template. It is also evident that models clearly have their strengths and weaknesses in terms of templates. GPT-4o gets 58\% on template 11, whereas other models like LLaMA and Mistral do not even cross 10\%. However, this dominance completely disappears in Templates 4, 7, and 9 where the GPT-4o fails to score more than 8\% (compared to other models that get up to 42\%). There are also unexpected trends across model generations. Gemma-2-27B scores a 52\% recall on template 14 which is only topped by GPT-4o and Qwen2.5 72B, both of which are relatively larger models. However, on the exact same template (14), Gemma-3-27B scores exactly half of its predecessor with a recall score of 26\%. This shows that newer generations of models are not necessarily improving their capabilities on this task across the board.

\noindent\textbf{(RQ10) In what cases do VLMs perform better than LLMs?} Recent advances in the image understanding and reasoning capabilities of VLMs make them a suitable candidate for the task at hand. VLMs also allow to skip the extraction and conversion to CSV process as they can directly consume the images of the tables from their source documents.
On average, VLMs score 25.5\% recall whereas text-only LLMs score 25\% as shown in (Table~\ref{tab:model_performance}). GPT4o when prompted using the text-modality achieves an average recall of 61.6\%, compared to 62.3\% when using the image-modality. This illustrates the progress vision models have made in terms of document understanding and reasoning tasks such as this one. We have established that \hcts inherently have visual characteristics that are crucial to understand the content in the table. A lot of those visual clues are lost when the \hct is converted to text format (regardless of which text format its represented in). VLMs are able to overcome this information loss by taking advantage of those visual cues. This is further shown by the top performing small sized VLM model, Pixtral-12B with a recall score of 45.9\% which is nearly 8\% better than the best performing text-only LLM of that size range (Qwen2.5-7B with recall 38\%). Furthermore, Microsoft's Phi-3-Vision-128k (image-modality), which is only a 4.2B parameter model outperforms every text-only LLM in the small size category (many of which are 2x in size). These results clearly highlight the potential of VLMs in visual-reasoning tasks such as HCT-QA. 
Another interesting observation is that VLMs struggle considerably more than LLMs with larger \gls{refNamePlural}. GPT4o performs almost identically in both the text and vision modalities except in the US CENSUS dataset which on average has 437 cells per table (nearly 3x more compared to the other datasets. This trend can also be seen in the smaller VLMs as most top performing models drop in performance when dealing with US CENSUS \hcts.

\begin{figure}[t]
    \centering
    \includegraphics[width=\linewidth]{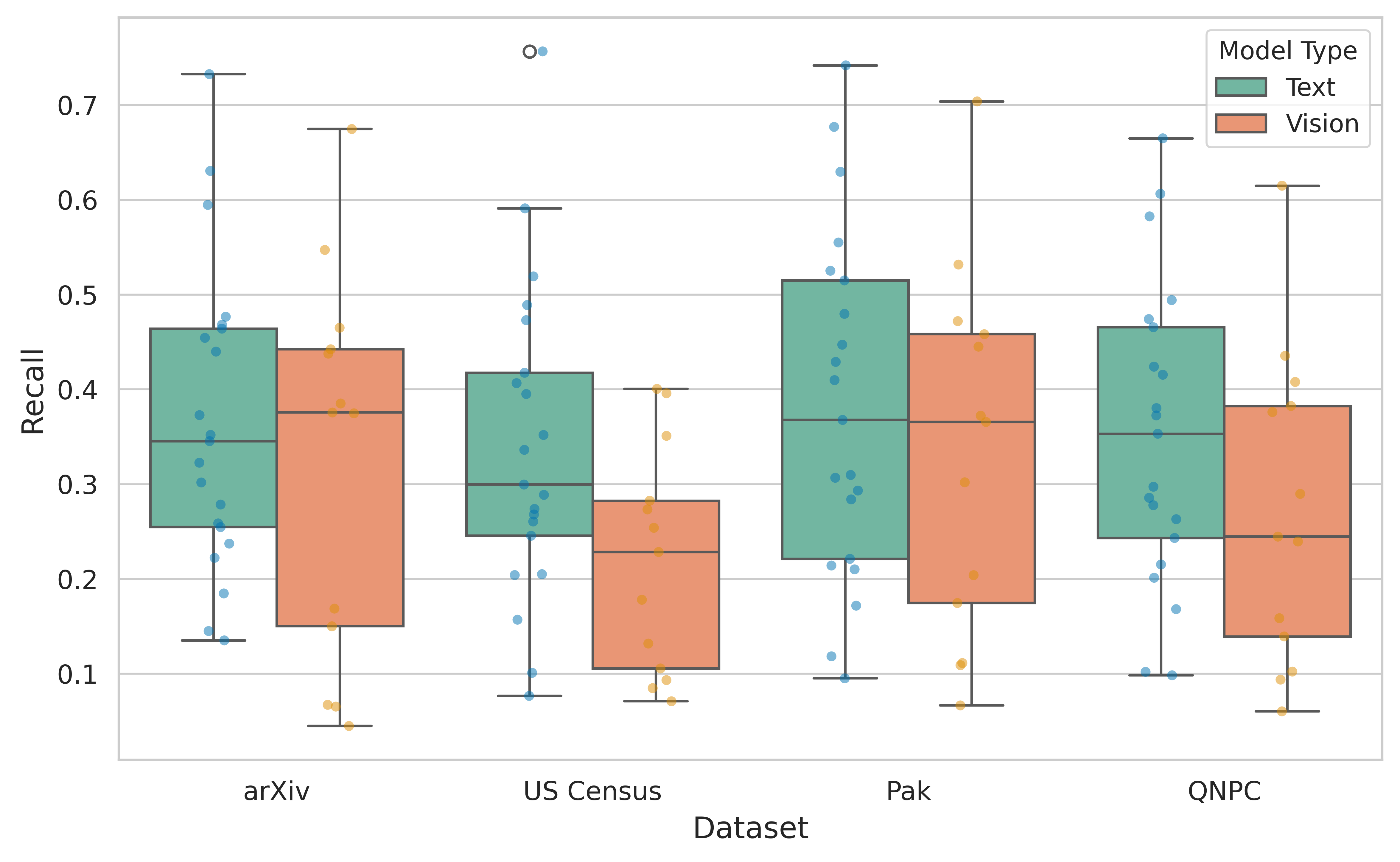}
    \caption{LLMs vs. VLMs on all Real World HCTs — Recall Score.}
    \label{fig:llm_vs_vlm_recall}
\end{figure}

\subsection{Prompting Details \& Experiments}
\label{appendix:additional_experiments_prompting_details}

For certain models (like Gemma-2) that do not accept a system prompt we add the system prompt as a prefix to the user prompt and send it as one prompt with the "user" role.

\subsubsection{Prompts}

Prompts are used with LLMs to give them context and role, and to format the results to allow automatic evaluation.

\begin{tcolorbox}[title=System Prompt for Inference, colback=gray!5, colframe=black!40]
You are a table question answering assistant. You are given a table and a question. Your task is to answer the question based on the information in the table. The table structure may be complex and not a standard relational table, so try to understand the structure of the table when answering the question.

If the question cannot be answered using information from the table, return `No Answer`. Do not provide any explanations, equations, code, or text explaining intermediate steps in figuring out the answer. Return only the final answer itself.

If there are multiple values, separate each value with the ` \texttt{||} ` token.
\end{tcolorbox}

\begin{tcolorbox}[title=User Prompt for Inference]
Answer the question given the input table. Do not provide any explanations, equations, code, or text explaining intermediate steps in figuring out the answer. Return only the final answer itself.\\\\
\# INPUTS: \\
Table: `\{table\}` \\
Question: `\{question\}`\\\\ 
\# OUTPUT:\\
Answer:
\end{tcolorbox}

\subsubsection{0-shot vs. 1-shot prompting}

The majority of the models we used in our experiments can handle context lengths of 8,000+ tokens. This means that it is possible to do 1-shot prompting with a full table-question-answer triplet as the shot. This approach could help leverage in-context learning to improve the model performance. We experimented on a small set of 1,000 queries using Real World \hcts from all four data sources. Our 1-shot prompt consisted of a fixed example for all the questions in the test set. We avoided testing the models given an example question on the same HCT the test question was on as that reflects a very unrealistic scenario. The results are shown in Table~\ref{tab:0shot_1shot_results}. We can see that adding the 1-shot example does not significantly affect the recall score as for gemma-2-9b-it it only increases by 1.3\% and for Llama-3.1-8B-Instruct it actually decreases by less than 1\%. Upon manual inspection, we found that the models were able to follow the output format stated in the prompt (values split by ` \texttt{||} ` and no explanations) in the 0-shot scenario almost perfectly. As the 1-shot example was on a different table, it provided no "learning" for the model to answer the test question it was being asked. Due to these results, our remaining experiments use 0-shot prompting. 

\begin{table}[h]
    \centering
    \caption{0-shot vs 1-shot Prompting - Recall Score}
    \label{tab:0shot_1shot_results}
    \begin{tabular}{lcc}
        \toprule
        \textbf{Model} & \textbf{0-Shot Recall} & \textbf{1-Shot Recall} \\
        \midrule
        gemma-2-9b-it & 0.367 & 0.373 \\
        Llama-3.1-8B-Instruct & 0.247 & 0.239 \\
        \bottomrule
    \end{tabular}
\end{table}


\subsection{Evaluation Metrics}
\label{appendix:additional_experiments_evaluation_metrics}

\begin{figure}[h]
    \centering
    \includegraphics[width=\linewidth]{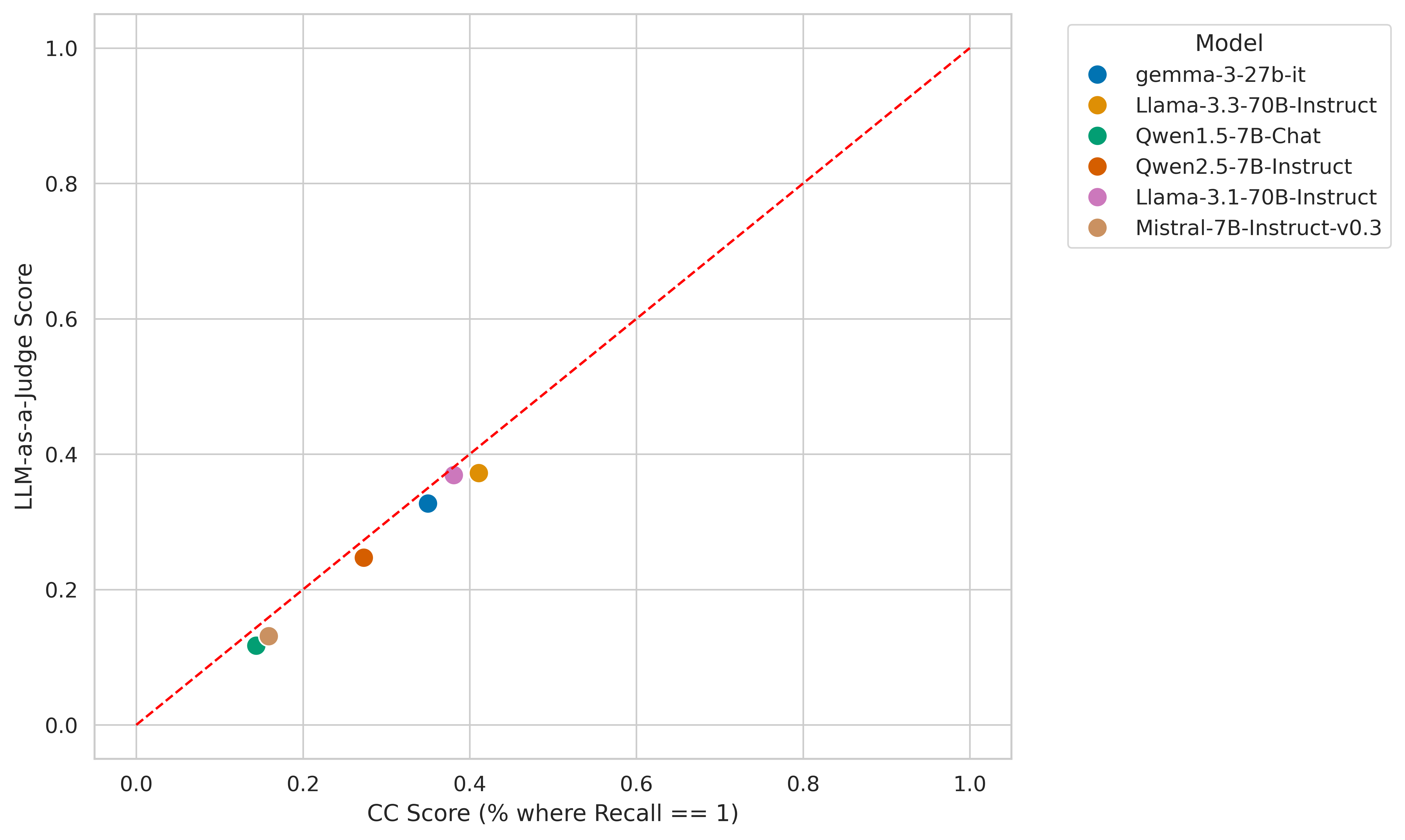}
    \caption{CC Score vs. LLM-as-a-Judge Score on Real World HCTs.}
    \label{fig:cc_vs_lmj_score}
\end{figure}

We experimented using \textit{Meta-Llama-3-70B} as judge model to gauge the correctness of a model's answer. The prompt given to the judge model  outlined a clear set of instructions: (1) minor spelling mistakes or typos in textual values can be ignored, (2) rounding numerical values is fine if done reasonably, (3) question must be answered fully in order to be "correct". We compared this model's "correct" labels to our CC Score (Complete Containment, which we defined as 1 when Recall Score == 1, else 0) on a subset of 500 real-world \hct QA pairs. The comparison is shown in Figure~\ref{fig:cc_vs_lmj_score}. It is evident that the model's judgment and the CC score closely align with each other. Since running an LLM-as-a-Judge is expensive in terms of computational resources, we opt to use Recall, Precision, and CC Score as our evaluation metrics.

\subsection{VLMs vs LLMs}
\label{appendix:additional_experiments_vlm_vs_llm}

We can see from the results in Table~\ref{tab:model_performance} that all VLMs are sensitive to table size. \hcts arXiv and Pak have relatively lower number of cells and number of words on average and vision models tend to do the best on these tables. It is evident that all vision models struggle with US Census tables (which are the largest tables on average with good margin). Gpt-4o scores 39.6\% recall on US Census which is less than majority of medium sized text-only models which are considerably smaller in size. 

\begin{figure}[t]
    \centering
    \includegraphics[width=\linewidth]{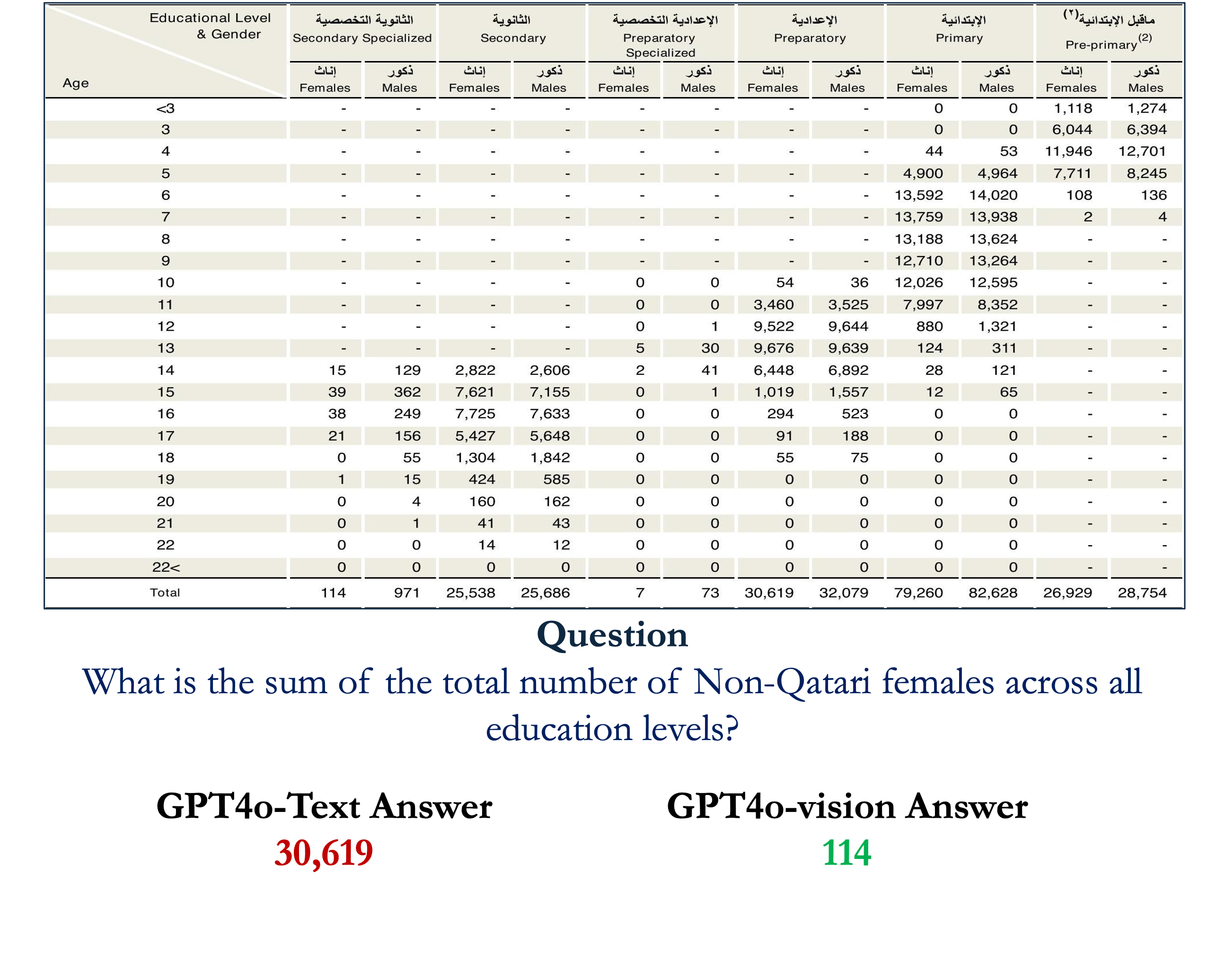}
    \caption{\textbf{gpt-4o-vision} is correct but \textbf{gpt-4o} text-only is wrong.}
    \label{fig:vlm_doing_good}
\end{figure}

\begin{figure}[t]
    \centering
    \includegraphics[width=\linewidth]{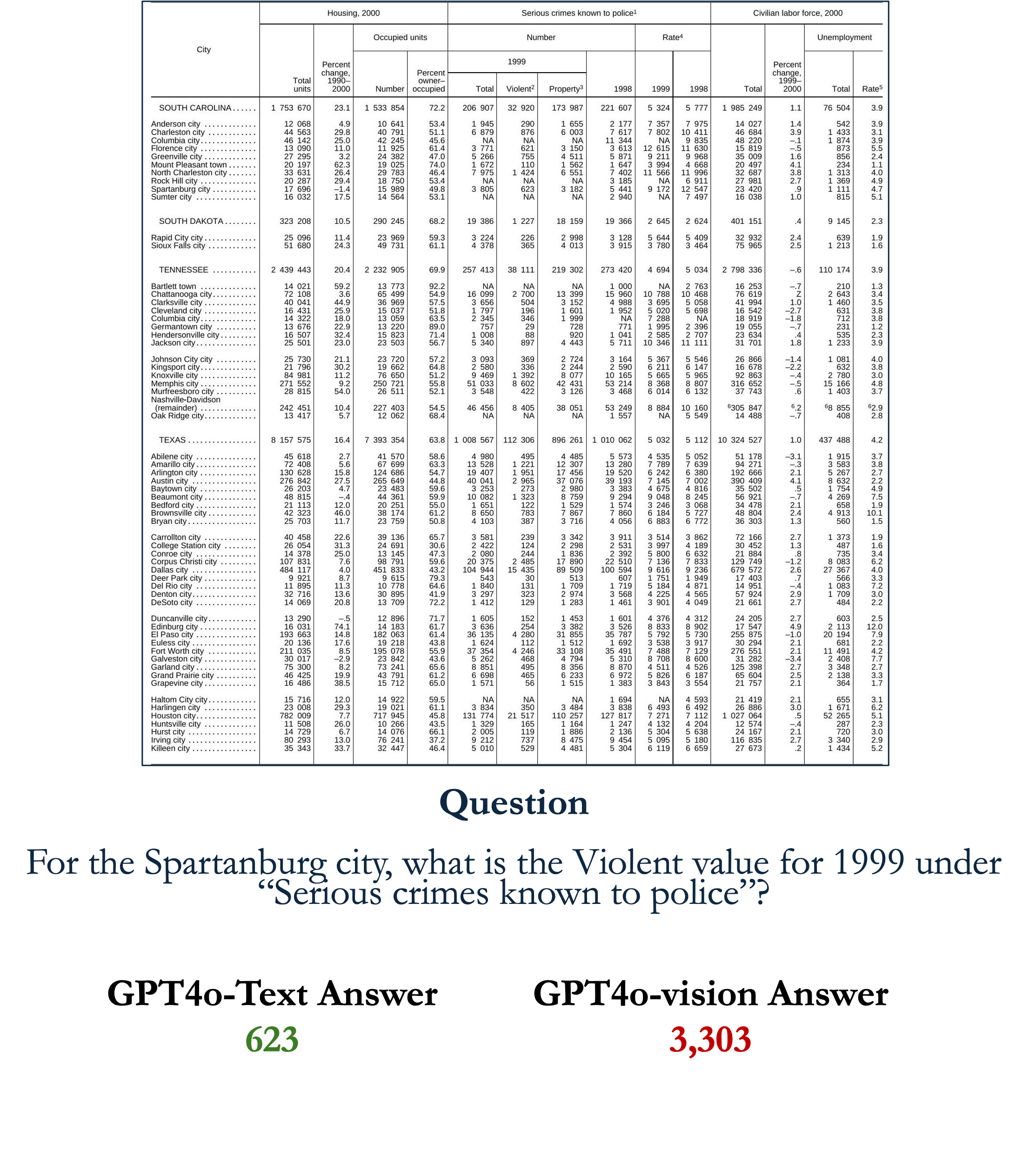}
    \caption{\textbf{gpt-4o-vision} is wrong but \textbf{gpt-4o} text-only is correct.}
    \label{fig:vlm_doing_bad}
\end{figure}

This is further clear by the fact the gpt-4o-vision which outperforms its text-only version (gpt-4o) in all 3 datasets (QNPC, Pak, arXiv) but only scores a recall of 39.6\% for US Census which is 20\% less than that of gpt-4o (text-only) at 59\%. Some examples of cases where gpt-4o gets answers a question correctly in one modality but gets it wrong in the other are provided in Figure~\ref{fig:vlm_doing_good} and Figure~\ref{fig:vlm_doing_bad}.

\subsection{Finetuning Details}
\label{appendix:additional_experiments_finetuning_details}

We fine-tuned 2 models (Meta-Llama-3-8B-Instruct and FanarC-9B) using LLaMA Factory\footnote{https://github.com/hiyouga/LLaMA-Factory} using the following hyperparameters:

\begin{tcolorbox}[title=Finetuning Configuration Summary, colback=gray!5, colframe=black!40]
\textbf{Finetuning Type:} Full-parameter finetuning \\
\textbf{Max Sequence Length:} 4096 tokens \\
\textbf{Precision:} bfloat16 (bf16) \\
\textbf{Training Epochs:} 1 \\
\textbf{Batch Size:} 2 per device \\
\textbf{Gradient Accumulation:} 2 steps \\
\textbf{Learning Rate:} 5e-7 \\
\textbf{Learning Rate Schedule:} Cosine with min LR = 5e-8
\end{tcolorbox}

The dataset itself, with the exact train + validation + test splits used in our finetuning experiments, can be found on HuggingFace\footnote{https://huggingface.co/datasets/qcri-ai/HCTQA}.

\end{document}